\documentclass[11pt, oneside, a4paper]{article}
\usepackage{mcite}
\usepackage{cite}
\usepackage{latexsym}
\usepackage{amssymb}
\usepackage{amsmath}
\usepackage{amsthm}
\usepackage[applemac]{inputenc}
\newcommand{\be}{\begin{eqnarray}}
            \newcommand{\ee}{\end{eqnarray}}
           \newcommand{\eel}[1]{\label{#1}\end{eqnarray}}
\newcommand{\e}[1]{\label{eq:#1}\end{eqnarray}}
     \newcommand{\eg}{{\em e.g.\ }}
            \newcommand{\ie}{{\em i.e.\ }}
            \newcommand{\ga}{{\gamma}}
 
            \newcommand{\la}{{\lambda}}

\newcommand{\del}{{\delta}}

 \newcommand{\cP}{{\cal{P}}}

\newcommand{\cW}{{\cal{W}}}
\newcommand{\cB}{{\cal{B}}}

           \newcommand{\ra}{{\rightarrow}}

\newcommand{\cL}{{\cal L}}

\newcommand{\cC}{{\cal C}}

            \newcommand{\beq}{\begin{quote}}
            \newcommand{\eq}{\end{quote}}
            
            \newcommand{\al}{\alpha}
            \newcommand{\ben}{\begin{enumerate}}
            \newcommand{\een}{\end{enumerate}}
            \newcommand{\bit}{\begin{itemize}}
            \newcommand{\ei}{\end{itemize}}
        \newcommand{\nn}{\nonumber}
            \newcommand{\rl}[1]{(\ref{eq:#1})}
            \newcommand{\edfl}[1]{\Label{#1}\end{df}}

\newcommand{\cF}{{\cal F}}
\newcommand{\cR}{{\cal R}}

\newcommand{\dif}{{\partial}}
\newcommand{\half}{\frac{1}{2}}

\begin{document}
\begin{titlepage}
%\today
\vspace*{20mm}
\begin{center}
{\LARGE\bf Lagrangian conformal \\higher spin theory}\end{center}
\vspace*{3 mm}
\begin{center}
\vspace*{3 mm}

\begin{center}
Robert Marnelius\footnote{E-mail: 
tferm@fy.chalmers.se}
 \\ \vspace*{7 mm} {\sl
Department of Fundamental Physics\\ Chalmers University of Technology\\
S-412 96  G\"{o}teborg, Sweden}\end{center}
\vspace*{22 mm}
\begin{abstract}
In a previous paper conformal gravity  was derived by means of a precise action principle on the hypercone in the conformal space. Here it is shown that the same technique used to construct conformal spin two theory as represented by linear conformal gravity may also be used for the construction of conformal higher spin theories. The basic ingredients in these constructions are gauge invariant field strengths (curvatures). In fact, their very existence as manifestly conformal fields requires the present conformal theory. The general form of the actions for the free theories on the hypercone are given. In spacetime these actions are shown to be expressed in terms of squares of generalized Weyl tensors. Conformal spin three and four are calculated explicitly. The theories are proposed to be equivalent to the free  theories given by Fradkin and Linetsky. Since the equations are of order $2s$ a consistent quantization is difficult to achieve but might exist.   Interactions are expected to exist and is the motivation behind this approach.  
\end{abstract}\end{center}\end{titlepage}
\tableofcontents\newpage

\setcounter{equation}{0}
\section{Introduction}
String theory is expected to be a consistent way to describe higher spin states and their interactions.
Since Witten's open string field theory \cite{Witten:1985cc} shows that it  can be formulated as a field theory although of a generalized type,  it is natural to expect that there  is also a way interacting higher spin theory can be formulated within a more conventional quantum field theory framework.
  However, if one tries to set up consistent quantum field theories for higher spins from scratch one encounters severe difficulties as the vast literature on the subject shows. (For reviews see \eg \cite{Vasiliev:2001ur,Sorokin:2004ie,Bouatta:2004kk}.)  Thirty years ago Fronsdal succeeded in writing down Lagrangians for free higher spin fields \cite{Fronsdal:1978rb,Fang:1978wz}. However, no interaction theory has been possible to set up in a standard local way.  If one wants to be successful in this approach it is therefore probable  that one has to consider  generalized or unconventional field theories and not  standard  local field theories with at most second order derivatives.  
  In fact,  Witten's open string field theory   has the features of a nonlocal quantum field theory   \cite{Eliezer:1989cr}.
It  contains also  an infinite number of higher spin states.
Now unconventional interacting higher spin field theories with an infinite number of different higher spins need not necessarily coincide with the string theory (see \eg the proposal in \cite{Craigie:1976ez}).  There is also
a wide spread view that the massive higher spin states like the masses in string theory should be generated by some spontaneous symmetry breaking mechanism. One approach could be to find how the unbroken phase of string theory looks like. (For specific approaches to the tensionless string, see \eg  \cite{Isberg:1993av,Savvidy:2003co}).  
  Another more general approach could be to continue the effort to find an interacting massless higher spin theory, and/or to  consider a conformally invariant higher spin theory as a candidate, which is the approach of the present paper.

Free higher spin fields and the equations and conditions they have to satisfy may be derived in many ways. Usually one considers a group theoretical derivation (see \eg \cite{Bekaert:2002dt,deMedeiros:2002nx,Bekaert:2003zq,Bandos:2005mb,Bekaert:2006py,Bekaert:2006ix}). One may also derive their possible forms by viewing the fields as wave functions in the quantum theory of a relativistic spinning particle model. This is the procedure I prefer myself. In fact, all free representations should be possible to derive by means of the procedure in \cite{Marnelius:1990de}, which also is a procedure to derive all possible spinning particle models. Notice that in this way one always obtain wave functions  that are gauge invariant which then are to be identified with field strengths (or curvatures). Such gauge invariant objects are also becoming increasingly more popular in the higher spin theories in the literature \cite{Francia:2002aa,Francia:2002pt, Francia:2005bu,Bekaert:2003az ,deMedeiros:2003dc,Bekaert:2005ka,Bandos:2005mb,Bekaert:2006ix}. Anyway, to find consistent sets of linear equations for higher spin fields are not too difficult. It is a much more difficult  to find a Lagrangian formulation for these equations.

In most of the literature on higher spin fields one uses a symmetric field of rank $s$ to describe spin $s$ for integer $s$. This is the case in this paper as well for $d=4$. (However, in dimensions $d>4$ fields with a more complex index structure enter.) The symmetric field is a gauge field generalizing the vector potential for spin one and the fluctuating metric for spin two. A gauge invariance is therefore necessary to impose. Fronsdal's Lagrangians \cite{Fronsdal:1978rb,Fang:1978wz} are expressed in terms of such fields and  have a restricted gauge invariance. However, there are natural gauge invariant field strengths or curvatures with larger gauge invariances. There are at least two different representations of  these gauge invariant objects: One is the deWit-Freedman curvature \cite{deWit:1979pe} and one is the Weinberg field strength \cite{Weinberg:1965rz}. They are equivalent since one representation  may be expressed in terms of the other (see appendix A). Which one to choose is therefore a matter of taste.   In the literature the deWit-Freedman curvature is the most commonly used one. However, here I choose the Weinberg representations as a starting point since they appear naturally in my approach. 

In the following I will present a Lagrangian approach to conformal higher spin theories that I  believe should allow for interactions. Although I only will describe the resulting properties of free fields with integer spins, the  properties of these fields contain an   unconventional feature already from the start. Due to the way I construct these models I am naturally led to theories where the fields satisfy higher order equations. Although this makes it easy to see that they are conformally invariant the concept of spins is \eg blurred and should therefore only be taken at an abstract level. However, a more serious objection is that fields satisfying higher order equations are difficult to turn into consistent unitary quantum theories. This problem will not be resolved but at the end I will point out my view on the problem which allows for some hope. The Lagrangians which I eventually arrive at are of the form
\be
&&\cL\propto C^2,
\e{01}
where $C$ is a generalized Weyl tensor of rank $2s$ expressed in terms a symmetric gauge field  of rank $s$. $C$ also contains $s$ derivatives on the gauge field. The equations of motion from \rl{01} has the form
\be
&&\dif^sC=0,
\e{02}
which means that the  gauge field satisfies an equation of order $2s$. The form \rl{01} for free integer spins was also proposed by Fradkin and Linetsky in $d=4$ \cite{Fradkin:1989md,Fradkin:1990ps} and was shown to be related to the Lagrangian given by Fradkin and Tseytlin  in  \cite{Fradkin:1985am}. In section 6 I argue that their approach is  related to the present one. This then strengthen my belief that there is an interacting theory at least in $d=4$, since Fradkin and Linetsky also proposed the existence of  cubic interactions in \cite{Fradkin:1989md,Fradkin:1990ps}. Furthermore, since they also have half-integer spins in their formulation this should also be  possible to have within  the present approach. (The free half-odd integer spin theory is  straight-forward to work out.)

In a previous paper \cite{Arvidsson:2006} the manifestly conformal formalism of Dirac \cite{Dirac:1936} was further developed. A precise action principle was given which also is applicable to gauge theories. The paper was divided into three parts: In part one the manifestly conformal particle in interaction with external symmetric fields of arbitrary ranks was shown to be consistent provided these external fields satisfy certain conditions. This indicated that interactions are possible provided these conditions are met. In part two the first quantization of the manifestly conformal spinning particle model in \cite{Martensson:1992ax} was set up and studied in details for spin one and two. The resulting wave functions were then  candidates for a manifestly conformal field theory. In part three the conditions for a manifestly conformally invariant Lagrangian field theory were given. The wave functions from part two had then in general to be modified  in order to become allowed manifestly conformal  fields. Although the conditions from part two for spin one could be retained, spin two was shown to require a modification. This modification allowed then for a manifestly conformal Lagrangian formulation for conformal gravity. 

In this paper I extend the treatment in \cite{Arvidsson:2006} for free spin two fields to higher spins. It turns out that the corresponding modifications of the gauge invariant field strengths  found necessary for spin two may be generalized to the field strengths for arbitrary integer spins in arbitrary even dimensions. The crucial conditions are that the gauge invariant field strengths exist as manifestly conformal fields. In the presentation given here I will, however, reverse the actual derivations. I will start with a spacetime treatment in $d=4$ (part I) and then turn to the manifest formulation for $d=4$ (part II) after which the generalization to arbitrary even dimensions $d$ is given (part III). Since I in general do not give general proofs,  the general properties are stated as proposals. Most of the proposals are supported by explicit calculations for spins 2, 3 and 4.

For other recent approaches to conformal higher spin theories, see \eg \cite{Vasiliev:2001zy,Shaynkman:2004vu,Vasiliev:2007yc}.

\part{The spacetime formulation}
   \setcounter{equation}{0}
\section{The structure of the conformal higher spin\\ models}
The forms and the equations determining free higher spin fields  may be derived in many ways. Usually one considers a group theoretic derivation directly within the field formulation (see \eg \cite{Bekaert:2002dt,deMedeiros:2002nx,Bekaert:2003zq,Bandos:2005mb,Bekaert:2006py,Bekaert:2006ix}). Personally I  prefer to derive the forms and properties of these field representations by viewing them as wave functions in a quantum theory of a spinning particle model.  In this way one will always find gauge invariant fields. Their expressions in terms of gauge fields or gauge potentials are determined through the equations they satisfy. Anyway this is a precise way to determine a representation. The problem is then to find what relativistic spinning particle models are possible. In \cite{Marnelius:1990de} a general procedure is given for the derivation of in principle all possible spinning particle models.  (In \cite{Edgren:2005gq,Edgren:2006un} it is \eg proved by this method  that Wigner's continuous spin representation follows from a definite {\em higher order} particle model.)     The most natural particle model yielding  a massless representation for arbitrary spins after quantization  is  the $O(2s)$ extended supersymmetric particle model, where $s$ is the spin.    (Its connection to higher spins was independently proposed  in \cite{Gershun:1979fb,Howe:1988ft,Marnelius:1988ab}.) The quantization performed in  \cite{Marnelius:1988ab}  yields here in a   natural  way a representation which seems to be  the Weinberg representation \cite{Weinberg:1965rz}.  For integer spins $s$ and in spacetime dimension $d=4$ the field strengths (or the wave functions) have $2s$ indices,
\be
&&F_{\mu_1\nu_1\mu_2\nu_2\cdots\mu_s\nu_s}(x),
\e{1}
with antisymmetry in the indices $(\mu_n,\nu_n)$ for all $n$ and symmetry under interchange of the pairs $(\mu_n,\nu_n)$ and $(\mu_m,\nu_m)$ for any $m$ and $n$. In addition, $F_{}$ satisfies
\be
&&F_{\mu_1\nu_1\mu_2\nu_2\cdots\mu_s\nu_s}(x)+ cycle(\mu_1,\nu_1,\mu_2)=0,\nn\\
&&\dif_{\rho}F_{\mu_1\nu_1\mu_2\nu_2\cdots\mu_s\nu_s}(x)+ cycle(\rho, \mu_1,\nu_1)=0,
\e{2}
and the traceless condition   
\be
&&\eta^{\mu_1\mu_2}F_{\mu_1\nu_1\mu_2\nu_2\cdots\mu_s\nu_s}(x)=0.
\e{3}
$F$ also satisfies the condition
\be
&&\dif^{\mu_1}F_{\mu_1\nu_1\mu_2\nu_2\cdots\mu_s\nu_s}(x)=0,
\e{31}
which together with the last relation in \rl{2} imply 
\be
&&\Box F_{\mu_1\nu_1\mu_2\nu_2\cdots\mu_s\nu_s}(x)=0.
\e{32}
This shows that the representation is massless. (For $s=0$ $F$ is a scalar and in this case \rl{32} is the only equation.)

The conditions \rl{2} may be solved in terms of a gauge field $\phi$,
\be
&&F_{\mu_1\nu_1\mu_2\nu_2\cdots\mu_s\nu_s}(x)=\dif_{[\mu_s}\cdots\dif_{[\mu_2}\dif_{[\mu_1}\phi_{\nu_1]\nu_2]\cdots\nu_s]}(x),
\e{4}
where the gauge field $\phi_{}$ is totally symmetric. With the notation on the right-hand side I mean antisymmetrization in ($\mu_k, \nu_k$) for $k=1,\ldots,s$. For instance, for spin one ($s=1$) we have the Maxwell field
\be
 &&F_{\mu\nu}=\dif_{[\mu}\phi_{\nu]}\equiv \dif_{\mu}\phi_{\nu}-\dif_{\nu}\phi_{\mu}.
\e{41}
 For  spin two ($s=2$)
\be
&&R_{\mu_1\nu_1\mu_2\nu_2}(x)=\half F_{\mu_1\nu_1\mu_2\nu_2}(x)
\e{6} 
is the linearized Riemann tensor.
  The field strength $F_{}$ in \rl{4} is obviously invariant under the gauge transformations
\be
&&\phi_{\nu_1\nu_2\cdots\nu_s}(x)\quad\ra\quad \phi_{\nu_1\nu_2\cdots\nu_s}(x)+\dif_{(\nu_1}\varepsilon_{\nu_2\cdots\nu_s)}(x),
\e{5}
where $\varepsilon_{\cdots}$ is an arbitrary symmetric function. In the following the representation given by \rl{3}-\rl{32} and \rl{4} will be referred to as the Weinberg representation  (see \cite{Weinberg:1965rz}). In appendix A it is shown that this representation is equivalent to the generalized curvature representation  by deWit and Freedman \cite{deWit:1979pe} (without the conditions \rl{31} and \rl{32}).

Although it is always possible to determine free fields and their equations for higher spins by various methods, it is not  easy to construct a Lagrangian theory for these equations as the history of the subject tells us.     Requiring second order equations for the symmetric gauge field Fronsdal has constructed consistent local Lagrangians  \cite{Fronsdal:1978rb,Fang:1978wz} with restricted gauge invariances. (Their forms have recently been extended to allow for full gauge invariance \cite{Francia:2002aa,Francia:2002pt, Francia:2005bu, Bekaert:2003az ,Bekaert:2005ka,Bandos:2005mb,Bekaert:2006ix}.)

When one turns to conformally invariant Lagrangian theories of the type considered in \cite{Arvidsson:2006}  it is very unnatural to impose the traceless condition \rl{3} for $s\geq2$  as was shown for $s=2$ in \cite{Arvidsson:2006} within the manifestly conformal formulation. Fortunately the procedure to modify the above equations for $s=2$ given in \cite{Arvidsson:2006} may  be generalized to $s>2$.  This means for  the spacetime treatment  that  I propose   the use of \rl{1} and \rl{4} (implied by \rl{2}) and no more for any $s$.  In place of \rl{3} I impose a  new gauge invariance which is not valid for the above conditions, namely the invariance under the generalized Weyl transformations given by Fradkin and Tseytlin in \cite{Fradkin:1985am}. These transformations  have the form
\be
&&\phi_{\mu_1\mu_2\cdots\mu_s}(x)\;\longrightarrow\;\phi_{\mu_1\mu_2\cdots\mu_s}(x)+\eta_{(\mu_1\mu_2}\la_{\mu_3\cdots\mu_s)}(x),
\e{61}
where $\eta$ is the Minkowski metric and $\la$ an arbitrary symmetric function.
 Since I also remove the equations \rl{31} and \rl{32}, $F$ will no longer strictly represent a spin $s$ particle. Still, but somewhat inappropriately, I will refer to this representation as the spin $s$ representation. The free field Lagrangians  I propose  are then of the form
\be
&&\cL\propto F_{}^2+\al(F'_{})^2+\beta(F''_{})^2+\cdots,
\e{7}
where $F_{}$ is the field strength \rl{2} given in the form \rl{4}, and where primes denotes traces of $F_{}$. Thus, the Lagrangian \rl{7} contains all possible traces of $F_{}$. I expect that the values of the real constants $\al, \beta, \cdots$ are uniquely determined by the invariance under the generalized Weyl transformations \rl{61}. For $s=1$ this describes the free Maxwell theory, and for $s=2$ it describes linear conformal gravity,
\be
&&\cL\propto R_{}^2-{4\over d-2}(R'_{})^2+{2\over(d-1)(d-2)}(R'')^2,
\e{8}
where $R'$ and $R''$ are the Ricci tensor and curvature scalar respectively as was shown in \cite{Arvidsson:2006}. I also propose  that  the Weyl invariant form of \rl{7} for any integer $s$ may be written as
\be
&&\cL\propto C^2,
\e{9}
where $C$ is a generalized Weyl tensor which is traceless and of rank $2s$. These proposals will be explicitly verified for the cases $s\leq4$.

 \setcounter{equation}{0}
 \section{The general procedure of construction}
 Let me define the following tensor which has the same rank ($2s$) and index symmetry  as $F$ in \rl{1}
 \be
 &&W\equiv F+\sum_k\al_kB_k,
 \e{101}
 where $\al_k$ are real numbers, and where $B_k$ is of the form
 \be
 &&B_k=(F_k\eta\cdots\eta)_{sym},
 \e{102}
 where in turn $F_k$ is one of the traces of $F$ and $\eta$ the metric tensor. The number of $\eta$-factors depends on the order of the trace $F_k$. By $sym$ is meant that the expression is symmetrized to the same index symmetry as $F$. Furthermore, I choose  $B_k$  to be normalized such that the following relation is true,
 \be
 &&B_k\cdot F=(F_k\eta\cdots\eta)_{sym}\cdot F=(F_k\eta\cdots\eta)\cdot F=F_k^2.
 \e{103}
 $W$ is required to contain all possible expressions of the form \rl{102} satisfying \rl{103}. 
 It is therefore crucial that one first determines all possible traces of $F$ as well as the relations between them, since this knowledge is necessary in order to determine {all} possible different $B_k$'s from \rl{102} and \rl{103}. As will be shown in the next section, for $s\geq3$ the number of different $B_k$ is not the same as the number of different traces of $F$. Due to the many different ways one may arrive at a higher order trace of $F$ one will often find that $B_k\neq B_l$ although $F_k=F_l$.

 All Lagrangians to be considered in part I of this paper are defined  in terms of the tensor $W$ by
  \be
 &&\cL\equiv W\cdot F=F^2+\sum_k\al_k F_k^2,
 \e{104}
 where the equality follows from \rl{101} and \rl{103}.  This relates the parameters in $W$ to the parameters in the Lagrangian (cf \rl{7}). 
 \beq
 \begin{tabular}{|p{114mm}|}
 \hline
  {\bf Theorem 1}: The equations of motion from this Lagrangian is $\dif^sW=0$ or explicitly
 \be
 &&\dif^{\mu_1}\dif^{\mu_2}\cdots\dif^{\mu_s}W_{\mu_1\nu_1\mu_2\nu_2\cdots\mu_s\nu_s}=0.
 \e{105}\\
   \hline
 \end{tabular}
 \eq
Proof:
 \be
 &&\del \cL=2\biggl(F\cdot \del F+\sum_k\al_k F_k\cdot\del F_k\biggr)=\nn\\
 &&=2\biggl( F+\sum_k\al_k B_k\biggr)\cdot\del F=\nn\\
 &&=2^{s+1}W\cdot(\dif^s\del\phi)=2^{s+1}(-1)^s(\dif^sW)\cdot\del\phi+\dif\cdot(\;\;\;)_\bullet
 \e{106}

 Notice that if one adds a term $J\cdot\phi$ to $\cL$, where $J$ is  an external current,  then the equations are
  \be
&&\dif^{\mu_1}\dif^{\mu_2}\cdots\dif^{\mu_s}W_{\mu_1\nu_1\mu_2\nu_2\cdots\mu_s\nu_s}=J_{\nu_1\nu_2\cdots\nu_s},
 \e{107}
 where consistency requires that the symmetric current $J$ is conserved. (This is also required by the gauge invariance under \rl{5}.) Notice also that \rl{105} and \rl{107} imply that the symmetric gauge field satisfies an equation of order $2s$. 
 
Due to the remark after \rl{103} the tensor $W$ determines the Lagrangian $\cL$ but not the other way around for $s\geq3$. 
 
 \beq
 \begin{tabular}{|p{114mm}|}
 \hline
  {\bf Theorem 2}: There is a unique choice of the parameters $\al_k$ in  \rl{101}  that makes $W$ traceless for any integer $s$.\\
 \hline
 \end{tabular}
 \eq
 For these values of $\al_k$   $W$ will be denoted $C$ in the following.\\
Proof: There is a general theorem that an arbitrary tensor ($F$) may be decomposed into its traceless part ($C$) and terms of the form $(B_k)_{\bullet}$

  \beq
 \begin{tabular}{|p{114mm}|}
 \hline
  {\bf Theorem 3}:  If $\al_k$ is chosen such that $W=C$ (\ie traceless) then the Lagrangian \rl{104} may be written as
 \be
 &&\cL=C^2.
 \e{108}\\
    \hline
 \end{tabular}
 \eq
 Proof: Since $C$ is traceless  the following equality is  valid:
 \be
 &&B_k\cdot C=0,\quad {\rm for\ all}\;\; k. 
 \e{109}
 Hence,
 \be
 &&C\cdot C=C\cdot W=C\cdot F.
 \e{110}
 The equivalence between \rl{104} and \rl{108} follows $_\bullet$
 
 This proof and theorem 1 implies
   \beq
 \begin{tabular}{|p{114mm}|}
 \hline
  {\bf Theorem 4}:  The equations of motion from the Lagrangian \rl{108} is
 \be
 &&\dif^{\mu_1}\dif^{\mu_2}\cdots\dif^{\mu_s}C_{\mu_1\nu_1\mu_2\nu_2\cdots\mu_s\nu_s}=0.
 \e{111}\\
    \hline
 \end{tabular}
 \eq
 
  \beq
   \begin{tabular}{|p{114mm}|}
 \hline
  {\bf  Proposal 1}:  There is a unique choice of the parameters $\al_k$ in $W$  given by \rl{101} that makes $W$ invariant under the generalized Weyl transformations defined in \rl{61}. For these values $W=C$, the traceless tensor in theorem 2.\\
  \hline
 \end{tabular}
 \eq 
  This will be explicitly proved for $s=1, 2, 3, 4$ in the following sections.
Notice that since the tensor $W$ starts with $F$, which is not Weyl invariant, the Weyl invariant tensor must involve a reduced number of components as compared to $F$. Tracelessness is therefore a natural property.  For another  general argument see section 6. Weyl tensors of the form proposed here are not entirely new. In fact, Damour and Deser \cite{Damour:1987vm} construct $C$ for spin 3 by essentially the same method as used here.

In the following $C$ will be called the generalized Weyl tensor.  It may either be defined  through proposal 1 or theorem 2. 

\beq
  \begin{tabular}{|p{114mm}|}
 \hline
  {\bf Proposal 2}:  For $W=C$ the theory given by the Lagrangian \rl{104}, or equivalently by \rl{108}, is conformally invariant in $d=4$.\\
 \hline
 \end{tabular}
 \eq
 \beq
\begin{tabular}{|p{110mm}|}
 \hline
  {\bf Proposal 3}:  If there exists a conformally invariant Lagrangian in the form \rl{104} then there also exists a Weyl tensor $C$  of the form \rl{101} in terms of which the Lagrangian may be written as  \rl{108}.\\
 \hline
 \end{tabular}
 \eq
 These statements are indirectly proved for $s=1, 2, 3, 4$ through the manifestly conformal invariant formulation later.

 \setcounter{equation}{0}
 \section{Generalized Weyl tensors $C$ for $s=1,2,3,4$ from the traceless condition}
 \subsection{$C$ for $s=1$}
 For $s=1$ the field strength $F$ in \rl{4}  is traceless in itself and there are no parameters  to be determined. Hence, one finds from \rl{101} and \rl{4},
$W_{\mu\nu}=C_{\mu\nu}=F_{\mu\nu}$, \ie just the Maxwell field \rl{41}.
 
  \subsection{$C$ for $s=2$}
  For $s=2$ the basic field strength  $F$ in \rl{4} is given by
  \be
  &&F_{\mu_1\nu_1\mu_2\nu_2}=\dif_{[\mu_1}\dif_{[\mu_2}\phi_{\nu_2]\nu_1]}.
  \e{202}
  This expression is directly related to the linearized Riemann tensor,
  \be
  &&R_{\mu_1\nu_1\mu_2\nu_2}=\half F_{\mu_1\nu_1\mu_2\nu_2}.
  \e{203}
  The symmetric gauge field $\phi$ is then connected to the metric tensor $g$ through
  \be
  &&g_{\nu_1\nu_2}=\eta_{\nu_1\nu_2}+\phi_{\nu_1\nu_2}.
  \e{204}
  The traces of $F$ may therefore be identified with the linearized Ricci and curvature tensors:
  \be
  &R_{\mu_1\mu_2}\equiv& \eta^{\nu_1\nu_2}R_{\mu_1\nu_1\mu_2\nu_2}=\half \eta^{\nu_1\nu_2}F_{\mu_1\nu_1\mu_2\nu_2},\nn\\
  &R\equiv& \eta^{\mu_1\mu_2}R_{\mu_1\mu_2}=   \eta^{\mu_1\mu_2} \eta^{\nu_1\nu_2}R_{\mu_1\nu_1\mu_2\nu_2}=\nn\\&&\half \eta^{\mu_1\mu_2} \eta^{\nu_1\nu_2}F_{\mu_1\nu_1\mu_2\nu_2}.
  \e{205}
  I turn now to the general procedure in the previous section. The ansatz for the general Weyl tensor \rl{101} yields here:
  \be
&&W_{\mu_1\nu_1\mu_2\nu_2}=R_{\mu_1\nu_1\mu_2\nu_2}+\al_1B_{1\;\mu_1\nu_1\mu_2\nu_2}+\al_2B_{2\;\mu_1\nu_1\mu_2\nu_2}.
\e{206}
where
\be
&&B_{1\;\mu_1\nu_1\mu_2\nu_2}\equiv(R_{\mu_1\mu_2}\eta_{\nu_1\nu_2})_{sym}={1\over4}R_{[\mu_1[\mu_2}\eta_{\nu_2]\nu_1]},\nn\\
&&B_{2\;\mu_1\nu_1\mu_2\nu_2}\equiv(R\eta_{\mu_1\mu_2}\eta_{\nu_1\nu_2})_{sym}={1\over4}R\;\eta_{[\mu_1[\mu_2}\eta_{\nu_2]\nu_1]}.
\e{207}
One may easily confirm the validity of the normalization \rl{103}.
The trace of the ansatz \rl{206} becomes then
\be
&&\eta^{\nu_1\nu_2}W_{\mu_1\nu_1\mu_2\nu_2}=c_0R_{\mu_1\mu_2}+c_1
\eta_{\mu_1\mu_2}R,
\e{208}
where
\be
&&c_0=1+{1\over 4}\al_1(d-2),\quad c_1={1\over4}\al_1+\half\al_2(d-1).
\e{2081}
Hence, $W$ is traceless if
\be
&&\al_1=-{4\over d-2},\quad \al_2={2\over(d-1)(d-2)}.
\e{209}
With these values of $\al_1$ and $\al_2$ the $W$-tensor \rl{206} is nothing but the linearized form of the standard Weyl tensor $C$. For these values the Lagrangian may therefore be written as
\be
&&\cL=C^{\mu_1\nu_1\mu_2\nu_2}C_{\mu_1\nu_1\mu_2\nu_2},
\e{210}
which equivalently may be written as
\be
&&\cL=R^{\mu_1\nu_1\mu_2\nu_2}R_{\mu_1\nu_1\mu_2\nu_2}+\al_1R^{\mu_1\mu_2}R_{\mu_1\mu_2}+\al_2R^2
\e{211}
with the values \rl{209} inserted. This is in agreement with \rl{8}, and it  is linearized conformal gravity which is a conformally invariant theory in $d=4$. Its conformal invariance  was  also proved directly  within the manifest formulation in \cite{Arvidsson:2006}, a proof which I partly recapitulate in subsection 10.2.

 \subsection{$C$ for $s=3$}
 For $s=3$ the basic field strength \rl{4} is given by
 \be
&&F_{\mu_1\nu_1\mu_2\nu_2\mu_3\nu_3}=\dif_{[\mu_1}\dif_{[\mu_2}\dif_{[\mu_3}\phi_{\nu_3]\nu_2]\nu_1]},
 \e{212}
 where $\phi$ is totally symmetric. I find the following traces
 \be
 &&F'_{\mu_1\mu_2\mu_3\nu_3}\equiv \eta^{\nu_1\nu_2}F_{\mu_1\nu_1\mu_2\nu_2\mu_3\nu_3},\nn\\
 &&F''_{\mu_3\nu_3}\equiv\half\eta^{\mu_1\mu_2}F'_{\mu_1\mu_2\mu_3\nu_3}=\half  \eta^{\mu_1\mu_2} \eta^{\nu_1\nu_2}F_{\mu_1\nu_1\mu_2\nu_2\mu_3\nu_3},
 \e{213}
 where $F'$ is symmetric in $\mu_1$ and $\mu_2$, and antisymmetric in $\mu_3$ and $\nu_3$. $F''$ is antisymmetric. A different way to arrive at $F''$ is
 \be
 &&F''_{\mu_2\nu_3}=\eta^{\mu_1\mu_3}F'_{\mu_1\mu_2\mu_3\nu_3}= \eta^{\nu_1\nu_2}\eta^{\mu_1\mu_3}F_{\mu_1\nu_1\mu_2\nu_2\mu_3\nu_3}.
 \e{214}
 That there are two different ways to obtain $F''$ implies that the tensor $W$ is not uniquely determined by the Lagrangian \rl{104} as is shown in \rl{224} below. 
 
 One may notice that the cyclicity properties \rl{2} of the field strength \rl{212} imply the following property of $F'$:
 \be
 &&F'_{\mu_1[\mu_2\mu_3]\nu_3}=-F'_{\mu_1\nu_3\mu_2\mu_3},
 \e{215}
 a property that will be used in the manipulations below. 
 
 The ansatz $W$ in \rl{101} for the generalized Weyl tensor $C$ is here
 \be
 &W_{\mu_1\nu_1\mu_2\nu_2\mu_3\nu_3}=&F_{\mu_1\nu_1\mu_2\nu_2\mu_3\nu_3}+\al_1B_{1\;\mu_1\nu_1\mu_2\nu_2\mu_3\nu_3}+\nn\\ &&+\al_2B_{2\;\mu_1\nu_1\mu_2\nu_2\mu_3\nu_3}+\al_3B_{3\;\mu_1\nu_1\mu_2\nu_2\mu_3\nu_3},
 \e{216}
 where
 \be
 &B_{1\;\mu_1\nu_1\mu_2\nu_2\mu_3\nu_3}\equiv&(F'_{\mu_1\mu_2\mu_3\nu_3}\eta_{\nu_1\nu_2})_{sym}={1\over4}F'_{[\mu_1[\mu_2\mu_3\nu_3}\eta_{\nu_2]\nu_1]}+\nn\\&&+{1\over4}F'_{[\mu_2[\mu_3\mu_1\nu_1}\eta_{\nu_3]\nu_2]}+{1\over4}F'_{[\mu_3[\mu_1\mu_2\nu_2}\eta_{\nu_1]\nu_3]},
\e{217}
 \be
 &B_{2\;\mu_1\nu_1\mu_2\nu_2\mu_3\nu_3}\equiv&(F''_{\mu_3\nu_3}\eta_{\mu_1\mu_2}\eta_{\nu_1\nu_2})_{sym}={1\over8}\eta_{[\mu_1[\mu_2}\eta_{\nu_2]\nu_1]}F''_{\mu_3\nu_3}+\nn\\&&+{1\over8}\eta_{[\mu_1[\mu_3}\eta_{\nu_3]\nu_1]}F''_{\mu_2\nu_2}+{1\over8}\eta_{[\mu_2[\mu_3}\eta_{\nu_3]\nu_2]}F''_{\mu_1\nu_1},
 \e{218}
 \be
 &B_{3\;\mu_1\nu_1\mu_2\nu_2\mu_3\nu_3}\equiv&(F''_{\mu_2\nu_3}\eta_{\nu_1\nu_2}\eta_{\mu_1\mu_3})_{sym}=-{1\over8}F''_{[\mu_2[\mu_3}\eta_{\nu_3][\mu_1}\eta_{\nu_1]\nu_2]}-\nn\\&&-{1\over8}F''_{[\mu_3[\mu_1}\eta_{\nu_1][\mu_2}\eta_{\nu_2]\nu_3]}-{1\over8}F''_{[\mu_1[\mu_2}\eta_{\nu_2][\mu_3}\eta_{\nu_3]\nu_1]}.
 \e{219}
 Notice that 
 \be
 &B_{1\;\mu_1\nu_1\mu_2\nu_2\mu_3\nu_3}F^{\mu_1\nu_1\mu_2\nu_2\mu_3\nu_3}&=(F'_{\mu_1\mu_2\mu_3\nu_3}\eta_{\nu_1\nu_2})F^{\mu_1\nu_1\mu_2\nu_2\mu_3\nu_3}=\nn\\&&=     F'_{\mu_1\mu_2\mu_3\nu_3}F'^{\;\mu_1\mu_2\mu_3\nu_3},\nn\\
 &B_{2\;\mu_1\nu_1\mu_2\nu_2\mu_3\nu_3}F^{\mu_1\nu_1\mu_2\nu_2\mu_3\nu_3}&=(F''_{\mu_3\nu_3}\eta_{\mu_1\mu_2}\eta_{\nu_1\nu_2})F^{\mu_1\nu_1\mu_2\nu_2\mu_3\nu_3}=\nn\\&&=    F''_{\mu_3\nu_3}F''^{\;\mu_3\nu_3},\nn\\
 &B_{3\;\mu_1\nu_1\mu_2\nu_2\mu_3\nu_3}F^{\mu_1\nu_1\mu_2\nu_2\mu_3\nu_3}&=(F''_{\mu_2\nu_3}\eta_{\nu_1\nu_2}\eta_{\mu_1\mu_3})F^{\mu_1\nu_1\mu_2\nu_2\mu_3\nu_3}=\nn\\&&=     F''_{\mu_2\nu_3}F''^{\;\mu_2\nu_3},
 \e{220}
 in agreement with the normalization \rl{103}. Notice also that the two different forms, $B_2$ and $B_3$, for $F''$ are directly related to \rl{213} and \rl{214} respectively. 
 
 The trace of the ansatz \rl{216} is
 \be
&\eta^{\nu_1\nu_2}W_{\mu_1\nu_1\mu_2\nu_2\mu_3\nu_3}=&c_0F'_{\mu_1\mu_2\mu_3\nu_3}+c_1\eta_{\mu_1\mu_2}F''_{\mu_3\nu_3}+\nn\\&&+c_2\biggl(\eta_{\mu_1[\mu_3}F''_{\nu_3]\mu_2}+\eta_{\mu_2[\mu_3}F''_{\nu_3]\mu_3}\biggr),
 \e{221}
 where
 \be
 &&c_0=1+{1\over4}d \al_1,\quad c_1=\half\al_1+{1\over4}(d-1)\al_2+\half\al_3,\nn\\&& c_2=-{1\over4}(\al_1+\al_2)-{1\over8}(d-1)\al_3.
 \e{222}
 Hence, the ansatz \rl{216} for $W$ is traceless if
 \be
 &&\al_1=-{4\over d},\quad \al_2=\al_3={8\over d(d+1)}.
 \e{223}
 (cf. the calculations given by Damour and Deser \cite{Damour:1987vm}.)
 
 The Lagrangian \rl{104} is here
 \be
 &&\cL=F^2+\al_1 {F'}^2+(\al_2+\al_3){F''}^2,
 \e{224}
 which for the values \rl{223} may be written as
 \be
 &&\cL=C^2,
 \e{225}
 where $C=W$ for the values \rl{223}.
 The equations of motion are
 \be
 &&\dif^3C=0.
 \e{2261}
 
 As a side remark one may notice that although Weyl invariance  under the general Weyl transformations \rl{61} of the Lagrangian 
 \be
  &&\cL=F^2+\beta_1 {F'}^2+\beta_2{F''}^2
  \e{2262}
  determines the two $\beta$-parameters, it does not determine the {\em three} $\al$-parameters in the ansatz \rl{216} for the Weyl tensor. A consequence of this is that the Lagrangian \rl{224}-\rl{225}
   may trivially be written as
 \be
 &&\cL=C^2+\beta V\cdot F,
 \e{226}
 where $\beta$ is an arbitrary parameter, and
 \be
 &&V\equiv B_3-B_2.
 \e{227}
This follows trivially since
 \be
 &&V\cdot F=(F'')^2-(F'')^2=0.
 \e{228}
 For the equations of motion from \rl{226}  this implies $\dif^3V=0$ or
 \be
 &&\dif^3C=0\quad\Leftrightarrow \dif^3(C+\beta V)=0.
 \e{229}
 Notice that the tensor $C+\beta V$ for nonzero $\beta$'s is neither traceless nor Weyl invariant (see next section ).

  \subsection{$C$ for $s=4$}
  For $s=4$ the basic field strength \rl{4} is
  \be
&&F_{\mu_1\nu_1\mu_2\nu_2\mu_3\nu_3\mu_4\nu_4}=\dif_{[\mu_1}\dif_{[\mu_2}\dif_{[\mu_3}\dif_{[\mu_4}\phi_{\nu_4]\nu_3]\nu_2]\nu_1]},
 \e{230}
 where the gauge field $\phi$ is totally symmetric. In this case I find the following five different traces
 \be
 &&F'_{\nu_1\nu_2\mu_3\nu_3\mu_4\nu_4}\equiv \eta^{\mu_1\mu_2}F_{\mu_1\nu_1\mu_2\nu_2\mu_3\nu_3\mu_4\nu_4},\nn\\
&&F''_{\mu_3\nu_3\mu_4\nu_4}\equiv\half\eta^{\nu_1\nu_2}F'_{\nu_1\nu_2\mu_3\nu_3\mu_4\nu_4}=
\half  \eta^{\mu_1\mu_2} \eta^{\nu_1\nu_2}F_{\mu_1\nu_1\mu_2\nu_2\mu_3\nu_3\mu_4\nu_4},\nn\\
 &&G''_{\nu_1\nu_2\nu_3\nu_4}\equiv\eta^{\mu_3\mu_4} F'_{\nu_1\nu_2\mu_3\nu_3\mu_4\nu_4}=\eta^{\mu_1\mu_2} \eta^{\mu_3\mu_4}    F_{\mu_1\nu_1\mu_2\nu_2\mu_3\nu_3\mu_4\nu_4}, \nn\\
 &&F'''_{\nu_3\nu_4}\equiv\eta^{\mu_3\mu_4}F''_{\mu_3\nu_3\mu_4\nu_4}=\half  \eta^{\nu_1\nu_2}\eta^{\mu_1\mu_2} \eta^{\mu_3\mu_4}  F_{\mu_1\nu_1\mu_2\nu_2\mu_3\nu_3\mu_4\nu_4},\nn\\
 &&U\equiv\eta^{\nu_3\nu_4}F'''_{\nu_3\nu_4}=\half  \eta^{\nu_1\nu_2} \eta^{\nu_3\nu_4}\eta^{\mu_1\mu_2} \eta^{\mu_3\mu_4}F_{\mu_1\nu_1\mu_2\nu_2\mu_3\nu_3\mu_4\nu_4},
 \e{231}
 where $F'$ is symmetric in $\nu_1$ and $\nu_2$ and antisymmetric in $\mu_3\nu_3$ and $\mu_4\nu_4$. It is also symmetric under interchange of the pairs $\mu_3\nu_3$ and $\mu_4\nu_4$. $F''$ is antisymmetric in $\mu_3\nu_3$ and $\mu_4\nu_4$ and symmetric under interchange of the pairs $\mu_3\nu_3$ and $\mu_4\nu_4$. $G''$ is symmetric in $\nu_1\nu_2$ and $\nu_3\nu_4$ and symmetric under interchange of the pairs $\nu_1\nu_2$ and $\nu_3\nu_4$.
 
 Here I find the following alternative relation
 \be
&&F''_{\nu_2\nu_3\mu_4\nu_4}=\eta^{\nu_1\mu_3}F'_{\nu_1\nu_2\mu_3\nu_3\mu_4\nu_4}=\eta^{\nu_1\mu_3}\eta^{\mu_1\mu_2}F_{\mu_1\nu_1\mu_2\nu_2\mu_3\nu_3\mu_4\nu_4},
 \e{232}
 which implies
 \be
&&F'''_{\nu_3\nu_4}=\eta^{\nu_2\mu_4}F''_{\nu_2\nu_3\mu_4\nu_4}=\eta^{\nu_2\mu_4}\eta^{\nu_1\mu_3}\eta^{\mu_1\mu_2}F_{\mu_1\nu_1\mu_2\nu_2\mu_3\nu_3\mu_4\nu_4},\nn\\
&&U=\eta^{\nu_3\nu_4}F'''_{\nu_3\nu_4}=\eta^{\nu_3\nu_4}\eta^{\nu_2\mu_4}\eta^{\nu_1\mu_3}\eta^{\mu_1\mu_2}F_{\mu_1\nu_1\mu_2\nu_2\mu_3\nu_3\mu_4\nu_4}.
\e{233}
In addition I find the property
\be
&&G''_{\nu_1[\nu_2\nu_3]\nu_4}=F''_{\nu_2\nu_3\nu_1\nu_4}.
\e{235}
This may also be viewed as still another relation for $F''$,
\be
&&F''_{\nu_2\nu_3\nu_1\nu_4}=G''_{\nu_1[\nu_2\nu_3]\nu_4}=(\eta^{\mu_1\mu_2}\eta^{\mu_3\mu_4}-
\eta^{\mu_1\mu_3}\eta^{\mu_2\mu_4})F_{\mu_1\nu_1\mu_2\nu_2\mu_3\nu_3\mu_4\nu_4}.\nn\\
\e{236}

The cyclicities in \rl{2} imply here
\be
&&F'_{\nu_1[\nu_2\mu_3]\nu_3\mu_4\nu_4}=-F'_{\nu_1\nu_3\nu_2\mu_3\mu_4\nu_4},\quad
F'_{\nu_1\nu_2\mu_3[\nu_3\mu_4]\nu_4}=-F'_{\nu_1\nu_2\mu_3\nu_4\nu_3\mu_4},\nn\\
&&F''_{\mu_3[\nu_3\mu_4]\nu_4}=-F''_{\mu_3\nu_4\nu_3\mu_4}.
\e{234}
That much about the properties of the traces.

The ansatz \rl{101} for $W$ is here given by
\be
&&W_{\mu_1\nu_1\mu_2\nu_2\mu_3\nu_3\mu_4\nu_4}=F_{\mu_1\nu_1\mu_2\nu_2\mu_3\nu_3\mu_4\nu_4}+\sum_{k=1}^9\al_kB_{k\;\mu_1\nu_1\mu_2\nu_2\mu_3\nu_3\mu_4\nu_4},\nn\\
\e{2361}
where the nine different $B_k$-fields are
\be
&B_{1\;\mu_1\nu_1\mu_2\nu_2\mu_3\nu_3\mu_4\nu_4}=&(F'_{\nu_1\nu_2\mu_3\nu_3\mu_4\nu_4}\eta_{\mu_1\mu_2})_{sym},\nn\\
&B_{2\;\mu_1\nu_1\mu_2\nu_2\mu_3\nu_3\mu_4\nu_4}=&\half(F''_{\mu_3\nu_3\mu_4\nu_4}\eta_{\mu_1\mu_2}\eta_{\nu_1\nu_2})_{sym},\nn\\
&B_{3\;\mu_1\nu_1\mu_2\nu_2\mu_3\nu_3\mu_4\nu_4}=&(F''_{\nu_2\nu_3\mu_4\nu_4}\eta_{\mu_1\mu_2}\eta_{\nu_1\mu_3})_{sym},\nn\\
&B_{4\;\mu_1\nu_1\mu_2\nu_2\mu_3\nu_3\mu_4\nu_4}=&(F''_{\nu_2\nu_3\nu_1\nu_4}\eta_{\mu_1\mu_2}\eta_{\mu_3\mu_4}-F''_{\nu_2\nu_3\nu_1\nu_4}\eta_{\mu_1\mu_3}\eta_{\mu_2\mu_4})_{sym},\nn\\
&B_{5\;\mu_1\nu_1\mu_2\nu_2\mu_3\nu_3\mu_4\nu_4}=&(G''_{\nu_1\nu_2\nu_3\nu_4}\eta_{\mu_1\mu_2}\eta_{\mu_3\mu_4})_{sym},\nn\\
&B_{6\;\mu_1\nu_1\mu_2\nu_2\mu_3\nu_3\mu_4\nu_4}=&\half(F'''_{\nu_3\nu_4}
\eta_{\mu_1\mu_2}\eta_{\nu_1\nu_2}\eta_{\mu_3\mu_4})_{sym},\nn\\
&B_{7\;\mu_1\nu_1\mu_2\nu_2\mu_3\nu_3\mu_4\nu_4}=&(F'''_{\nu_3\nu_4}\eta_{\mu_1\mu_2}\eta_{\nu_1\mu_3}\eta_{\nu_2\mu_4})_{sym},            \nn\\
&B_{8\;\mu_1\nu_1\mu_2\nu_2\mu_3\nu_3\mu_4\nu_4}=&\half U(\eta_{\mu_1\mu_2}\eta_{\nu_1\nu_2}\eta_{\mu_3\mu_4}\eta_{\nu_3\nu_4})_{sym},\nn\\
&B_{9\;\mu_1\nu_1\mu_2\nu_2\mu_3\nu_3\mu_4\nu_4}=&U(\eta_{\mu_1\mu_2}\eta_{\nu_1\mu_3}\eta_{\nu_2\mu_4}\eta_{\nu_3\nu_4})_{sym}. 
\e{237}
(These expressions are  explicitly written down  in appendix B.) Notice  that the three different forms, $B_2$, $B_3$ and $B_4$ of $F''$ are directly related to \rl{231}, \rl{232} and \rl{236} respectively, and that the two different forms, $B_6$ and $B_7$ of $F'''$ as well as the two different forms, $B_8$ and $B_9$ of $U$ are directly related to \rl{231} and \rl{233} respectively. (This explains why I have introduced a factor one-half in $B_2$, $B_6$, and $B_8$ above.)

The trace of the ansatz \rl{2361} for $W$ will be expressed in terms $F'$, the higher traces of $F$ and $\eta$'s combined in expressions with the same symmetry as $F'$. We have 
\be
&&\eta^{\mu_1\mu_2}W_{\mu_1\nu_1\mu_2\nu_2\mu_3\nu_3\mu_4\nu_4}=c_0F'_{\nu_1\nu_2\mu_3\nu_3\mu_4\nu_4}+\sum_{l=1}^{10}c_lE_{l\;\nu_1\nu_2\mu_3\nu_3\mu_4\nu_4},\nn\\
\e{238}
where $E_l$  are  different for different $l$'s all with the same index symmetry as $F'$. There are ten such expressions.  They may \eg be written as (without choosing a particular normalization)
\be
&E_{1\;\nu_1\nu_2\mu_3\nu_3\mu_4\nu_4}=&\eta_{\nu_1\nu_2}F''_{\mu_3\nu_3\mu_4\nu_4},\nn\\
&E_{2\;\nu_1\nu_2\mu_3\nu_3\mu_4\nu_4}=&\eta_{[\mu_3[\mu_4}F''_{\nu_4]\nu_1\nu_3]\nu_2}+
\eta_{[\mu_3[\mu_4}F''_{\nu_4]\nu_2\nu_3]\nu_1}+\nn\\
&&+\eta_{[\mu_4[\mu_3}F''_{\nu_3]\nu_1\nu_4]\nu_2}+
\eta_{[\mu_4[\mu_3}F''_{\nu_3]\nu_2\nu_4]\nu_1}=\nn\\
&&=2\biggl(\eta_{[\mu_3[\mu_4}F''_{\nu_4]\nu_1\nu_3]\nu_2}+
\eta_{[\mu_3[\mu_4}F''_{\nu_4]\nu_2\nu_3]\nu_1}\biggr),\nn\\
&E_{3\;\nu_1\nu_2\mu_3\nu_3\mu_4\nu_4}=&\eta_{\nu_1[\mu_3}F''_{\nu_3]\nu_2\mu_4\nu_4}+\eta_{\nu_2[\mu_3}F''_{\nu_3]\nu_1\mu_4\nu_4}+\nn\\&&+\eta_{\nu_1[\mu_4}F''_{\nu_4]\nu_2\mu_3\nu_3}+\eta_{\nu_2[\mu_4}F''_{\nu_4]\nu_1\mu_3\nu_3},\nn\\
&E_{4\;\nu_1\nu_2\mu_3\nu_3\mu_4\nu_4}=&\eta_{[\mu_3[\mu_4}G''_{\nu_4]\nu_3]\nu_1\nu_2},\nn\\
&E_{5\;\nu_1\nu_2\mu_3\nu_3\mu_4\nu_4}=&\eta_{\nu_1\nu_2}\eta_{[\mu_3[\mu_4}F'''_{\nu_4]\nu_3]},\nn\\
&E_{6\;\nu_1\nu_2\mu_3\nu_3\mu_4\nu_4}=&\eta_{[\mu_3[\mu_4}\eta_{\nu_4]\nu_3]}F'''_{\nu_1\nu_2},\nn\\
&E_{7\;\nu_1\nu_2\mu_3\nu_3\mu_4\nu_4}=&\eta_{\nu_1[\mu_3}\eta_{\nu_2[\mu_4}F'''_{\nu_4]\nu_3]}+\eta_{\nu_2[\mu_3}\eta_{\nu_1[\mu_4}F'''_{\nu_4]\nu_3]},\nn\\
&E_{8\;\nu_1\nu_2\mu_3\nu_3\mu_4\nu_4}=&\eta_{\nu_1[\mu_3}\eta_{\nu_3][\mu_4}F'''_{\nu_4]\nu_2}+\eta_{\nu_2[\mu_3}\eta_{\nu_3][\mu_4}F'''_{\nu_4]\nu_1}+,\nn\\
&&\eta_{\nu_1[\mu_4}\eta_{\nu_4][\mu_3}F'''_{\nu_3]\nu_2}+\eta_{\nu_2[\mu_4}\eta_{\nu_4][\mu_3}F'''_{\nu_3]\nu_1},\nn\\
&E_{9\;\nu_1\nu_2\mu_3\nu_3\mu_4\nu_4}=&\eta_{\nu_1\nu_2}\eta_{[\mu_3[\mu_4}\eta_{\nu_4]\nu_3]}U,\nn\\
&E_{10\;\nu_1\nu_2\mu_3\nu_3\mu_4\nu_4}=&\biggl(\eta_{\nu_1[\mu_3}\eta_{\nu_3][\mu_4}\eta_{\nu_4]\nu_2}+\eta_{\nu_2[\mu_3}\eta_{\nu_3][\mu_4}\eta_{\nu_4]\nu_1}\biggr)U.\nn\\
\e{239}
Since the traces of the $B_k$-fields are (suppressing indices)
\be
&&B'_1={1\over24}\biggl((d+2)F'+2E_1-E_3+E_4\biggr),\nn\\
&&B'_2={1\over48}\biggl(2(d-1)E_1-2E_3+E_6\biggr),\nn\\
&&B'_3={1\over96}\biggl(8E_1+E_2-dE_3-E_8\biggr),\nn\\
&&B'_4={1\over48}\biggl(\half(d-3)E_2-3E_3+2E_5-E_7\biggr),\nn\\
&&B'_5={1\over48}\biggl(-\half E_2-E_3+dE_4+2E_5+E_7\biggr),\nn\\
&&B'_6={1\over192}\biggl(2(d-1)E_5+(d-2)E_6+4E_7-2E_8+E_9\biggr),\nn\\
&&B'_7={1\over192}\biggl(6E_5+2E_6+dE_7-(d-2)E_8-E_{10}\biggr),\nn\\
&&B'_8={1\over96}\biggl(2(d-1)E_9-4E_{10}\biggr),\nn\\
&&B'_9={1\over48}\biggl(2E_9-(d-1)E_{10}\biggr).
\e{240}
I find \rl{238} with the following values of the coefficients $c_l$
\be
&&c_0=1+{(d+2)\over24}\al_1,\nn\\
&&c_1={1\over12}\biggl(\al_1+{(d-1)\over2}\al_2+\al_3\biggr),\nn\\
&&c_2={1\over96}\biggl(\al_3+(d-3)\al_4-\al_5\biggr),\nn\\
&&c_3=-{1\over24}(\al_1+\al_2)-{d\over96}\al_3-{1\over16}\al_4-{1\over48}\al_5,\nn\\
&&c_4={1\over24}\biggl(\al_1+{d\over2}\al_5\biggr),\nn\\
&&c_5={1\over24}(\al_4+\al_5)+{(d-1)\over96}\al_6+{1\over32}\al_7,\nn\\
&&c_6={1\over48}\biggl(\al_2+{(d-2)\over4}\al_6+{1\over2}\al_7\biggr),\nn\\
&&c_7={1\over48}\biggl(-\al_4+\al_5+\al_6+{d\over4}\al_7\biggr),\nn\\
&&c_8=-{1\over96}(\al_3+\al_6)-{(d-2)\over192}\al_7,\nn\\
&&c_9={1\over192}\al_6+{(d-1)\over48}\al_8+{1\over24}\al_9,\nn\\
&&c_{10}=-{1\over192}\al_7-{1\over24}\al_8-{(d-1)\over48}\al_9.
\e{241}
Hence, a traceless $W$ ($W=C$)  requires
\be
&&\al_1=-{24\over d+2},\quad\quad\quad\al_2={48\over(d+2)(d+3)},\nn\\&&\al_3={96\over(d+2)(d+3)},\quad
\al_4=-{48\over d(d+2)(d+3)},\nn\\&&\al_5={48\over d(d+2)},\quad\quad\quad\quad\quad\al_6=\al_7=-{192\over d(d+2)(d+3)},\nn\\
&&\al_8=\al_9={48\over d(d+1)(d+2)(d+3)}.
\e{242}
The Lagrangian is here
\be
&&\cL=C^2=F^2+\al_1({F'})^2+(\al_2+\al_3+\al_4)({F''})^2+\al_5({G''})^2+\nn\\&&+(\al_6+\al_7)({F'''})^2+(\al_8+\al_9)U^2,
\e{243}
where $\al_1$ and $\al_5$ are given in \rl{242} and where
\be
&&\al_2+\al_3+\al_4={48(3d-1)\over d(d+2)(d+3)},\nn\\
&&\al_6+\al_7=-{384\over d(d+2)(d+3)},\nn\\
&&\al_8+\al_9={96\over d(d+1)(d+2)(d+3)},
\e{244}
from \rl{242}. 

As a side remark one may notice that although Weyl  invariance of  the Lagrangian 
\be
&&\cL=F^2+\beta_1({F'})^2+\beta_2({F''})^2+\beta_3({G''})^2+\nn\\&&+\beta_4({F'''})^2+\beta_5U^2
\e{2441}
determines the five $\beta$-parameters, it
does not determine the nine $\al$-parameters in the ansatz for the Weyl tensor due to the expression \rl{243}.
This implies  also that one may trivially replace the Lagrangian \rl{243} by
\be
&&\cL=C^2+\sum_{i=1}^4\beta_i V_i\cdot F,
\e{245}
where $\beta_i$ are arbitrary parameters and 
\be
&&V_1\equiv B_3-B_2,\quad V_2\equiv B_4-B_2,\nn\\
&&V_3\equiv B_7-B_6,\quad V_4\equiv B_9-B_8.
\e{246}
The reason is that  the normalization \rl{103} trivially  implies
\be
&&V_i\cdot F=0,\;\;i=1,2,3,4.
\e{247}
For the equations of motion from \rl{245}  this implies $\dif^4V_i=0$ or
\be
&&\dif^4C=0\quad\Leftrightarrow\quad\dif^4(C+\sum_i\beta_iV_i)=0.
\e{248}
The tensor $C+\sum_i\beta_iV_i$ is neither traceless nor Weyl invariant (see next section) for nonzero $\beta_i$-parameters.
However, the form $\cL=W^2$ requires  $W=C$.

\setcounter{equation}{0}
\section{The generalized Weyl tensors $C$ from generalized Weyl invariance}
In \cite{Fradkin:1985am} the generalized Weyl transformations for free fields of arbitrary integer spins $s$, represented by symmetric  gauge fields of rank $s$, are defined by
\be
&&\phi_{\mu_1\mu_2\cdots\mu_s}(x)\;\longrightarrow\;\phi_{\mu_1\mu_2\cdots\mu_s}(x)+\eta_{(\mu_1\mu_2}\la_{\mu_3\cdots\mu_s)}(x),
\e{301}
where $\eta$ is the flat Minkowski metric and $\la$ an arbitrary symmetric function with $s-2$ indices.
This should be viewed as an additional linear gauge transformation to \rl{5}. For the field strength $F$ in \rl{4} the transformations  \rl{301} imply
\be
&F_{\mu_1\nu_1\mu_2\nu_2\cdots\mu_s\nu_s}(x)\;\longrightarrow\;&
F_{\mu_1\nu_1\mu_2\nu_2\cdots\mu_s\nu_s}(x)+ \nn\\
&&+\sum_{{\rm sym\{\ pairs\}} (\mu_i\nu_i)}P_{\mu_1\nu_1\mu_2\nu_2}\Lambda_{\mu_3\nu_3\cdots\mu_s\nu_s}(x),\nn\\
\e{302}
where $\Lambda$ is a "field strength" constructed out of the functions $\la$ in \rl{301}, \ie
\be
&&\Lambda_{\mu_1\nu_1\mu_2\nu_2\cdots\mu_k\nu_k}(x)=\dif_{[\mu_k}\cdots\dif_{[\mu_2}\dif_{[\mu_1}\la_{\nu_1]\nu_2]\cdots\nu_k]}(x).
\e{303}
$P$ in \rl{302} is the second order differential operator
\be
&&P_{\mu_1\nu_1\mu_2\nu_2}\equiv\eta_{[\mu_1[\mu_2}\dif_{\nu_2]}\dif_{\nu_1]}.
\e{304}
The symmetrization in \rl{302} is in the pairs ($\mu_i\nu_i$). The definition \rl{304} of $P$  implies
that $P$ is symmetric under interchanges of the pairs  ($\mu_1\nu_1$) and ($\mu_2\nu_2$) which in turn makes  the symmetrizations in \rl{301} and \rl{302}  similar. Below we treat the cases $s=2,3,4$ explicitly. ($\la=0$ for $s=1$.)

\subsection{$C$ for $s=2$}
For $s=2$ we have the Weyl transformation
\be
&&\phi_{\mu_1\mu_2}(x)\;\longrightarrow\;\phi_{\mu_1\mu_2}(x)+\eta_{\mu_1\mu_2}\la(x).
\e{305}
It may be viewed as an infinitesimal scale transformation of the metric $g_{\mu_1\mu_2}$ in \rl{204}.
For the linearized Riemann tensor \rl{203} this implies 
\be
&&R_{\mu_1\nu_1\mu_2\nu_2}(x)\;\longrightarrow\;R_{\mu_1\nu_1\mu_2\nu_2}+\half P_{\mu_1\nu_1\mu_2\nu_2}\la(x)
\e{306}
from \rl{302}. For the Ricci tensor and the curvature scalar this implies in turn
\be
&&R_{\mu_1\mu_2}(x)\;\longrightarrow\;R_{\mu_1\mu_2}+\half P'_{\mu_1\mu_2}\la(x),\nn\\
&&R(x)\;\longrightarrow\;R+\half P''\la(x),
\e{307}
where
\be
&&P'_{\mu_1\mu_2}\equiv\eta^{\nu_1\nu_2}P_{\mu_1\nu_1\mu_2\nu_2}=\eta_{\mu_1\mu_2}\Box+(d-2)\dif_{\mu_1}\dif_{\mu_2},\nn\\
&&P''\equiv\eta^{\mu_1\mu_2}P'_{\mu_1\mu_2}=2(d-1)\Box.
\e{308}
For the expressions $B_1$ and $B_2$ in \rl{207} the transformations \rl{307} induce the transformations
\be
&B_{1\;\mu_1\nu_1\mu_2\nu_2}(x)\;\longrightarrow\;&B_{1\;\mu_1\nu_1\mu_2\nu_2}(x)+{1\over4}\eta_{[\mu_1[\mu_2}\eta_{\nu_2]\nu_1]}\Box\la(x)+\nn\\
&&+{1\over8}(d-2)P_{\mu_1\nu_1\mu_2\nu_2}\la(x),\nn\\
&B_{2\;\mu_1\nu_1\mu_2\nu_2}(x)\;\longrightarrow\;&B_{2\;\mu_1\nu_1\mu_2\nu_2}(x)+{1\over4}(d-1)\eta_{[\mu_1[\mu_2}\eta_{\nu_2]\nu_1]}\Box\la(x).\nn\\
\e{309}
It follows therefore that the ansatz $W$ in  \rl{206} for the Weyl tensor $C$ transforms as follows under \rl{305}
\be
&W_{\mu_1\nu_1\mu_2\nu_2}(x)\;\longrightarrow\;&W_{\mu_1\nu_1\mu_2\nu_2}(x)+\biggl(\half+{1\over8}(d-2)\al_1\biggr)P_{\mu_1\nu_1\mu_2\nu_2}\la(x)+\nn\\
&&+\biggl({1\over8}\al_1+{1\over4}(d-1)\al_2\biggr)\eta_{[\mu_1[\mu_2}\eta_{\nu_2]\nu_1]}\Box\la(x).
\e{310}
Hence, the ansatz $W$ is invariant under the Weyl transformation \rl{305} for the $\la$-values
\be
&&\al_1=-{4\over(d-2)},\quad\al_2={2\over(d-1)(d-2)},
\e{311}
in agreement with the result of the traceless condition in subsection 4.1. For these values $W=C$, the linearized Weyl tensor.

\subsection{$C$ for $s=3$}
The generalized Weyl transformations \rl{301} becomes for $s=3$
\be
&&\phi_{\mu_1\mu_2\mu_3}(x)\;\longrightarrow\;\phi_{\mu_1\mu_2\mu_3}(x)+\eta_{\mu_1\mu_2}\la_{\mu_3}(x)+\eta_{\mu_1\mu_3}\la_{\mu_2}(x)+\eta_{\mu_2\mu_3}\la_{\mu_1}(x),\nn\\
\e{312}
which for the field strength \rl{212} implies the similar transformation
\be
&F_{\mu_1\nu_1\mu_2\nu_2\mu_3\nu_3}\;\longrightarrow\;&F_{\mu_1\nu_1\mu_2\nu_2\mu_3\nu_3}+
P_{\mu_1\nu_1\mu_2\nu_2}\Lambda_{\mu_3\nu_3}(x)+\nn\\&&+P_{\mu_1\nu_1\mu_3\nu_3}\Lambda_{\mu_2\nu_2}(x)+P_{\mu_2\nu_2\mu_3\nu_3}\Lambda_{\mu_1\nu_1}(x),
\e{313}
where
\be
&&\Lambda_{\mu_1\nu_1}(x)\equiv\dif_{\mu_1}\la_{\nu_1}-\dif_{\nu_1}\la_{\mu_1}.
\e{314}
Eq.\rl{313} yields the following transformations for the traces \rl{213}
\be
&F'_{\mu_1\mu_2\mu_3\nu_3}\;\longrightarrow\;&F'_{\mu_1\mu_2\mu_3\nu_3}+P'_{\mu_1\mu_2}\Lambda_{\mu_3\nu_3}(x)+P_{\mu_3\nu_3\mu_1}^{\phantom{\mu_3\nu_3\mu_1}\nu_2}\Lambda_{\mu_2\nu_2}(x)+\nn\\&&+P_{\mu_3\nu_3\mu_2}^{\phantom{\mu_3\nu_3\mu_2}\nu_1}\Lambda_{\mu_1\nu_1}(x)=F'_{\mu_1\mu_2\mu_3\nu_3}+d\dif_{\mu_1}\dif_{\mu_2}\Lambda_{\mu_3\nu_3}(x)+\nn\\
&&+\eta_{\mu_1\mu_2}\Box\Lambda_{\mu_3\nu_3}(x)+\eta_{\mu_1[\mu_3}\dif_{\nu_3]}(\dif_{\mu_2}\dif^{\rho}\la_{\rho}(x)-\Box\la_{\mu_2}(x))+\nn\\
&&+\eta_{\mu_2[\mu_3}\dif_{\nu_3]}(\dif_{\mu_1}\dif^{\rho}\la_{\rho}(x)-\Box\la_{\mu_1}(x)),\nn\\
&F''_{\mu_3\nu_3}\;\longrightarrow\;&F''_{\mu_3\nu_3}+(d+1)\Box\Lambda_{\mu_3\nu_3}(x).
\e{315}
These transformations together with \rl{313} inserted into  the $B_k$-expressions \rl{217}-\rl{219} imply that the ansatz $W$ in \rl{216} of the generalized Weyl tensor $C$ transforms as
\be
&W_{\mu_1\nu_1\mu_2\nu_2\mu_3\nu_3}\;\longrightarrow\;&W_{\mu_1\nu_1\mu_2\nu_2\mu_3\nu_3}+
\biggl(1+{d\over4}\al_1\biggr)(P_{\mu_1\nu_1\mu_2\nu_2}\Lambda_{\mu_3\nu_3}(x)+\nn\\&&+P_{\mu_1\nu_1\mu_3\nu_3}\Lambda_{\mu_2\nu_2}(x)+P_{\mu_2\nu_2\mu_3\nu_3}\Lambda_{\mu_1\nu_1}(x))+\nn\\&&+\biggl({1\over4}\al_1+{1\over8}(d+1)\al_2\biggr)(\eta_{[\mu_1[\mu_2}\eta_{\nu_2]\nu_1]}\Box\Lambda_{\mu_3\nu_3}(x)+\nn\\&&+\eta_{[\mu_2[\mu_3}\eta_{\nu_3]\nu_2]}\Box\Lambda_{\mu_1\nu_1}(x)+\eta_{[\mu_1[\mu_3}\eta_{\nu_3]\nu_1]}\Box\Lambda_{\mu_2\nu_2}(x))-\nn\\&&-\biggl({1\over4}\al_1+{1\over8}(d+1)\al_3\biggr)(\eta_{[\mu_1[\mu_2}\eta_{\nu_2][\mu_3}\Box\Lambda_{\nu_3]\nu_1]}+\nn\\&&+\eta_{[\mu_2[\mu_1}\eta_{\nu_1][\mu_3}\Box\Lambda_{\nu_3]\nu_2]}+\eta_{[\mu_1[\mu_3}\eta_{\nu_3][\mu_2}\Box\Lambda_{\nu_2]\nu_1]}).
\e{316}
Hence, $W$ is invariant ($W=C$) for the $\al$-values \rl{223} in agreement with the result from the traceless condition in subsection 4.3.

\subsection{$C$ for $s=4$}
For $s=4$ the symmetric gauge field transform as
\be
&\phi_{\mu_1\mu_2\mu_3\mu_4}(x)\;\longrightarrow\;&\phi_{\mu_1\mu_2\mu_3\mu_4}(x)+\eta_{\mu_1\mu_2}\la_{\mu_3\mu_4}(x)+\eta_{\mu_3\mu_4}\la_{\mu_1\mu_2}(x)+\nn\\
&&+\eta_{\mu_1\mu_3}\la_{\mu_2\mu_4}(x)+\eta_{\mu_2\mu_4}\la_{\mu_1\mu_3}(x)+\nn\\
&&+\eta_{\mu_2\mu_3}\la_{\mu_1\mu_4}(x)+\eta_{\mu_1\mu_4}\la_{\mu_2\mu_3}(x)
\e{317}
under generalized Weyl transformations. This yields 
 for the field strength \rl{230}  the similar transformation
\be
&F_{\mu_1\nu_1\mu_2\nu_2\mu_3\nu_3\mu_4\nu_4}\;\longrightarrow\;&F_{\mu_1\nu_1\mu_2\nu_2\mu_3\nu_3\mu_4\nu_4}+
P_{\mu_1\nu_1\mu_2\nu_2}\Lambda_{\mu_3\nu_3\mu_4\nu_4}(x)+\nn\\
&&+
P_{\mu_3\nu_3\mu_4\nu_4}\Lambda_{\mu_1\nu_1\mu_2\nu_2}(x)+P_{\mu_1\nu_1\mu_3\nu_3}\Lambda_{\mu_2\nu_2\mu_4\nu_4}(x)+\nn\\
&&+
P_{\mu_2\nu_2\mu_4\nu_4}\Lambda_{\mu_1\nu_1\mu_3\nu_3}(x)+P_{\mu_2\nu_2\mu_3\nu_3}\Lambda_{\mu_1\nu_1\mu_4\nu_4}(x)+\nn\\
&&+
P_{\mu_1\nu_1\mu_4\nu_4}\Lambda_{\mu_2\nu_2\mu_3\nu_3}(x),
\e{318}
where
\be
&&\Lambda_{\mu_1\nu_1\mu_2\nu_2}\equiv\dif_{[\mu_1}\dif_{[\mu_2}\la_{\nu_2]\nu_1]}.
\e{319}
This implies that the five traces \rl{231} transform under \rl{317} as
\be
 &F'_{\nu_1\nu_2\mu_3\nu_3\mu_4\nu_4}(x)\;\longrightarrow\;&F'_{\nu_1\nu_2\mu_3\nu_3\mu_4\nu_4}(x)+P_{\mu_3\nu_3\mu_4\nu_4}\Lambda'_{\nu_1\nu_2}(x)+\nn\\
&&+(d+2)\dif_{\nu_1}\dif_{\nu_2}\Lambda_{\mu_3\nu_3\mu_4\nu_4}+\eta_{\nu_1\nu_2}\Box\Lambda_{\mu_3\nu_3\mu_4\nu_4}-\nn\\
&&-\eta_{\nu_1[\mu_3}\dif_{\nu_3]}\dif_{[\mu_4}\Lambda'_{\nu_4]\nu_2}-\eta_{\nu_2[\mu_3}\dif_{\nu_3]}\dif_{[\mu_4}\Lambda'_{\nu_4]\nu_1}-\nn\\
&&-\eta_{\nu_1[\mu_4}\dif_{\nu_4]}\dif_{[\mu_3}\Lambda'_{\nu_3]\nu_2}-\eta_{\nu_2[\mu_4}\dif_{\nu_4]}\dif_{[\mu_3}\Lambda'_{\nu_3]\nu_1},\nn\\
 &F''_{\mu_3\nu_3\mu_4\nu_4}(x)\;\longrightarrow\;&F''_{\mu_3\nu_3\mu_4\nu_4}(x)+\half P_{\mu_3\nu_3\mu_4\nu_4}\Lambda''(x)+\nn\\&&+(d+3)\Box\Lambda_{\mu_3\nu_3}(x),\nn\\
&G''_{\nu_1\nu_2\nu_3\nu_4}(x)\;\longrightarrow\;&G''_{\nu_1\nu_2\nu_3\nu_4}(x)+(d+2)\dif_{\nu_3}\dif_{\nu_4}\Lambda'_{\nu_1\nu_2}(x)+\nn\\
&&+(d+2)\dif_{\nu_1}\dif_{\nu_2}\Lambda'_{\nu_3\nu_3}(x)-\dif_{\nu_1}\dif_{\nu_3}\Lambda'_{\nu_2\nu_4}(x)  -\nn\\
&&-\dif_{\nu_2}\dif_{\nu_3}\Lambda'_{\nu_1\nu_4}(x)-\dif_{\nu_1}\dif_{\nu_4}\Lambda'_{\nu_2\nu_3}(x)  -\nn\\&&-\dif_{\nu_2}\dif_{\nu_4}\Lambda'_{\nu_1\nu_3}(x)+
\eta_{\nu_1\nu_2}\Box\Lambda'_{\nu_3\nu_4}(x)+\nn\\&&+\eta_{\nu_3\nu_4}\Box\Lambda'_{\nu_1\nu_2}(x)+\eta_{\nu_1\nu_3}\Box\Lambda'_{\nu_2\nu_4}(x)+\nn\\
&&+\eta_{\nu_2\nu_4}\Box\Lambda'_{\nu_1\nu_3}(x)+\eta_{\nu_2\nu_3}\Box\Lambda'_{\nu_1\nu_4}(x)+\nn\\&&+\eta_{\nu_1\nu_4}\Box\Lambda'_{\nu_2\nu_3}(x)-\nn\\
&&-\half\eta_{\nu_2\nu_4}\dif_{\nu_1}\dif_{\nu_3}\Lambda''(x)-\half\eta_{\nu_1\nu_4}\dif_{\nu_2}\dif_{\nu_3}\Lambda''(x)-\nn\\
&&-\half\eta_{\nu_2\nu_3}\dif_{\nu_1}\dif_{\nu_4}\Lambda''(x)-\half\eta_{\nu_1\nu_3}\dif_{\nu_2}\dif_{\nu_4}\Lambda''(x),\nn\\
 &F'''_{\nu_3\nu_4}(x)\;\longrightarrow\;&F'''_{\nu_3\nu_4}(x)+\half(d-2)\dif_{\nu_3}\dif_{\nu_4}\Lambda''(x)+\nn\\&&+\half\eta_{\nu_3\nu_4}\Box\Lambda''(x)+
(d+3)\Box\Lambda'_{\nu_3\nu_4}(x),\nn\\
 &U(x)\;\longrightarrow\;&U(x)+2(d+1)\Box\Lambda''(x),
 \e{320}
 where
 \be
&&\Lambda'_{\mu_1\mu_2}\equiv\eta^{\nu_1\nu_2}\Lambda_{\mu_1\nu_1\mu_2\nu_2}=\dif_{\mu_1}\dif_{\mu_2}\la'+\Box\la_{\mu_1\mu_2}-\dif_{\mu_1}\dif^{\rho}\la_{\rho\mu_2}-\dif_{\mu_2}\dif^{\rho}\la_{\rho\mu_1},\nn\\
&&\Lambda''\equiv\eta^{\mu_1\mu_2}\Lambda'_{\mu_1\mu_2}=2(\Box\la'-\dif^{\rho_1}\dif^{\rho_2}\la_{\rho_1\rho_2}),\nn\\
&&\la'\equiv\eta^{\mu_1\mu_2}\la_{\mu_1\mu_2},
\e{321}
from \rl{319}. The transformations in \rl{320} induce transformations for the nine $B_k$-functions in appendix B. These and \rl{318} inserted into the ansatz $W$ in \rl{2361} for the generalized Weyl tensor  implies that $W$ transforms as
\be
&W_{\mu_1\nu_1\mu_2\nu_2\mu_3\nu_3\mu_4\nu_4}\;\longrightarrow\;&W_{\mu_1\nu_1\mu_2\nu_2\mu_3\nu_3\mu_4\nu_4}+\sum_{i=1}^{11}c_iM_{i\;\mu_1\nu_1\mu_2\nu_2\mu_3\nu_3\mu_4\nu_4},\nn\\
\e{322}
where $M_i$ depends on the metric tensor $\eta$ and the $\la$-functions  in the generalized Weyl transformations \rl{317}. $M_i$ are explicitly given in appendix C. The coefficients $c_i$ depend on the $\al$'s in the ansatz \rl{2361} for $W$ and are given by
\be
&&c_1=1+{(d+2)\over 24}\al_1,\nn\\
&&c_2={1\over24}\biggl(\al_1+{(d+3)\over2}\al_2\biggr),\nn\\
&&c_3={1\over24}\biggl(\al_1+{(d+3)\over4}\al_3\biggr),\nn\\
&&c_4={1\over24}\biggl(\al_5+(d+3)\al_4\biggr),\nn\\
&&c_5={1\over24}\biggl(\al_1+{d\over2}\al_5\biggr),\nn\\
&&c_6={1\over48}\biggl(\al_5+{(d+3)\over4}\al_6\biggr),\nn\\
&&c_7=-{1\over48}\biggl(\al_5+{(d+3)\over4}\al_7\biggr),\nn\\
&&c_8={1\over48}\biggl((d+1)\al_8+{1\over4}\al_6\biggr),\nn\\
&&c_9=-{1\over24}\biggl((d+1)\al_9+{1\over4}\al_7\biggr),\nn\\
&&c_{10}={1\over48}\biggl(\al_4+\half\al_2+{(d-2)\over8}\al_6\biggr),\nn\\
&&c_{11}={1\over96}\biggl(\al_3-\al_4-\al_5+{(d-2)\over4}\al_7\biggr),\nn\\
\e{323}
Hence, it follows that  $W=C$ for the $\al$-values \rl{244} in agreement with the result of the traceless condition in subsection 4.4.

\setcounter{equation}{0}
\section{Relations to the  proposals by Fradkin, Linetsky and Tseytlin}
 A conformal higher spin theory in a Lagrangian form yielding equations of the same order as those presented here seems first to have been proposed by   Fradkin and Tseytlin \cite{Fradkin:1985am}. (They only considered  free theories for arbitrary spins.) The structure considered was very implicit. The Lagrangian for free integer  spins was given in  the form 
\be
&&\cL=\phi_{\mu_1\mu_2\cdots\mu_s}\Box^sP^{{\mu_1\mu_2\cdots\mu_s}}_{{\nu_1\nu_2\cdots\nu_s}}\phi^{\nu_1\nu_2\cdots\nu_s},
\e{701}
where $\phi$ is totally symmetric and where the spin projector $P$ satisfies
\be
&&\eta_{\mu_i\mu_k}P^{{\mu_1\mu_2\cdots\mu_s}}_{{\nu_1\nu_2\cdots\nu_s}}=0,\quad\dif_{\mu_i}P^{{\mu_1\mu_2\cdots\mu_s}}_{{\nu_1\nu_2\cdots\nu_s}}=0,\quad{\rm any}\;i,k(i\neq k)= 1,2,\ldots,s.\nn\\
\e{702}
These conditions obviously make the action to \rl{701}
 invariant under the same gauge transformations  as those considered here, \ie
\be
&&\del\phi_{\mu_1\mu_2\cdots\mu_s}(x)=\dif_{(\mu_1}\varepsilon_{\mu_2\cdots\mu_s)}(x)+\eta_{(\mu_1\mu_2}\la_{\mu_3\cdots\mu_s)}(x)
\e{703}
for arbitrary symmetric functions $\varepsilon$ and $\la$.
%(In fact, the generalized Weyl transformation ($\la$) was introduced in this paper.) 
They noted that Maxwells theory and linear conformal gravity have the form \rl{701} for $s=1$ and $s=2$.

Fradkin and Linetsky \cite{Fradkin:1989md,Fradkin:1990ps} further elaborated this approach. By means of a geometric/algebraic method they even proposed a  version with interactions up to cubic order. For free integer spins they proposed an alternative form to \rl{701} (see appendix in \cite{Fradkin:1989md}, and section 3 in \cite{Fradkin:1990ps}):
\be
&&\cL(x)=(-1)^sC_{\mu_1\mu_2\cdots\mu_s{\nu_1\nu_2\cdots\nu_s}}(x)C^{\mu_1\mu_2\cdots\mu_s{\nu_1\nu_2\cdots\nu_s}}(x),
\e{704}
where the linearized Weyl tensor $C$ has the form (symmetry in indices with the same letter)
\be
&&C_{\mu_1\mu_2\cdots\mu_s{\nu_1\nu_2\cdots\nu_s}}=\cP_{\mu_1\mu_2\cdots\mu_s,{\nu_1\nu_2\cdots\nu_s}}^{\phantom{\mu_1\mu_2\cdots\mu_s,{\nu_1\nu_2\cdots\nu_s}}\rho_1\rho_2\cdots\rho_s,{\sigma_1\sigma_2\cdots\sigma_s}}\dif_{\rho_1}\dif_{\rho_2}\cdots\dif_{\rho_s}\phi_{\sigma_1\sigma_2\cdots\sigma_s},\nn\\
\e{705}
where the Young projector $\cP_s$ is associated with the irreducible traceless tableau\\
\beq
  \begin{tabular}{|c|c|c|}
 \hline
$\mu_1$&$\cdots$&$\mu_s$\\
\hline
$\nu_1$&$\cdots$&$\nu_s$\\
  \hline
 \end{tabular}\\
 \eq
This means that $\cP_s$ is required to satisfy
\be
&&\cP_{\mu_1\mu_2\cdots\mu_s,{\nu_1\nu_2\cdots\nu_s}}^{\phantom{\mu_1\mu_2\cdots\mu_s,{\nu_1\nu_2\cdots\nu_s}}\rho_1\rho_2\cdots\rho_s,{\sigma_1\sigma_2\cdots\sigma_s}}=\cP_{\phantom{\rho_1\rho_2\cdots\rho_s,{\sigma_1\sigma_2\cdots\sigma_s}}\mu_1\mu_2\cdots\mu_s,{\nu_1\nu_2\cdots\nu_s}}^{\rho_1\rho_2\cdots\rho_s,{\sigma_1\sigma_2\cdots\sigma_s}} \nn\\&&\equiv\cP_{\mu_1\mu_2\cdots\mu_s,{\nu_1\nu_2\cdots\nu_s}}^{\rho_1\rho_2\cdots\rho_s,{\sigma_1\sigma_2\cdots\sigma_s}}
\e{706}
and 
\be
&&\cP_{\mu_1\mu_2\cdots\mu_s,{\nu_1\nu_2\cdots\nu_s}}^{\al_1\al_2\cdots\al_s,{\beta_1\beta_2\cdots\beta_s}}\cP_{\al_1\al_2\cdots\al_s,{\beta_1\beta_2\cdots\beta_s}}^{\rho_1\rho_2\cdots\rho_s,{\sigma_1\sigma_2\cdots\sigma_s}}=\cP_{\mu_1\mu_2\cdots\mu_s,{\nu_1\nu_2\cdots\nu_s}}^{\rho_1\rho_2\cdots\rho_s,{\sigma_1\sigma_2\cdots\sigma_s}}.
\e{707}
($\cP_s$ is a sum of terms involving a real constant and products of Kronecker delta's and $\eta$'s.)
Furthermore, $\cP_s$ is required to satisfy the conditions
\be
&&
%\sum_{{\rm sym\ in\ all\ }\sigma's}
\cP_{\mu_1\mu_2\cdots\mu_s,{\nu_1\nu_2\cdots\nu_s}}^{\rho_1\rho_2\cdots\rho_{s-1},\sigma_{s+1},{\sigma_1\sigma_2\cdots\sigma_s}}+cycle({\rm all\ } \sigma's)=0,
\e{708}
and
\be
&&\eta^{\mu_k\mu_l}\cP_{\mu_1\mu_2\cdots\mu_s,{\nu_1\nu_2\cdots\nu_s}}^{\rho_1\rho_2\cdots\rho_s,{\sigma_1\sigma_2\cdots\sigma_s}}=0,\quad k,l (k\neq l)= 1,2,\ldots,s.
\e{709}
The last two conditions make the Weyl tensor   in \rl{705} invariant under the gauge transformations \rl{703}.
Integration by parts relates the Lagrangians \rl{701} and \rl{704}:
\be
&&\Box^sP^{{\mu_1\mu_2\cdots\mu_s}}_{{\nu_1\nu_2\cdots\nu_s}}=
\cP_{\mu_1\mu_2\cdots\mu_s,{\rho_1\rho_2\cdots\rho_s}}^{\nu_1\nu_2\cdots\nu_s,{\sigma_1\sigma_2\cdots\sigma_s}}\dif^{\rho_1}\dif^{\rho_2}\cdots\dif^{\rho_s}\dif_{\sigma_1}\dif_{\sigma_2}\cdots\dif_{\sigma_s}.
\e{710}
Comment: Notice that this relation implies that the equations of motion may be written as (cf theorem 4):
\be
&&\dif^{\mu_1}\dif^{\mu_2}\cdots\dif^{\mu_s}C_{\mu_1\mu_2\cdots\mu_s{\nu_1\nu_2\cdots\nu_s}}=0.
\e{711}

The Weyl tensors defined in section 3 have a similar form to \rl{705}. However the condition \rl{708} is not satisfied. On the other hand it is clear that the difference is due to a different 
choice of representation of the above Young tableau.  I would have found \rl{705} if I had used the deWit-Freedman curvatures instead of the Weinberg representation $F$ (cf property \rl{x3} in appendix A). The above Young projector is therefore related to the deWit-Freedman representation. In my case the generalized Weyl tensors seem to have the form
\be
&&C^{(W)}_{\mu_1\nu_1\mu_2\nu_2\cdots\mu_s\nu_s}=\cP_{\mu_1\nu_1\mu_2\nu_2\cdots\mu_s\nu_s}^{(W)\;\rho_1\sigma_1\rho_2\sigma_2\cdots\rho_s\sigma_s}\dif_{\rho_1}\dif_{\rho_2}\cdots\dif_{\rho_s}\phi_{\sigma_1\sigma_2\cdots\sigma_s}\nn\\
\e{712}
with antisymmetry in $\mu_k$ and $\nu_k$ for $k=1,2,\ldots,s$, and symmetry under interchange of pairs $\mu_k\nu_k$ and $\mu_l\nu_l$, which is the symmetry relevant for the Weinberg representation. The conditions \rl{708} and \rl{709} should then be replaced by
\be
&&
%\sum_{{\rm antisym\ in\ any\ three\ indices\ }\mu_k\nu_k\mu_l}
\cP_{\mu_1\nu_1\mu_2\nu_2\cdots\mu_s\nu_s}^{(W)\;\rho_1\sigma_1\rho_2\sigma_2\cdots\rho_s\sigma_s}+cycle({\rm any\ three\ indices\ }\mu_k \nu_k \mu_l)=0,
\e{713}
and 
\be
&&\eta^{\mu_k\mu_l}\cP_{\mu_1\nu_1\mu_2\nu_2\cdots\mu_s\nu_s}^{(W)\;\rho_1\sigma_1\rho_2\sigma_2\cdots\rho_s\sigma_s}=0,
\e{714}
which makes the expression \rl{712} invariant under the gauge transformations \rl{703}. The condition \rl{713} is identical to the first relation in \rl{2}.
Also here we have the relations \rl{710} and \rl{711} with $\cP$ and $C$ replaced by $\cP^{(W)}$ and $C^{(W)}$. The similarity with the results of the  construction in previous sections is obvious. 
In fact, in order to prove equivalence it remains only to prove that
my construction satisfies  condition \rl{713} which I guess is the case. 
\beq
 \begin{tabular}{|p{114mm}|}
 \hline
  {\bf Proposal 4}:   The Weyl tensors \rl{712} are equal to the  Weyl tensors  defined in section 3 for all spins $s$.\\
 \hline
 \end{tabular}
 \eq
 
\vspace{2cm}

\part{The manifest formulation}
\setcounter{equation}{0}
\section{Manifestly conformally invariant theories}
There is a large class of massless theories that are conformally invariant. In $d>2$ this means that one has apart from Poincar\'e invariance also invariance under scaling, $x^{\mu}\ra \la x^{\mu}$, and under the special conformal transformations
\be
&&x^{\mu}\;\;\ra\;\;x^{\prime\mu}=\frac{x^{\mu}+b^{\mu}x^2}{1+2b\cdot x+b^2x^2}.
\e{401}
All these transformations form the conformal group. Although the transformations are nonlinear the group is simply $SO(d,2)$.
This led Dirac \cite{Dirac:1936} to propose a manifestly conformally covariant formulation for conformal theories. He did this explicitly for free massless scalar, spinor, and vector fields at the level of the equations of motion in $d=4$. This formulation is given on a space of dimensions $d+2$  called the {\em conformal space}. The coordinates on the conformal space are denoted $y^A=(y^\mu,y^{d+1},y^{d+2})$, and involve  two time-like directions ($y^0$, $y^{d+2}$).  The indices are raised and lowered by the flat metric
\be
&&\eta_{AB}={\rm diag}(-1,1,\ldots,1,-1).
\e{402}
 $SO(d,2)$ acts linearly on the coordinates $y^A$ and leave the scalar products invariant.  Following Dirac the ordinary spacetime theory is required to live on a $(d+1)$-dimensional hypercone in the conformal space.
This hypercone is defined by 
\be
&&y^2 \equiv \eta_{AB} y^A y^B = 0,
\e{403}
which by definition  is invariant under SO($d$,2) transformations. Due to the projectiveness of this relation the dimensions of $y^A$ may be chosen freely. Choosing dimensionless coordinates $y^A$ on the conformal space the Minkowski coordinates $x^{\mu}$ are reached by the nonlinear point transformation
\be
x^{\mu} & = & \frac{y^{\mu}}{y^-}R, \qquad y^-\equiv y^{d+2}-y^{d+1}, \nn\\
\ga & = & y^-,  \nn\\
\rho& = & {y^2}, 
\e{404}
where $R$ is a constant with dimension length.  ( I choose $x^{\mu}$ to have dimension length, and the additional variables $\ga$ and $\rho$ to be dimensionless.) This transformation is invertible and the inverse is
\be
y^{\mu} & = & \frac{\ga}{R}x^{\mu}, \nn \\
y^-&=&\ga,\nn\\
y^+&=&\frac{\rho}{2\ga}-\frac{\ga}{2R^2}\eta_{\mu\nu}x^{\mu}x^{\nu},\qquad y^+\equiv \half\bigl(y^{d+2}+y^{d+1}\bigr).
\e{405}
This transformation is well defined for $y^-\neq 0$ ($\ga\neq 0$), and for such values the flat metric \rl{402} induces  an invertible metric $g_{\mu\nu}(x)$ in the $x^{\mu}$-coordinates given by
\be
g_{\mu \nu}(x) & = & \ga^2 \eta_{\mu \nu}.
\e{406}

On the hypercone $y^2=0$ ($\rho=0$)  the extra variable  $\ga$ in \rl{404} acts as a projective parameter. For the metric above we find a reduction to the Minkowski metric for $\ga=\pm 1$. However, as \eg the conformal particle models show \cite{Marnelius:1980}, $\ga$ may be chosen to be an arbitrary function of $x^{\mu}$ in which case the projected space is turned into an arbitrary conformally flat space. Conformal invariance in the sense of invariance under 
\be
&&g_{\mu\nu}\;\ra\;\la(x) g_{\mu\nu}
\e{407}
is obviously automatic for all conformal theories which are derivable from the manifestly conformally covariant formulation.

 \setcounter{equation}{0}
\section{The action principle for manifestly conformal theories}
Manifestly conformal fields are fields defined on the above conformal space of dimension $d+2$  coordinatized by $y^A$, $A=0, 1, \ldots,d+2$. The fields are either tensor or spinor fields on $SO(d,2)$. 
In order for these fields to correspond to spacetime fields they have to satisfy some conditions. These conditions are homogeneity and transversality. Thus, 
all fields are required to have a definite degree of homogeneity which means that they satisfy equations of the type
\be
&&(y\cdot\dif-n)\phi_{AB\cdots}(y)=0,
\e{502}
where $\dif_A$ is a differential with respect to $y^A$. The constant $n$ is the degree of homogeneity of the $\phi$ field. This condition is necessary in order to project the field to spacetime (see next section). Furthermore, all basic fields with tensor indices are required to satisfy the following transversality conditions
\be
&&y^A\phi_{AB\cdots}(y)=0,\quad y^B\phi_{AB\cdots}(y)=0,\quad etc.
\e{503}
These conditions remove unphysical components from the tensor fields.  Although the properties \rl{502} and \rl{503} only are necessary on the hypercone $y^2=0$, the action principle requires \rl{502} and \rl{503} to be valid on the whole conformal space \cite{Arvidsson:2006}.

The action for the manifestly conformal fields has the form \cite{Marnelius:1980,Arvidsson:2006}.
\be
&&A=\int dy \del(y^2)\cL(y),
\e{504}
where $dy$ is the flat $O(d,2)$ measure, and where $\cL(y)$ is a local expression of the involved conformal fields. $\cL$ must transform as a scalar under $O(d,2)$. Furthermore, it must have the degree of homogeneity $-d$, \ie it must satisfy
\be
&&(y\cdot\dif+d)\cL(y)=0,
\e{505}
This implies that the theory is invariant  under $y\ra\la y$. This condition  also  restricts the form of $\cL$.

In order to have a conformally invariant theory the action \rl{504}, and consequently the corresponding equations of motion,  must also be invariant under transformations of the type
\be
&&\phi_{AB\cdots}(y)\;\;\ra\;\;\phi'_{AB\cdots}(y)=\phi_{AB\cdots}(y)+y^2\tilde{\phi}_{AB\cdots}(y)
\e{506}
for all fields involved in $\cL$. This implies that only the part of the fields on the hypercone $y^2=0$ is physically relevant. 
The invariance under \rl{506} allows
us to choose  $\phi$ freely outside the hypercone $y^2=0$ and to make all manipulations off the cone allowed. However, for the action \rl{504} invariance under \rl{506} is required only for restricted $\tilde{\phi}$: Here we are free to choose $\tilde{\phi}$ arbitrary except that it must be   such that $\phi\;\ra\;\phi'$ preserves the homogeneity and the transversality properties \rl{502} and \rl{503}. Thus, $\tilde{\phi}$ is transverse and has the degree of homogeneity $n-2$ if $\phi$ has the degree $n$ as in \rl{502}.

The results for $s=2$ and $s=3$ in section 10 further on suggest that
 \beq
 \begin{tabular}{|p{114mm}|}
 \hline
  {\bf Proposal 5}:  The actions are invariant under \rl{506} for arbitrary functions  $\tilde{\phi}_{AB\cdots}$ if they are invariant under \rl{506} for arbitrary infinitesimal  functions $\tilde{\phi}_{AB\cdots}$.\\
 \hline
 \end{tabular}
 \eq
 One implication of the invariance under \rl{505} seems also to be
\beq
 \begin{tabular}{|p{114mm}|}
 \hline
  {\bf Proposal 6}:   If the action is invariant under the special gauge transformations \rl{506} then the equations of motion from the action \rl{504} have the form
\be
&&\del(y^2)(\cdots)=0,
\e{507}
\ie they are entirely defined on the hypercone $y^2=0$. No derivatives on the delta function $\del(y^2)$ appear.\\
 \hline
 \end{tabular}
 \eq
 Notice that in the derivation of equations of motion all divergences of the form 
 \be
 &&\int dy\dif_A\Big(\del(y^2)f^A(y)\Big)
 \e{508}
 are assumed to vanish. 
Under this assumption the property in proposal 6 has been checked for the models considered in \cite{Arvidsson:2006}.
Proposal 6 implies that the entirely  theory is defined on the hypercone $y^2=0$. Both the action \rl{504} as well as the resulting equations of motion. This provided \rl{504} is invariant under the special gauge transformations \rl{506} which in turn makes the fields off the hypercone irrelevant.

  \setcounter{equation}{0}
 \section{The reduction  of consistent manifestly conformal field theories in $d+2$ dimensions to $d$ dimensional spacetime}
 The fields, the actions and the equations of motion in the above defined manifestly conformally covariant formulation in $d+2$ dimensions  may be reduced to $d$-dimensional spacetime by means of the general procedure given in \cite{Mack:1969}.
Consider a general field $F^i(y)$ defined on the conformal space where the superscript $i$ denotes tensor or/and spinor indices depending on the type of field. Note that $i$ should be consistent with a linear representation of the conformal group SO($d$,2). First the field is required to be a homogeneous function of the coordinates $y^A$ in the sense 
\be
&&(y^A \dif_A-n)F^i(y) = 0,
\e{homo_y}
where $n$ denotes the degree of homogeneity.
From the relations \rl{404} and \rl{405} one finds 
\be
&&y^A\dif_A= \ga \frac{\partial}{\partial \ga} + 2 \rho \frac{\partial}{\partial \rho},\quad \ga\equiv y^-,\quad \rho\equiv y^2.
\e{hom}
Now since the  gauge invariance under the special gauge transformations \rl{506} always allows for the choice of a $\rho$-independent field 
$F^i(y)$, \ie
\be
&&\left.F^i(y)=F^i(y)\right |_{\rho=0},\quad \left.\dif_{\rho}F^i(y)\right |_{\rho=0}=0,\nn\\
&&\left.\dif^2_{\rho}F^i(y)\right |_{\rho=0}=0, \;\;{\rm etc.},\quad \rho\equiv y^2,
\e{concond}
 one finds from \rl{homo_y} and \rl{hom}
\be
&&\ga \frac{\partial}{\partial \ga} \left.F^i(y)\right |_{\rho=0} = n \left.F^i(y)\right |_{\rho=0}.
\e{homo_ga}
 This relation allows then for a definition of  a field $\tilde{f}^i(x)$ which only depends on the $d$-dimensional coordinates $x^{\mu}$. This definition is
\be
&&\tilde{f}^i(x) \equiv \ga^{-n} \left.F^i(y)\right |_{\rho=0}.
\e{phi_tilde}
To be precise one should notice that
 when $i$ contains tensor indices we can no longer consistently impose \rl{concond} due to the strong transversality conditions in \rl{503}. Here, as explained in \cite{Arvidsson:2006} and the remarks below, one must split the field  in terms of auxiliary fields and the explicit coordinates $y^A$. One may then require  weak transversality of the auxiliary fields which  then permit us to impose \rl{concond} on these auxiliary fields separately leaving the $y^2$ dependence in the explicit coordinates $y^A$. This splitting does not change the degrees of freedom. It is just a way to specify how to get down to $d$ dimensions.

For non-scalar fields, which allow for the above procedure,  \rl{phi_tilde} is not enough to produce a consistent spacetime field. The reason for this is the following: $J_{AB}$ are the generators of $SO(d,2)$-transformations.     It has an orbital, differential part $L_{AB}$ and a spin part $S_{AB}$.
The operator that generates translations, $P_{\mu}$, which is equal to $J_{+ \mu}$, is supposed to act only differentially on the spacetime field. Now this latter operator may be divided into two parts as well
\be
&&J_{+ \mu} = L_{+ \mu} + S_{+ \mu}.
\e{translation}
In this expression, $L_{+ \mu}$ is the differential (orbital) piece while $S_{+ \mu}$ is the intrinsic (spin) piece. The intrinsic piece is non-differential and should, therefore, be removed. This may be done by   defining the true field on the $d$-dimensional spacetime as follows \cite{Mack:1969}:
\be
&&f^i(x) \equiv V(x) \tilde{f}^i(x) = \ga^{-n} V(x) \left.F^i(y)\right |_{\rho=0},
\e{fields}
where the operator $V(x)$ is defined by
\be
&&V(x) \equiv \exp(- i x^\mu S_{+ \mu}).
\e{V}
The field $f^i(x)$ behaves then as expected under translations, and depends only on the spacetime coordinates $x^\mu$.

Now it should be mentioned that the projected field $f^i(x)$ in \rl{fields} for tensor fields will yield unphysical components in addition to the expected physical ones.
Although the transversality conditions \rl{503} remove some of the unphysical components (essentially half of them) the remaining ones should be removed as well. These components  may \eg be  projected out by the additional condition \cite{Mack:1969}, 
\be
&&(S_{-\mu} f)^i = 0,\;\;\; {\rm for \ all}\;\; \mu\;\;{\rm and} \;\;i.
\e{unphysical}
\beq
 \begin{tabular}{|p{114mm}|}
 \hline
  {\bf Proposal 7}:  The remaining unphysical components projected out by the condition \rl{unphysical} automatically disappear  from the consistent actions and their equations. No additional conditions like \rl{unphysical} are therefore required for such theories.\\
 \hline
 \end{tabular}
 \eq
 This statement is true for the models $s=1$ and $s=2$ as was carefully demonstrated in \cite{Arvidsson:2006}, and the results of this paper indirectly prove this to be true also for $s=3$ and $s=4$.

{\bf Remarks}: In order for tensor fields $\phi_{AB\cdots}$ to be completely arbitrary off the hypercone $y^2=0$ the functions $\tilde{\phi}_{AB\cdots}$ in the special gauge transformations \rl{506} must be completely arbitrary. However, the strong transversality conditions \rl{503} are also required for $\tilde{\phi}_{AB\cdots}$. A way to avoid this restriction is to express $\phi_{AB\cdots}$ in terms of temporarily defined auxiliary fields which only satisfy weak transversality. For instance, in the following I consider symmetric fields $\phi_{AB\cdots}$ of rank $s$. For them one has  to write
\be
&\phi_{A_1A_2\cdots A_s}\equiv &V_{A_1A_2\cdots A_s}+y_{(A_1}K_{1\;A_2\cdots A_s)}+y_{(A_1}y_{A_2}K_{2\;A_3\cdots A_s)}+\cdots\nn\\&&\cdots+K_sy_{A_1}y_{A_2}\cdots y_{A_s},
\e{VK}
where $V$ and $K_i$ are auxiliary symmetric fields with the homogeneity $n$ and $n-i$ respectively. They are recursively related by weak transversalities of the type
\be
&&y\cdot V=y^2K_{1},\quad y\cdot K_r=y^2K_{r+1}\;\;\; {\rm for}\;\; r=1,2,\ldots,s-1,
\e{VKtransv}
such that $\phi$ in \rl{VK} satisfies strong transversalities. ($K_s$ is a scalar and satisfies no transversality.) For  $V$ and $K_i$ one may then apply the above procedure.   Notice, however,  that  the original field $\phi$ does not depend entirely on $x^{\mu}$ when  this procedure is followed due to the explicit $y$'s. Details for $s=1$ and $s=2$ for the fields in section 10 are given in \cite{Arvidsson:2006}. 

\noindent{\bf Reduction of the equations of motion}:\\
The equations of motion \rl{507} contains a delta function $\del(\rho)$. Thus, it has to be integrated over $\rho$ which then only picks out the $\rho=0$ part of the manifest equations. (Only the equations on the hypercone are relevant \cite{Dirac:1936}.) From the required homogeneity \rl{505} of the Lagrangian $\cL(y)$ it follows that the manifest equations  multiplying $\del(y^2)$ have the homogeneity $-d-n$ if the field in question has homogeneity $n$. One may then follow the  procedure above to impose \rl{homo_ga} and use the definition \rl{phi_tilde} to produce well defined spacetime equations for any finite value of $\ga$. Since the equations are zero the appropriate operator $V(x)$ in \rl{V}   may be multiplied.

When the basic field is a tensor field, $\phi_{AB\cdots}$, then one has to be more careful due to the strong transversality conditions. For instance, a variation of a Lagrangian of the type to be considered in the next section, \rl{609}, yields
\be
&&\del(\del(y^2)\cL)=\del(y^2)\dif^s(\;\;)\cdot\del\phi+\dif\cdot(\;\;).
\e{del}
In this case one finds the equations of motion, $E=0$, where
\be
&&E\equiv\del(y^2)\dif^s(\;\;)+N,
\e{E}
where in turn $N$ satisfies
the identity $N\cdot\del\phi\equiv 0$, which only may yield a nontrivial $N$ if $\phi$ satisfies a restriction
like strong transversality. In fact, strong transversality implies that $N$ may be quite arbitrary if only it contains explicit $y^A$-factors since $y\cdot\del\phi\equiv 0$. In order to reduce the equations of motion one has first to determine $N$. In fact, the part of $\dif^s(\;\;)$ that contains explicit $y$'s from the rewriting in \rl{VK}  determines $N$ from $E=0$. What remains is then the spacetime equations after integration over $\rho$ and multiplication of $\ga^{n+d}$. (All details for $s=1$ and $s=2$ are given \cite{Arvidsson:2006}. In these cases $N$ is expressed in terms if the unphysical components projected out by the condition \rl{unphysical}.) 

\noindent{\bf Reductions of Lagrangians}:\\
The Lagrangians may be reduced to $d$ dimensions as follows: From the form of the action \rl{504} one has
\be
&&\cL(y)=\cL(y,y^2=0).
\e{a010}
From \rl{505}, \ie
\be
&&(y\cdot\dif+d)\cL(y)=0,
\e{a011}
it follows then from the steps \rl{homo_ga} and \rl{phi_tilde} that
\be
&&\cL(x)=\ga^{-d}\cL(y,y^2=0).
\e{a012}
 Since the Lagrangian  $\cL(y)$ is a scalar there are no further modifications. 
 
 \noindent{\bf Reductions of actions}:\\
 The action becomes from \rl{a010}-\rl{a012}
\be
&&A=\int d y \del(y^2)\cL(y)=\int d y \del(y^2)\cL(y,y^2=0)=c\int dx \cL(x),\nn\\
\e{a013}
where
\be
&&c\equiv\half\int {d \ga\over \ga}.
\e{a014}
This constant is badly defined. In fact, the projection procedure \cite{Mack:1969} which is followed here is not well defined in the limits $\ga\rightarrow \pm 0$ and  $\ga\rightarrow \pm \infty$.  However, since \rl{a011} is a consistent spacetime Lagrangian it follows that one may  consistently  treat $c$ as a finite constant. This should be studied further (see \cite{Arvidsson:2006}). (One may also notice that the projected spacetime should be a compactified Minkowski space \cite{Penrose:1965am}.)
\beq
 \begin{tabular}{|p{114mm}|}
 \hline
  {\bf Proposal 8}:  The reduced Lagrangians \rl{a012} (and actions \rl{a013}) produce equations of motion which are equal to the reduced equations of motion.\\
 \hline
 \end{tabular}
 \eq
 This is true for the $s=1$ and $s=2$ theories treated in \cite{Arvidsson:2006}, and the results of this paper indirectly prove this to be true also for $s=3$ and $s=4$.

 \setcounter{equation}{0}
 \section{The construction of manifestly conformally invariant  theories for arbitrary integer $s$}
I follow here the procedure in \cite{Arvidsson:2006} for $s=1,2$. This procedure requires   fields on the conformal space which are  analogous to  the spacetime fields already considered. The spacetime field strength \rl{4} is therefore just formally  lifted to a  field strength on the $d+2$-dimensional conformal space, \ie
 \be
 &&F_{A_1B_1A_2B_2\cdots A_sB_s}(y)=\dif_{[A_s}\cdots\dif_{[A_2}\dif_{[A_1}\phi_{B_1]B_2]\cdots B_s]}(y).
 \e{601}
 This is the starting point here. Notice that the actual relation between \rl{601} and \rl{4} is highly nontrivial (see  next section). 
In addition to the properties in the spacetime treatment  one has  in conformal space to require the conditions of transversality and homogeneity. First, one has to impose the transversality \rl{503} for $\phi$, \ie
 \be
 &&y^{B_1}\phi_{B_1B_2\cdots B_s}(y)=0.
 \e{602}
 I would also like the field strength \rl{601} to be a satisfactory field satisfying transversality. Now 
the transversality of the field strength \rl{601}, \ie
 \be
  &&y^{A_1}F_{A_1B_1A_2B_2\cdots A_sB_s}(y)=0,
 \e{603}
 requires
 \be
 &&\dif_{A_s}(y^{B_s}G_{A_1B_1A_2B_2\cdots A_{s-1}B_{s-1}B_s}(y))-\nn\\&&-(y\cdot\dif+1)G_{A_1B_1A_2B_2\cdots A_{s-1}B_{s-1}A_s}(y)=0,
 \e{604}
 where
 \be
 &&G_{A_1B_1A_2B_2\cdots A_{s-1}B_{s-1}C}\equiv\dif_{[A_{s-1}}\cdots\dif_{[A_2}\dif_{[A_1}\phi_{B_1]B_2]\cdots B_{s-1}]C}(y).\nn\\
 \e{605}
 When the condition \rl{602} is inserted into \rl{604} one finds the condition
  \be
 &&(y\cdot\dif+1)G_{A_1B_1A_2B_2\cdots A_{s-1}B_{s-1}A_s}(y)=0,
 \e{606}
 which in turn from \rl{605} requires
 \be
 &&(y\cdot\dif-s+2)\phi_{B_1B_2\cdots B_s}(y)=0,
 \e{607}
 since a derivative has the degree of homogeneity $-1$. Remarkably enough the homogeneity property \rl{607} is exactly what is required for a symmetric external field in the action for the conformal particle ( see \cite{Arvidsson:2006}). (Also found in a different setting in \cite{Bars:2001um}.) It is the correct homogeneity for a vector potential $s=1$, and it agrees with the homogeneity found for $s=2$ in \cite{Arvidsson:2006}.
 
 In the following I require the homogeneity \rl{607} for the basic, symmetric gauge field $\phi$ for arbitrary integer $s$. As a consequence the field strength, $F$, in \rl{601} is  transversal for arbitrary $s$, \ie it  satisfies the transversality condition \rl{603}. Now the homogeneity property \rl{607} also implies (through \rl{601} and \rl{606})
 \be
 &&(y\cdot\dif+2)F_{A_1B_1A_2B_2\cdots A_sB_s}(y)=0.
 \e{608}
 Hence a manifestly conformal Lagrangian satisfying the condition \rl{505} for $d=4$ must have the form
 \be
 &&\cL(y)=F^2(y)+\al F'^2(y)+\beta F''^2(y)+\ldots,
 \e{609}
 where $F', F''$ etc are various traces of $F$.  $\al, \beta,\ldots$ are real constants. Notice that since the traces do not affect homogeneities we have
 \be
 &&(y\cdot\dif+2)F'_{\cdots}=0,\quad (y\cdot\dif+2)F''_{\cdots}=0, \quad etc.
 \e{610}
 The form of the Lagrangian \rl{609} is exactly what we have considered in spacetime but now formally lifted to the  two dimensions higher conformal space. In order to be able to reduce the action corresponding to \rl{609} to spacetime we have to require invariance under the special gauge transformations \rl{506} (see  next section). As we shall see this invariance determines the parameters  $\al, \beta,\ldots$ in \rl{609} (cf  \cite{Arvidsson:2006}). Below the constants are determined for $s=1,2,3,4.$
 
 \subsection{The $s=1$ case. The Maxwell theory.}
In this caes there are no constants to be determined.
 Free spin one in $d=4$ is in the manifest formulation given by the Lagrangian
 \be
 &&\cL={1\over4}F_{AB}F^{AB},\quad F_{AB}=\dif_AA_B-\dif_BA_A.
 \e{6101}
 Under the special gauge transformations,
 \be
 &&A_A\quad\longrightarrow\quad A_A+y^2\tilde{A}_A,
 \e{6102}
 the Lagrangian \rl{6101} transforms as \cite{Arvidsson:2006} (I ignore $y^2$-terms due to the delta function $\del(y^2)$ in the action \rl{504})
 \be
&&\cL\quad\longrightarrow\quad\cL+\tilde{\cL}_1+\tilde{\cL}_2,
 \e{6103}
 where
 \be
 &&\tilde{\cL}_1\equiv F^{AB}(y_A\tilde{A}_B-y_B\tilde{A}_A)
 \e{6104}
 is linear in $\tilde{A}_A$, and
 \be
 &&\tilde{\cL}_2\equiv-2(y_A\tilde{A}_A)^2
 \e{6105}
 is quadratic in $\tilde{A}_A$. $\tilde{\cL}_1=0$ since $F_{AB}$ is transversal, and $\cL_2=0$ since $\tilde{A}_A$ is transversal. 

 \subsection{The $s=2$ case. Linear conformal gravity.}
 For  $s=2$ we have the linear Riemann tensor
 \be
 &&R_{ABCD}(y)=\half F_{ABCD}(y)=\half \dif_{[C}\dif_{[A}H_{B]D]}(y),
 \e{611}
 where $H_{AB}$ is the deviation of the metric tensor $G_{AB}$ from $\eta_{AB}$, \ie
 \be
 &&G_{AB}(y)=\eta_{AB}+H_{AB}(y).
 \e{612}
 (I choose here the notation $H_{AB}$ instead of $\phi_{AB}$ in accordance with the treatment in \cite{Arvidsson:2006}.)
 Since
 \be
 &&y\cdot\dif G_{AB}(y)=0\quad\Rightarrow\quad y\cdot\dif H_{AB}(y)=0,
 \e{613}
 and
 \be
 &&y^AH_{AB}(y)=0,
 \e{614}
 the linear Riemann tensor \rl{611} satisfies the transversality properties \rl{603}
 \be
 &&y^AR_{ABCD}(y)=0,
 \e{615}
 and the homogeneity
 \be
 &&(y\cdot\dif+2)R_{ABCD}(y)=0.
 \e{616}
 A Lagrangian satisfying the condition \rl{505} is therefore
 \be
 &&\cL(y)=R_{ABCD}(y)R^{ABCD}(y)+\al R_{AB}(y)R^{AB}(y)+\beta R^2(y),\nn\\
 \e{617}
 where $\al$ and $\beta$ are real constants, and where $R_{AB}$ and $R$ are the linear Ricci tensor and curvature scalar respectively, \ie
 \be
 &&R_{AC}(y)\equiv \eta^{BD}R_{ABCD}(y),\quad R(y)\equiv \eta^{AC}R_{AC}(y).
 \e{618}
 The real constants $\al$ and $\beta$ are determined if we require invariance under the special gauge transformations 
 \be
 &&H_{AB}(y)\;\longrightarrow\;H_{AB}(y)+y^2 \tilde{H}_{AB}(y),
 \e{619}
 where $\tilde{H}_{AB}$ satisfies
\be
&&y^A\tilde{H}_{AB}(y)=0,\quad (y\cdot \dif+2)\tilde{H}_{AB}(y)=0,
\e{620}
which secures the properties \rl{613} and \rl{614}. The linear Riemann and Ricci tensors as well as the curvature scalar transform under \rl{619} in arbitrary dimensions $d$ as follows: (I ignore terms multiplied by $y^2$ since they vanish in the action  \rl{504} due to the delta function $\del(y^2)$ in the measure.)
\be
&&R_{ABCD}(y)\;\;\ra\;\;R_{ABCD}(y)+\tilde{R}_{ABCD}(y),\nn\\
&&\tilde{R}_{ABCD}(y)\equiv
\eta_{AC}\tilde{H}_{BD}(y)+\eta_{BD}\tilde{H}_{AC}(y)-\nn\\&&-\eta_{BC}\tilde{H}_{AD}(y)-\eta_{AD}\tilde{H}_{BC}(y)+\nn\\&&+y_A(\dif_C\tilde{H}_{BD}(y)-\dif_D\tilde{H}_{BC}(y))+\nn\\&&+y_B(\dif_D\tilde{H}_{AC}(y)-\dif_C\tilde{H}_{AD}(y))+\nn\\&&+y_C(\dif_A\tilde{H}_{BD}(y)-\dif_B\tilde{H}_{AD}(y))+\nn\\&&+y_D(\dif_B\tilde{H}_{AC}(y)-\dif_A\tilde{H}_{BC}(y)),
\e{621}
which implies
\be
&&R_{BD}(y)\;\;\ra\;\;R_{BD}(y)+\tilde{R}_{BD}(y),\nn\\
&&\tilde{R}_{BD}(y)\equiv\eta^{AC}\tilde{R}_{ABCD}(y)=(d-2)\tilde{H}_{BD}(y)+\eta_{BD}\tilde{H}(y)+\nn\\&&+y_B\dif_D\tilde{H}(y)+y_D\dif_B\tilde{H}(y)-\nn\\&&-y_B\dif_C\tilde{H}^C_{\;\;D}(y)-y_D\dif_C\tilde{H}^C_{\;\;B}(y),
\e{622}
and
\be
&&R(y)\;\;\ra\;\;R(y)+\tilde{R}(y),\nn\\
&&\tilde{R}(y)\equiv\eta^{BD}\tilde{R}_{BD}(y)=2(d-1)\tilde{H}(y),
\e{623}
where $\tilde{H}\equiv\eta^{AB}\tilde{H}_{AB}$. When these transformations are inserted into the Lagrangian \rl{617} I find (ignoring terms with explicit $y^2$-dependence, and using the transversality of $R$ in \rl{615})
\be
&&\cL(y)\;\longrightarrow\; \cL(y)+\tilde{\cL}_1(y)+\tilde{\cL}_2(y),
\e{624}
where $\tilde{\cL}_1(y)$ and $\tilde{\cL}_2(y)$ depend linearly and quadratically on $\tilde{H}_{AB}$.
They are explicitly
\be
&\tilde{\cL}_1(y)=&4\biggl(4+\al(d-2)\biggr)R_{AB}(y)\tilde{H}^{AB}(y)+\nn\\&&+4\biggl(\al+2\beta(d-1)\biggr)
R(y)\tilde{H}(y),
\e{625}
\be
&\tilde{\cL}_2(y)=&(d-2)\biggl(4+\al(d-2)\biggr)\tilde{H}_{AB}\tilde{H}^{AB}+\nn\\&&+\biggl(4+\al(3d-4)+4\beta(d-1)^2\biggr)\tilde{H}^2(y).
\e{626}
The linear condition,
\be
&&\tilde{\cL}_1(y)=0,
\e{627}
yields
\be
&&\al=-{4\over(d-2)},\quad\beta={2\over(d-1)(d-2)}.
\e{628}
Also the quadratic condition,
 \be
&&\tilde{\cL}_2(y)=0,
\e{629}
requires \rl{628}. Hence, the Lagrangian \rl{617} is conformally invariant in $d=4$ for 
\be
&&\al=-2,\quad\beta={1\over3}.
\e{630}
The corresponding Lagrangian \rl{617} corresponds then to linear conformal gravity \cite{Arvidsson:2006}.

 \subsection{The $s=3$ case.}
 The basic field strength in conformal space is here
 \be
 &&F_{A_1B_1A_2B_2A_3B_3}(y)=\dif_{[A_3}\dif_{[A_2}\dif_{[A_1}\phi_{B_1]B_2]B_3]}(y)
 \e{631}
 from \rl{601}. The symmetric gauge field $\phi_{ABC}$ satisfies
 \be
 &&y^A\phi_{ABC}(y)=0,\quad(y\cdot\dif-1)\phi_{ABC}(y)=0,
 \e{632}
 according to \rl{602} and \rl{607}.
This implies
 \be
 &&y^{A_1}F_{A_1B_1A_2B_2A_3B_3}(y)=0,\quad(y\cdot\dif+2)F_{A_1B_1A_2B_2A_3B_3}(y)=0.\nn\\
 \e{633}
 In order to make the ansatz \rl{609} for the Lagrangian  I have to determine the independent traces of $F$. However, this was already done in the spacetime case. Hence, 
I find the following traces from subsection 4.3
 \be
 &&F'_{A_1A_2A_3B_3}\equiv \eta^{B_1B_2}F_{A_1B_1A_2B_2A_3B_3},\nn\\
 &&F''_{A_3B_3}\equiv\half\eta^{A_1A_2}F'_{A_1A_2A_3B_3}=\half  \eta^{A_1A_2} \eta^{B_1B_2}F_{A_1B_1A_2B_2A_3B_3},\nn\\
 \e{634}
 where $F'$ is symmetric in $A_1$ and $A_2$, and antisymmetric in $A_3$ and $B_3$, and where $F''$ is antisymmetric. A different way to arrive at $F''$ is
 \be
 &&F''_{A_2B_3}=\eta^{A_1A_3}F'_{A_1A_2A_3B_3}= \eta^{B_1B_2}\eta^{A_1A_3}F_{A_1B_1A_2B_2A_3B_3}.
 \e{635}
Also here the field strength \rl{631} satisfies cyclicity properties like \rl{2} which for $F'$ imply,
 \be
 &&F'_{A_1[A_2A_3]B_3}=-F'_{A_1B_3A_2A_3},
 \e{636}
 which is a  property used in the following.
 
 The general ansatz for the Lagrangian in conformal space is then
 \be
 &\cL(y)=&F_{A_1B_1A_2B_2A_3B_3}F^{A_1B_1A_2B_2A_3B_3}+\nn\\&&+\al F'_{A_1A_2A_3B_3}F^{\prime\;A_1A_2A_3B_3}+\beta F''_{A_2B_3}F^{\prime\prime\;A_2B_3}.
 \e{637}
 
 It remains to investigate the possibility of $\cL$ to be  invariant under the special gauge transformations,
 \be
 &&\phi_{ABC}(y)\;\;\longrightarrow\;\;\phi_{ABC}(y)+y^2\tilde{\phi}_{ABC}(y),\nn\\
 &&y^A\tilde{\phi}_{ABC}(y)=0,\quad(y\cdot\dif+1)\tilde{\phi}_{ABC}(y)=0.
 \e{638}
These transformations induce  the following transformation properties of the field and its traces ignoring explicit $y^2$-terms
 \be
 &F_{A_1B_1A_2B_2A_3B_3}(y)\;\longrightarrow\;&F_{A_1B_1A_2B_2A_3B_3}(y)+\tilde{F}_{A_1B_1A_2B_2A_3B_3}(y),\nn\\
 \e{639}
 where $\tilde{F}$  is 
 the $\tilde{\phi}$-dependent part.  $\tilde{F}$ may be split into two parts, $\tilde{F}=\tilde{F}|_0+\tilde{F}|_y$, where $\tilde{F}|_0$ has no explicit $y$-dependence, and where $\tilde{F}|_y$ has explicit $y$-dependence. These parts are explicitly
 \be
 &\tilde{F}_{A_1B_1A_2B_2A_3B_3}(y)|_0=&2\biggl(\eta_{[A_2[A_1}\tilde{L}_{B_1]B_2]A_3B_3}(y)+
\eta_{[A_3[A_1}\tilde{L}_{B_1]B_3]A_2B_2}(y)+\nn\\&&+\eta_{[A_3[A_2}\tilde{L}_{B_2]B_3]A_1B_1}(y)\biggr),
 \e{640}
 \be
 &\tilde{F}_{A_1B_1A_2B_2A_3B_3}(y)|_y=&2\biggl(y_{[A_1}\tilde{K}_{B_1]A_2B_2A_3B_3}(y)+
 y_{[A_2}\tilde{K}_{B_2]A_1B_1A_3B_3}(y)+\nn\\&&+y_{[A_3}\tilde{K}_{B_3]A_1B_1A_2B_2}(y)\biggr),
 \e{641}
 where in turn
 \be
 &&\tilde{L}_{B_1B_2A_3B_3}(y)\equiv\dif_{[A_3}\tilde{\phi}_{B_3]B_1B_2}(y),\nn\\
&&\tilde{K}_{B_1A_2B_2A_3B_3}(y)\equiv\dif_{[A_2}\dif_{[A_3}\tilde{\phi}_{B_3]B_2]B_1}(y)=\tilde{K}_{B_1A_3B_3A_2B_2}(y).\nn\\
 \e{642}
$\tilde{L}_{B_1B_2A_3B_3}$ is transverse in all indices, while  $\tilde{K}_{B_1A_2B_2A_3B_3}$ is transverse only in the first index. For the other indices I find
\be
&&y^{A_3}\tilde{K}_{B_1A_2B_2A_3B_3}(y)=-\tilde{L}_{B_1B_3A_2B_2}(y),\nn\\&&y^{A_2}\tilde{K}_{B_1A_2B_2A_3B_3}(y)=-\tilde{L}_{B_1B_2A_3B_3}(y).
\e{643}
The expression \rl{639} implies
\be
&&F'_{B_1B_2A_3B_3}(y)\;\longrightarrow\;F'_{B_1B_2A_3B_3}(y)+\tilde{F}'_{B_1B_2A_3B_3}(y),
\e{644}
where again  the $\tilde{\phi}$-dependent terms, $\tilde{F}'$, may be split into two parts, $\tilde{F}'=\tilde{F}'|_0+\tilde{F}'|_y$, where
\be
&\tilde{F}'_{B_1B_2A_3B_3}(y)|_0\equiv &2\biggl(d\tilde{L}_{B_1B_2A_3B_3}(y)+\eta_{B_1B_2}\tilde{L}'_{A_3B_3}(y)-\nn\\
&&-\eta_{B_1[A_3}\tilde{M}_{B_3]B_2}(y)-\eta_{B_2[A_3}\tilde{M}_{B_3]B_1}(y)\biggr),\nn\\
&\tilde{F}'_{B_1B_2A_3B_3}(y)|_y\equiv&2\biggl(y_{A_3}\tilde{N}_{B_3]B_1B_2}(y)-\nn\\&&-y_{B_1}\tilde{K}'_{B_2A_3B_3}(y)-y_{B_2}\tilde{K}'_{B_1A_3B_3}(y)\biggr),
\e{645}
where in turn
\be
&\tilde{L}'_{A_3B_3}(y)\equiv&\eta^{A_1B_1}\tilde{L}_{A_1B_1A_3B_3}(y)=\dif_{[A_3}\tilde{\phi}_{B_3]}(y),\nn\\
&\tilde{M}_{B_2B_3}(y)\equiv &\eta^{A_2A_3}\tilde{L}_{A_2B_2A_3B_3}(y)=\dif^{A_2}\tilde{\phi}
_{A_2B_2B_3}(y)-\dif_{B_3}\tilde{\phi}_{B_2}(y),  \nn\\
&\tilde{K}'_{A_2B_2A_3}(y)\equiv&\eta^{A_1B_1}\tilde{K}_{A_1B_1A_2B_2A_3}(y)=\nn\\
&&\dif_{[B_2}\dif^{A}\tilde{\phi}_{AA_2A_3]}(y)-\dif_{A_2}\dif_{[B_2}\tilde{\phi}_{A_3]}(y),\nn\\
&\tilde{N}_{A_1A_2A_3}(y)\equiv &\eta^{B_1B_2}\tilde{K}_{A_1B_1A_2B_2A_3}(y)=\Box\tilde{\phi}_{A_1A_2A_3}(y)+\nn\\&&+\dif_{A_2}\dif_{A_3}\tilde{\phi}_{A_1}(y)-\dif_{A_2}\dif^A\tilde{\phi} _{AA_3A_1}(y)-\dif_{A_3}\dif^A\tilde{\phi} _{AA_2A_1}(y). \nn\\
&\tilde{\phi}_{A}(y)\equiv&\eta^{BC}\tilde{\phi}_{ABC}(y)
\e{646}
The expression \rl{644} implies in turn
\be
&&F''_{A_3B_3}(y)\;\longrightarrow\;F''_{A_3B_3}(y)+\tilde{F}''_{A_3B_3}(y),
\e{647}
where $\tilde{F}''_{A_3B_3}=\tilde{F}''_{A_3B_3}|_0+\tilde{F}''_{A_3B_3}|_y$ with
\be
&&\tilde{F}''_{A_3B_3}(y)|_0=2(d+1)\tilde{L}'_{A_3B_3}(y),\nn\\
&&\tilde{F}''_{A_3B_3}(y)(y)|_y=y_{[A_3}\tilde{N}_{B_3]}(y).
\e{648}
$\tilde{N}_{A}$ is given by
\be
&&\tilde{N}_{A_1}(y)\equiv\eta^{A_2A_3}\tilde{N}_{A_1A_2A_3}(y)=2\bigl(\Box\tilde{\phi}_{A_1}(y)-\dif^B\dif^C\tilde{\phi}_{CBA_1}(y)\bigr).\nn\\
\e{6481}

The transformations \rl{639}, \rl{644} and \rl{647} yield for the Lagrangian \rl{637} (up to terms with $y^2$-dependence) 
\be
&&\cL(y)\;\;\longrightarrow\;\;\cL(y)+\tilde{\cL}_1(y)+\tilde{\cL}_2(y),
\e{649}
where 
\be
&&\tilde{\cL}_1(y)\equiv2\bigl(F\cdot\tilde{F}+\al F'\cdot\tilde{F}'+\beta F''\cdot\tilde{F}''\bigr),\nn\\
&&\tilde{\cL}_2(y)\equiv\tilde{F}^2+\al(\tilde{F}')^2+\beta(\tilde{F}'')^2.
\e{6491}
$\tilde{\cL}_1(y)$ is linear in $\tilde{\phi}$, and $\tilde{\cL}_2(y)$ is quadratic.  The invariance of the action requires therefore the vanishing of both $\tilde{\cL}_1(y)$ and $\tilde{\cL}_2(y)$. Since $F$, $F'$ and $F''$ are transverse the vanishing of the linear terms  requires
\be
&&\half\tilde{\cL}_1(y)=F\cdot\tilde{F}|_0+\al F'\cdot\tilde{F}'|_0+\beta F''\cdot\tilde{F}''|_0=0.
\e{650} 
This condition is explicitly
\be
&&(24+2\al d)F^{\prime\;A_1B_1A_2B_2}(y)\tilde{L}_{A_1B_1A_2B_2}(y)+\nn\\&&+(8+2\beta(d+1))F^{\prime\prime\;A_3B_3}(y)\tilde{L}'_{A_3B_3}(y)=0,
\e{651}
with the solution
\be
&&\al=-{12\over d},\quad \beta={48\over d(d+1)}.
\e{652}
The vanishing of the quadratic terms in $\tilde{\phi}$ in the Lagrangian \rl{637} requires $\tilde{\cL}_2(y)=0$
which according to appendix D  becomes
\be
&&(12+\al d)(4d\tilde{L}^2-16\tilde{M}^2)+
4(36+\al(7d+4)+\beta(d+1)^2)(\tilde{L}')^2=0.\nn\\
\e{654}
This condition is satisfied for exactly the same values as the linear condition \rl{650}, \ie \rl{652}. This is in agreement with proposal 5 in section 8.

Let me finally point out that $F''_{AB}$ has imilar properties to the Maxwell field for $S=1$. I notice the following relations
\be
&&F''_{AB}=\dif_{[A}B_{B]},\quad B_B\equiv\Box\eta^{AC}\phi_{ABC}-\dif^A\dif^B\phi_{ABC}.
\e{6541}
$y^AF''_{AB}=0$ implies $y^AB_A=0$ which also is consistent with $y^A\phi_{ABC}=0$. The gauge invariance under ($V_{AB}$ is symmetric)
\be
&&\phi_{ABC}\;\;\longrightarrow\;\;\phi_{ABC}+\dif_AV_{BC}+\dif_BV_{CA}+\dif_CV_{AB},
\e{6542}
implies
\be
&&B_A\;\;\longrightarrow\;\;B_A+\dif_AV,\quad V\equiv(\Box\eta^{AB}-\dif^A\dif^B)V_{AB}.
\e{6543}
However, the special gauge transformation \rl{638} implies
\be
&&B_A\;\;\longrightarrow\;\;B_A+y^2\tilde{B}_A+2(d+1)\tilde{\phi}_A,
\e{6544}
where $\tilde{B}_A$ is $B_A$ in \rl{6541} with $\phi$ replaced by $\tilde{\phi}$. ($\tilde{N}_A=s\tilde{B}_A$ in \rl{6481}.)

  \subsection{The $s=4$ case.}
  For $s=4$ the basic field strength \rl{601} is
  \be
&&F_{A_1B_1A_2B_2A_3B_3A_4B_4}(y)=\dif_{[A_1}\dif_{[A_2}\dif_{[A_3}\dif_{[A_4}\phi_{B_4]B_3]B_2]B_1]}(y),
 \e{655}
 where $\phi$ is totally symmetric. From the spacetime treatment  in subsection 4.4 I have the following five different traces
 \be
 &&F'_{B_1B_2A_3B_3A_4B_4}\equiv \eta^{A_1A_2}F_{A_1B_1A_2B_2A_3B_3A_4B_4},\nn\\
&&F''_{A_3B_3A_4B_4}\equiv\half\eta^{B_1B_2}F'_{B_1B_2A_3B_3A_4B_4}=
\half  \eta^{A_1A_2} \eta^{B_1B_2}F_{A_1B_1A_2B_2A_3B_3A_4B_4},\nn\\
 &&G''_{B_1B_2B_3B_4}\equiv\eta^{A_3A_4} F'_{B_1B_2A_3B_3A_4B_4}=\eta^{A_1A_2} \eta^{A_3A_4}    F_{A_1B_1A_2B_2A_3B_3A_4B_4}, \nn\\
 &&F'''_{B_3B_4}\equiv\eta^{A_3A_4}F''_{A_3B_3A_4B_4}=\half  \eta^{B_1B_2}\eta^{A_1A_2} \eta^{A_3A_4}  F_{A_1B_1A_2B_2A_3B_3A_4B_4},\nn\\
 &&U\equiv\eta^{B_3B_4}F'''_{B_3B_4}=\half  \eta^{B_1B_2} \eta^{B_3B_4}\eta^{A_1A_2} \eta^{A_3A_4}F_{A_1B_1A_2B_2A_3B_3A_4B_4},
 \e{656}
 where $F'$ is symmetric in $A_1$ and $A_2$ and where both $F'$ and $F''$ are antisymmetric in $A_3B_3$ and $A_4B_4$ as well as symmetric under interchange of the pairs $A_3B_3$ and $A_4B_4$.   $G''$ is symmetric in $B_1B_2$ and $B_3B_4$, and symmetric under interchange of the pairs $B_1B_2$ and $B_3B_4$.
 
For $F''$ I find the following alternative relation
 \be
&&F''_{B_2B_3A_4B_4}=\eta^{B_1A_3}F'_{B_1B_2A_3B_3A_4B_4}=\eta^{B_1A_3}\eta^{A_1A_2}F_{A_1B_1A_2B_2A_3B_3A_4B_4},\nn\\
 \e{657}
 which implies
 \be
&&F'''_{B_3B_4}=\eta^{B_2A_4}F''_{B_2B_3A_4B_4}=\eta^{B_2A_4}\eta^{B_1A_3}\eta^{A_1A_2}F_{A_1B_1A_2B_2A_3B_3A_4B_4},\nn\\
&&U=\eta^{B_3B_4}F'''_{B_3B_4}=\eta^{B_3B_4}\eta^{B_2A_4}\eta^{B_1A_3}\eta^{A_1A_2}F_{A_1B_1A_2B_2A_3B_3A_4B_4},\nn\\
\e{658}
In addition I find the property
\be
&&G''_{B_1[B_2B_3]B_4}=F''_{B_2B_3B_1B_4}.
\e{659}
This may also be viewed as still another relation for $F''$,
\be
&&F''_{B_2B_3B_1B_4}=G''_{B_1[B_2B_3]B_4}=\nn\\
&&=(\eta^{A_1A_2}\eta^{A_3A_4}-
\eta^{A_1A_3}\eta^{A_2A_4})F_{A_1B_1A_2B_2A_3B_3A_4B_4}.\nn\\
\e{660}

The cyclicities \rl{2} imply here
\be
&&F'_{B_1[B_2A_3]B_3A_4B_4}=-F'_{B_1B_3B_2A_3A_4B_4},\nn\\
&&F'_{B_1B_2A_3[B_3A_4]B_4}=-F'_{B_1B_2A_3B_4B_3A_4},\nn\\
&&F''_{A_3[B_3A_4]B_4}=-F''_{A_3B_4B_3A_4}.
\e{661}

 The general ansatz for the Lagrangian in conformal space is then
 \be
 &\cL(y)=&F_{A_1B_1A_2B_2A_3B_3A_4B_4}F^{A_1B_1A_2B_2A_3B_3A_4B_4}+\nn\\&&+\al F'_{B_1B_2A_3B_3A_4B_4}F^{\prime\;B_1B_2A_3B_3A_4B_4}+\beta F''_{A_3B_3A_4B_4}F^{\prime\prime\;A_3B_3A_4B_4}+\nn\\&&+\ga G''_{B_1B_2B_3B_4}G^{\prime\prime\;B_1B_2B_3B_4}+\la F'''_{B_3B_4}F^{\prime\prime\prime\;B_3B_4}+\rho U^2,
 \e{662}
 where $\al,\beta,\ga,\la$ and $\rho$ are real constants.
 
 It remains to investigate the possible invariance of $\cL$ under the special gauge transformations,
 \be
 &&\phi_{ABCD}(y)\;\;\longrightarrow\;\;\phi_{ABCD}(y)+y^2\tilde{\phi}_{ABCD}(y),\nn\\
 &&y^A\tilde{\phi}_{ABCD}(y)=0,\quad y\cdot\dif\tilde{\phi}_{ABCD}(y)=0.
 \e{663}
  Ignoring explicit $y^2$-terms I find the following transformation properties of the field \rl{655}
 \be
&&F_{A_1B_1A_2B_2A_3B_3A_4B_4}(y)\;\longrightarrow\;\nn\\
&&F_{A_1B_1A_2B_2A_3B_3A_4B_4}(y)+\tilde{F}_{A_1B_1A_2B_2A_3B_3A_4B_4}(y),\nn\\
 \e{664}
 where $\tilde{F}$ is the $\tilde{\phi}$-dependent part.  $\tilde{F}$ as well as all its traces are given explicitly in \rl{c1}-\rl{c10} in appendix E. These transformations imply for the Lagrangian \rl{662} 
\be
&&\cL(y)\;\;\longrightarrow\;\;\cL(y)+\tilde{\cL}_1(y)+\tilde{\cL}_2(y),
\e{665}
where 
\be
&&\tilde{\cL}_1(y)\equiv2\bigl(F\cdot\tilde{F}+\al F'\cdot\tilde{F}'+\beta F''\cdot\tilde{F}'' +\ga G''\cdot\tilde{G}''+   \la F'''\cdot\tilde{F}'''+  \rho U\tilde{U} \bigr),\nn\\
&&\tilde{\cL}_2(y)\equiv\tilde{F}^2+\al(\tilde{F}')^2+\beta(\tilde{F}'')^2+\ga(\tilde{G}'')^2+\la(\tilde{F}''')^2+\rho\tilde{U}^2.
\e{666}
Invariance under the special gauge transformations \rl{663} requires the vanishing of both $\tilde{\cL}_1$ and $\tilde{\cL}_2$. Since $F$, $F'$, $F''$, $G''$, and $F'''$ all  are transverse the vanishing of the linear terms  requires
\be
&&\half\tilde{\cL}_1(y)=F\cdot\tilde{F}|_0+\al F'\cdot\tilde{F}'|_0+\beta F''\cdot\tilde{F}''|_0+\nn\\&&+\ga G''\cdot\tilde{G}''|_0+   \la F'''\cdot\tilde{F}'''|_0+  \rho U\tilde{U}=0.
\e{667} 
From formulas \rl{c12}-\rl{c17} in appendix E I find the following explicit form
\be
&&c_1F'^{B_1B_2A_3B_3A_4B_4}(y)\dif_{[A_3}\dif_{[A_4}\tilde{\phi}_{B_4]B_3]B_2B_1}+\nn\\&&+c_2 F''^{A_3B_3A_4B_4}(y)\dif_{[A_3}\dif_{[A_4}\tilde{\phi}_{B_4]B_3]}+\nn\\&&+c_3 G''^{B_1B_2B_3B_4}(y)\bigl(\Box\tilde{\phi}_{B_1B_2B_3B_4}+\dif_{B_1}\dif_{B_2}\tilde{\phi}_{B_3B_4}-\nn\\ &&-\dif_{B_1}\dif^A\tilde{\phi}_{AB_2B_3B_4}-\dif_{B_2}\dif^A\tilde{\phi}_{AB_1B_3B_4}\bigr)+\nn\\&&+c_4 F'''^{B_3B_4}\bigl(\Box\tilde{\phi}_{B_3B_4}-\dif^A\dif^B\tilde{\phi}_{ABB_3B_4}\bigr)+\nn\\&&+c_5 F'''^{B_3B_4}(y)\bigl(\Box\tilde{\phi}_{B_3B_4}+\dif_{B_3}\dif_{B_4}\tilde{\phi}-\dif_{B_3}\dif^A\tilde{\phi}_{AB_4}-\dif_{B_4}\dif^A\tilde{\phi}_{AB_3}\bigr)+\nn\\&&+c_6 U(y)\bigl(\Box\tilde{\phi}-\dif^A\dif^B\tilde{\phi}_{AB}\bigr)=0,
\e{668}
where
\be
&&c_1\equiv  48+2(d+2)\al ,\nn\\
&&c_2\equiv 12\al+2(d+3)\beta +2\ga  ,\nn\\
&&c_3\equiv  8\al+4d\ga ,\nn\\
&&c_4\equiv  8\beta+2(d-2)\la-8\ga ,\nn\\
&&c_5\equiv  2(d+3)\la+16\ga ,\nn\\
&&c_6\equiv 2\la+8(d+1)\rho.
\e{669}
Hence, the vanishing of the linear part $\tilde{\cL}_1$ requires
\be
&&\al=-{24\over d+2},\quad\beta={48(3d-1)\over d(d+2)(d+3)},\quad\ga={48\over d(d+2)},\nn\\
&&\la=-{384\over d(d+2)(d+3)},\quad \rho={96\over d(d+1)(d+2)(d+3)}.
\e{670}
Notice that although the number of conditions \rl{669} exceeds the number of parameters by one, there is one unique solution which is agreement with the solutions in subsections 4.4 and 5.3.
 
  \setcounter{equation}{0}
\section{A search for generalized Weyl tensors within the manifestly conformal formulation}
It is clear that also in the conformal space there exist generalized Weyl tensors of the corresponding form to those given in spacetime as presented in sections 3-6. They are determined by means of  the ansatz
\be
&&W_{A_1B_1A_2B_2\cdots A_sB_s}(y)\equiv F_{A_1B_1A_2B_2\cdots A_sB_s}(y)+\nn\\&&+\sum_k\al_kB_{k\;A_1B_1A_2B_2\cdots A_sB_s}(F_k)(y),
\e{801}
where $F$ is the field strength \rl{601}, $F_k$ its traces, and where $B_k$ here are expressed in terms of the traces  $F_k$ and the metric tensor \rl{402} in such a way that it has the same index symmetry as $F$. $B_k$ is also normalized as in section 3:
\be
&&B_k\cdot F=F_k^2.
\e{8010}
Tracelessness, or generalized Weyl invariance under (cf \rl{61})
\be
&&\phi_{A_1A_2\cdots A_s}(y)\;\;\longrightarrow\;\;\phi_{A_1A_2\cdots A_s}(y)+\eta_{(A_1A_2}{\la}_{A_1A_2\cdots A_s)}(y),
\e{8011}
where $\la$ is symmetric,
determines then the parameters $\al_k$ in \rl{801}. However, if the resulting tensor, $C$, is used to define the Lagrangian by 
\be
&&\cL=C^2,
\e{8012}
then one finds that this Lagrangian does not represent a conformally invariant theory. The reason is that although the parameters $\al_k$  have the same form  as in sections 4 and 5, $d$ is everywhere replaced by $d+2$. This means that the tensor $C$  does {\em not} represent a useful generalized Weyl tensors within the manifestly conformally invariant formulation. 
 Now, the transformations \rl{8011} have never entered the manifest construction so far. Are they really relevant here? Notice that \rl{8011} actually is inconsistent with strong transversality of $\phi$.

The appropriate question to ask oneself is "what corresponds to the generalized Weyl transformations in spacetime within the manifest formulation?"  The only gauge transformation one has is the special gauge transformations off the hypercone:
\be
&&\phi_{A_1A_2\cdots A_s}(y)\;\;\longrightarrow\;\;\phi_{A_1A_2\cdots A_s}(y)+y^2\tilde{\phi}_{A_1A_2\cdots A_s}(y).
\e{802}
Has this transformation anything to do with the generalized Weyl transformations? The answer is yes!  One may \eg notice the following peculiar property: Choose $\tilde{\phi}$ in \rl{802} to be of the restricted form 
\be
&&\tilde{\phi}_{A_1A_2\cdots A_s}(y)=\dif_{(A_1}\dif_{A_2}{\la}_{A_3\cdots A_s)}(y),
\e{803}
and insert this into the transformation formulas  \rl{802}. One finds then that the field strength in \rl{601} transforms exactly in the same way as under \rl{8011} except for terms proportional to $y^2$. Also the traces of $F$ transforms in the the same way up to $y^2$-terms, {\em except} that $d$ is everywhere replaced by $d-2$! This is exactly what one would like to have. However, this is inconsistent. The reason is that although \rl{803} implies that $\la$ has the same degree of homogeneity as $\phi$ (\ie $s-2$) in agreement with \rl{8011}, the strong transversality of $\tilde{\phi}$ implies a different degree of homogeneity for $\la$. 
Although inconsistent 
these properties  make me view the special gauge transformations \rl{802} as the appropriate generalization of the generalized Weyl transformations in spacetime within the manifest formulation in the conformal space.  

Consider the ansatz $W$ in \rl{801}, \ie
\be
&&W_{A_1B_1A_2B_2\cdots A_sB_s}(y)\equiv F_{A_1B_1A_2B_2\cdots A_sB_s}(y)+\nn\\&&+\sum_k\al_kB_{k\;A_1B_1A_2B_2\cdots A_sB_s}(F_k)(y).
\e{804}
Obviously  this tensor must transform in a particular simple way under the special gauge transformations \rl{506} since the invariant Lagrangian of the form \rl{609} in section 10 may be written as
\be
&&\cL=W\cdot F.
\e{8041}
However, at the same time this form of the Lagrangian does not seem to allow $W$ to be invariant under \rl{506}. I propose
 \beq
 \begin{tabular}{|p{114mm}|}
 \hline
  {\bf Proposal 9}:  The tensor $W$ yields only terms with explicit y-dependence under the special gauge transformation \rl{506} for the $\al$-values that make the Lagrangian \rl{8041} invariant.\\
   \hline
 \end{tabular}
 \eq
 This should at least help to make the Lagrangian \rl{8041} invariant up to $y^2$ dependent terms since $F$ is transversal. The statement is proved for $s=2$ below.
 
 \noindent
 {\bf Example:} $s=2$.\\
 Here \rl{804} reduces to
 \be
 &&W_{A_1B_1A_2B_2}=R_{A_1B_1A_2B_2}+\al_1B_{1\;A_1B_1A_2B_2}(R_1)+\al_2B_{2\;A_1B_1A_2B_2}(R_2),\nn\\
  &&W_{ABCD}=R_{ABCD}+\al_1B_{1\;ABCD}(R_1)+\al_2B_{2\;ABCD}(R_2),
  \e{8042}
  where (cf subsection 10.2)
 \be
 &&R_{ABCD}(y)=\half F_{ABCD}(y)=\half \dif_{[C}\dif_{[A}H_{B]D]}(y),
 \e{8043}
 and
 \be
&&B_{1\;ABCD}\equiv{1\over4}R_{[A[C}\eta_{D]B]},\nn\\
&&B_{2\;ABCD}\equiv{1\over4}R\;\eta_{[A[C}\eta_{D]B]}.
\e{8044}
  Under the special gauge transformations
  \be
 &&H_{AB}(y)\;\longrightarrow\;H_{AB}(y)+y^2 \tilde{H}_{AB}(y),
 \e{8045}
 $W$ in \rl{8042} transforms as
 \be
 &&W_{ABCD}\;\longrightarrow\;W_{ABCD}+y^2\tilde{W}_{ABCD}+y_{[A}\dif_{[C}\tilde{H}_{D]B]}+y_{[C}\dif_{[A}\tilde{H}_{B]D]}+\nn\\
 &&+\Big(1+{1\over4}(d-2)\al_1\Big)\eta_{[A[C}\tilde{H}_{D]B]}+\Big({1\over4}\al_1+\half\al_2(d-1)\Big)\tilde{H}\eta_{[A[C}\eta_{D]B]},\nn\\
 \e{8046}
 where $\tilde{W}$ is equal to $W$ with $H_{AB}$ replaced by $\tilde{H}_{AB}$. Only for the $\al$-values
 \rl{209}/\rl{628} do the transformation of $W$ reduces to
  \be
 &&W_{ABCD}\;\longrightarrow\;W_{ABCD}+y^2\tilde{W}_{ABCD}+y_{[A}\dif_{[C}\tilde{H}_{D]B]}+y_{[C}\dif_{[A}\tilde{H}_{B]D]}.\nn\\
  \e{8047}
  Thus, for the appropriate $\al$-values the transformed part of $W$ consists only of terms with explicit $y$-dependence in accordance with proposal 9 above.
  
  Notice that $W$ in \rl{8042} for the $\al$-values  \rl{209}/\rl{628} is {\em not} transvers:
  \be
  &&y^AW_{ABCD}=\al_1{1\over4}y_{[C}R_{BD]}+\al_2\half R y_{[C}\eta_{D]B},
  \e{8048}
  where
  \be
  &&R_{BD}=\eta^{AC}R_{ABCD},\quad R=\eta^{BD}R_{BD}.
  \e{8049}
 Notice also that $W$ is {\em not} traceless
 \be
 &&W_{BD}=\eta^{AC}W_{ABCD}=\Big(1+{d\over4}\al_1\Big)R_{BD}+\Big({1\over4}\al_1+\half (d+1)\al_2\Big)R\eta_{BD},\nn\\
 &&W=\eta^{BD}W_{BD}=\Big\{1+\half (d+1)\al_1+\half(d+1)(d+2)\al_2\Big\}R.
 \e{80491}
 However, one remarkable property satisfied by the traces \rl{80491} is that for $d=4$ and for the $\al$-values \rl{209}/\rl{628} I find 
 \be
 &&W_{BD}W^{BD}=R_{BD}R^{BD}, \quad W^2=R^2.
 \e{80492}
 
 In order to find tensors that transform still simpler under the special gauge transformations \rl{506} than the $W$ above one has to consider tensors of higher ranks.
The quantization of the conformal spinning particle in \cite{Arvidsson:2006} led to wave functions of the form ($d=4$) 
$$\cF_{A_1B_1C_1A_2B_2C_2\cdots A_sB_sC_s}(y)$$
with antisymmetry in $A_k$,$B_k$,$C_k$ for $k=1,2,\ldots,s$, and symmetry under interchange of any triplets ($A_kB_kC_k$) and ($A_lB_lC_l$). In terms of symmetric tensors $\phi$, $\cF$ has the form
(in \cite{Arvidsson:2006} the notation $\cF$ also contained a delta function $\del(y^2)$)
\be
&&\cF_{A_1B_1C_1A_2B_2C_2\cdots A_sB_sC_s}(y)=\nn\\&&=\sum_{{\rm antisym}\;A_kB_kC_k}y_{A_1}y_{A_2}\cdots y_{A_s}\dif_{B_1}\dif_{B_2}\cdots\dif_{B_s}\phi_{C_1C_2\cdots C_s}(y).
\e{805}
Notice that this tensor is not just invariant under the gauge transformations
\be
&&\phi_{A_1A_2\cdots A_s}(y)\;\;\longrightarrow\;\;\phi_{A_1A_2\cdots A_s}(y)+\dif_{(A_1}{\varepsilon}_{A_2A_3\cdots A_s)}(y),
\e{806}
but also under
\be
&&\phi_{A_1A_2\cdots A_s}(y)\;\;\longrightarrow\;\;\phi_{A_1A_2\cdots A_s}(y)+y_{(A_1}{\omega}_{A_2A_3\cdots A_s)}(y),
\e{807}
for arbitrary symmetric functions $\varepsilon$ and $\omega$. If one could use such tensors in an action principle then one would not have to bother about the technicalities in the reduction procedure mentioned in section 9 (\rl{VK}, \rl{VKtransv}). However, it is unclear whether or not they are useful here since they really appeared in the second order formulation

The expression \rl{805} implies that the $\cF$-fields also may be expressed in terms of the $F$-fields in \rl{601} as follows
\be
&&\cF_{A_1B_1C_1A_2B_2C_2\cdots A_sB_sC_s}(y)=\nn\\&&=\sum_{{\rm antisym}\;A_kB_kC_k}F_{A_1B_1A_2B_2\cdots A_sB_s}(y)y_{C_1}y_{C_2}\cdots y_{C_s}.
\e{808}
Asuming that this tensor can be used also in the present formulation, which has less restrictions, I
consider now the following tensor 
\be
&&\cW_{A_1B_1C_1A_2B_2C_2\cdots A_sB_sC_s}(y)\equiv\cF_{A_1B_1C_1A_2B_2C_2\cdots A_sB_sC_s}(y)+\nn\\&&+\sum_k\beta_k\cB_{k\;A_1B_1C_1A_2B_2C_2\cdots A_sB_sC_s}(\cF_k)(y),
\e{809}
where $\cB_k$ is of the form
\be
&&\cB_k(\cF_k)=(\cF_k\eta\cdots\eta)_{{\rm sym}},
\e{810}
where $\cF_k$ is one of the traces of $\cF$, and $\eta$ the metric tensor $\eta_{AB}$ in \rl{302}. 
$\cB_k(\cF_k)$ is normalized such that
\be
&&\cB_k\cdot\cF=(\cF_k)^2.
\e{8091}
$\cW$ contains {\em all} different $\cB_k$ of the form \rl{810} satisfying \rl{8091}. The constructions are analogous to the construction in section 3.

\beq
 \begin{tabular}{|p{114mm}|}
 \hline
  {\bf Theorem 5}:  There is a unique choice of the parameters $\beta_k$ in \rl{809} which makes $\cW$ traceless.\\
   \hline
 \end{tabular}
 \eq
  For this choice $\cW$ is denoted $\cC_0$.\\
Proof: The general theorem that an arbitrary tensor ($\cF$) may be decomposed in terms of its traceless part ($\cC_0$) and terms of the form $\cB_k\bullet$

\beq
 \begin{tabular}{|p{114mm}|}
 \hline
  {\bf Definition}:  $\cC(\beta)$ is a tensor of the form $\cW$ in \rl{809} which is traceless up to terms containing  $y^2$-factors.\\
   \hline
 \end{tabular}
 \eq
 Thus, $\cC(\beta)$ has some dependence on the $\beta$-parameters and contains the traceless tensor $\cC_0$ in theorem 5 above.
\beq
 \begin{tabular}{|p{114mm}|}
 \hline
  {\bf Proposal 10}:  The  tensor $\cC(\beta)$  is invariant under the special gauge transformations \rl{802} up to terms proportional to $y^2$  and is determined by this condition. It also satisfies weak transversality, \ie $y^{A_1}\cC_{A_1B_1C_1A_2\cdots B_sC_s}(\beta)=y^2(\cdots)$ and the property $y_{[D_1}\cC_{A_1B_1C_1]A_2\cdots B_sC_s}(\beta)=y^2(\cdots)$.\\
   \hline
 \end{tabular}
 \eq
 I have no general proof of this statement. However, below I show that it is valid for $s=2$. Notice that 
 invariance of the ansatz  $\cW$  under the special gauge transformations \rl{802} up to $y^2$-terms  does {\em not} determine the parameters $\beta_k$ uniquely, since some $\cB_k$ will be proportional to $y^2$ (see below).  
  
 \subsection{The properties of $\cW$ for $s=2$.}
 Instead of $\cF$ I choose here as in \cite{Arvidsson:2006} to consider $\cR$ as the basic field strength. The relation between the two is simply $\cF=2\cR$ (cf \rl{202}). $\cR$ has the following traces
 \be
 &&\cR'_{B_1C_1B_2C_2}(y)\equiv\eta^{A_1A_2}\cR_{A_1B_1C_1A_2B_2C_2}(y),\nn\\
&&\cR''_{C_1C_2}(y)\equiv\eta^{B_1B_2}\cR'_{B_1C_1B_2C_2}(y)=\eta^{A_1A_2}\eta^{B_1B_2}\cR_{A_1B_1C_1A_2B_2C_2}(y),\nn\\
&&\cR'''(y)\equiv\eta^{C_1C_2}\cR''_{C_1C_2}(y)=\eta^{A_1A_2}\eta^{B_1B_2}\eta^{C_1C_2}\cR_{A_1B_1C_1A_2B_2C_2}(y).\nn\\
\e{811}
$\cR$ may be expressed in terms of the linearized Riemann tensor $R$ as follows:
\be
&\cR_{A_1B_1C_1A_2B_2C_2}(y)=&\biggl(\Big(y_{A_1}y_{A_2}R_{B_1C_1B_2C_2}(y)+cycle(A_1B_1C_1)\Big)\nn\\&&+cycle(A_2B_2C_2)\biggr).
\e{812}
This implies for the traces \rl{811}
\be
&&\cR'_{B_1C_1B_2C_2}(y)=y_{[B_1}y_{[B_2}R_{C_2]C_1]}(y)+y^2R_{B_1C_1B_2C_2}(y),\nn\\
&&\cR''_{C_1C_2}(y)=y_{C_1}y_{C_2}R(y)+2y^2R_{C_1C_2}(y),\nn\\
&&\cR'''(y)=3y^2R(y),
\e{813}
where $R_{C_1C_2}$ and $R$ are the Ricci and curvature tensors defined in \rl{618}.

Consider now the following ansatz for the generalized Weyl tensor in the manifest formulation:
\be
&&\cW_{A_1B_1C_1A_2B_2C_2}(y)=\cR_{A_1B_1C_1A_2B_2C_2}(y)+\beta_1\cB_{1\;A_1B_1C_1A_2B_2C_2}(\cR')(y)+\nn\\&&+\beta_2\cB_{2\;A_1B_1C_1A_2B_2C_2}(\cR'')(y)+\beta_3\cB_{3\;A_1B_1C_1A_2B_2C_2}(\cR''')(y),
\e{814}
where
\be
&&\cB_{1\;A_1B_1C_1A_2B_2C_2}(\cR')(y)={1\over9}(\eta_{A_1A_2}\cR'_{B_1C_1B_2C_2}(y))_{sym}=\nn\\&&
={1\over9}\Big((\eta_{A_1A_2}\cR'_{B_1C_1B_2C_2}(y)+cycle(A_1B_1C_1))+cycle(A_2B_2C_2)\Big),\nn\\
&&\cB_{2\;A_1B_1C_1A_2B_2C_2}(\cR'')(y)={1\over18}(\eta_{A_1A_2}\eta_{B_1B_2}\cR''_{C_1C_2}(y))_{sym}=\nn\\&&=
{1\over36}\half\Big((\eta_{[A_1[A_2}\eta_{B_2]B_1]}\cR''_{C_1C_2}(y)+cycle(A_1B_1C_1))+cycle(A_2B_2C_2)\Big),\nn\\
&&\cB_{3\;A_1B_1C_1A_2B_2C_2}(\cR''')(y)={1\over6}(\eta_{A_1A_2}\eta_{B_1B_2}\eta_{C_1C_2}\cR'''(y))_{sym}=\nn\\&&=
{1\over36}\Big((\eta_{[A_1[A_2}\eta_{B_2]B_1]}\eta_{C_1C_2}+cycle(A_1B_1C_1))+cycle(A_2B_2C_2)\Big)\cR'''(y).\nn\\
\e{815}

For the trace of the ansatz \rl{814} I find the expression
\be
&&\eta^{A_1A_2}\cW_{A_1B_1C_1A_2B_2C_2}(y)=\Big(1+{1\over9}(d-2)\beta_1\Big)\cR'_{B_1C_1B_2C_2}(y)
+\nn\\&&+{1\over9}\Big(\beta_1+\half(d-1)\beta_2\Big)\eta_{[B_1[B_2}\cR''_{C_2]C_1]}(y)+\nn\\&&+{1\over18}\Big(\half\beta_2+d\beta_3\Big)\eta_{[B_1[B_2}\eta_{C_2]C_1]}\cR'''(y).\nn\\
\e{816}
Since $\cR'''$ is proportional to $y^2$,
$\cW$ is weakly  traceless for the following $\beta$-values:
\be
&&\beta_1=-{9\over d-2},\nn\\
&&\beta_2={18\over(d-1)(d-2)}.
\e{817}
These values inserted into \rl{814} determines $\cC(\beta)$ here. ($\beta_3$ is still a free  parameter.)
Notice that these values of $\beta_1$ and $\beta_2$ contain the same $d$-dependent factors as the values of $\al_1$ and $\al_2$ in the spacetime treatment of $s=2$ in subsection 4.2. $\cW$ is traceless for the $\beta$-values \rl{817} and
\be
&&\beta_3=-{9\over d(d-1)(d-2)},
\e{8171}
which determines $\cC_0$.

Under the special gauge transformations ($H\equiv\phi$)
\be
&&H_{A_1A_2}(y)\;\;\longrightarrow\;\;H_{A_1A_2}(y)+y^2\tilde{H}_{A_1A_2}(y),
\e{818}
$\cR$ and its traces transform as follows
\be
&&\cR_{\cdots}(y)\;\;\longrightarrow\;\;\cR_{\cdots}(y)+\tilde{\cR}_{\cdots}(y),\quad etc,
\e{819}
where $\tilde{\cR}$ is $\cR$ with $R$ replaced by $\tilde{R}$ here given by ($y^2$-terms included) (cf. \rl{621}-\rl{623})
\be
&&\tilde{R}_{A_1B_1A_2B_2}(y)=\eta_{[A_1[A_2}\tilde{H}_{B_2]B_1]}(y)+y_{[A_1}\dif_{[A_2}\tilde{H}_{B_2]B_1]}+\nn\\&&+y_{[A_2}\dif_{[A_1}\tilde{H}_{B_1]B_2]}+\half y^2\dif_{[A_1}\dif_{[A_2}\tilde{H}_{B_2]B_1]}(y),\nn\\
&&\tilde{R}_{B_1B_2}(y)=(d-2)\tilde{H}_{B_1B_2}(y)+\eta_{B_1B_2}\tilde{H}(y)+(y_{B_1}\dif_{B_2}+y_{B_2}\dif_{B_1})\tilde{H}(y)-\nn\\&&-y_{B_1}\dif_{A}\tilde{H}^A_{\;\;B_2}(y)     -y_{B_2}\dif_A\tilde{H}^A_{\;\;B_1}(y)+\half y^2\Big(
\Box\tilde{H}_{B_1B_2}(y)+\dif_{B_1}\dif_{B_2}\tilde{H}(y)-\nn\\&&-\dif_{B_1}\dif^A\tilde{H}_{AB_2}(y)-\dif_{B_2}\dif^A\tilde{H}_{AB_1}(y)\Big),\nn\\
&&\tilde{R}(y)=2(d-1)\tilde{H}(y)+y^2(\Box\tilde{H}(y)-\dif^A\dif^B\tilde{H}_{AB}(y)),\nn\\
&&\tilde{H}(y)\equiv \eta^{AB}\tilde{H}_{AB}(y).
\e{820}
This implies that the ansatz \rl{814} transforms as follows
\be
&&\cW_{A_1B_1C_1A_2B_2C_2}(y)\;\;\longrightarrow\;\;\cW_{A_1B_1C_1A_2B_2C_2}(y)+\tilde{\cW}_{A_1B_1C_1A_2B_2C_2}(y),\nn\\
&&\tilde{\cW}_{A_1B_1C_1A_2B_2C_2}(y)\equiv (1+\beta_1(d-2))\Big((\eta_{A_1A_2}y_{[B_1}y_{[B_2}\tilde{H}_{C_2]C_1]}(y)+\nn\\&&+cycle(A_1B_1C_1))+cycle(A_2B_2C_2)\Big)+\nn\\&&+
(\beta_1+(d-1)\beta_2)\Big(y_{C_1}y_{C_2}\eta_{[A_1[A_2}\eta_{B_2]B_1]}\tilde{H}(y)+cycle(A_1B_1C_1))+\nn\\&&+cycle(A_2B_2C_2)\Big)+
y^2(\cdots).
\e{821}
Thus, the special gauge invariance of the ansatz \rl{814} reproduces the values \rl{817} of $\beta_1$ and $\beta_2$. However, the $y^2$-dependent terms consists of 17 different sets of terms and cannot be cancelled by a choice of $\beta_3$. Consequently the parameter  $\beta_3$ is not determined here. 
$\cW$ is invariant only up to $y^2$-terms or in other words $\cW$ is only weakly invariant under the special gauge transformations. Notice that weak tracelessness yielded identical result.

Let me now turn to the transversality of $\cW$. Since the linear Riemann tensor $R$ satisfies strong transversality, $\cR$ satisfies weak transversality. Explicitly
\be
&&y^{A_1}\cR_{A_1B_1C_1A_2B_2C_2}(y)=y^2(R_{B_1C_1B_2C_2}y_{A_2}+cycle(A_2B_2C_2)).\nn\\
\e{822}
For the $\cB_k$ in $\cW$ I find
\be
&&y^{A_1}\cB_{1\;A_1B_1C_1A_2B_2C_2}(y)=y^2\Big(\eta_{A_2[B_1}R_{C_1][B_2}(y)y_{C_2]}+cycle(A_2B_2C_2)\Big),\nn\\
&&y^{A_1}\cB_{2\;A_1B_1C_1A_2B_2C_2}(y)=2y^2\Big(y_{A_2}\eta_{[B_2[B_1}R_{C_1]C_2]}(y)+cycle(A_2B_2C_2)\Big),\nn\\
&&y^{A_1}\cB_{3\;A_1B_1C_1A_2B_2C_2}(y)={3\over2}y^2\Big(y_{A_2}\eta_{[B_2[B_1}\eta_{C_1]C_2]}+cycle(A_2B_2C_2)\Big)R(y).\nn\\
\e{823}
Hence, the whole ansatz $\cW$ in \rl{814} satisfies weak transversality.

The relation $y_{[D_1}\cW_{A_1B_1C_1]A_2B_2C_2}=y^2(\cdots)$ follows from the expressions \rl{813} and \rl{815}. Notice that $\cR$ in \rl{812} satisfies $y_{[D_1}\cR_{A_1B_1C_1]A_2B_2C_2}=0$.

{\bf Final remark}: Although the above properties of $\cC(\beta)$ ($\cW$ in \rl{814} with the values \rl{817}) are very nice, this tensor does not determine the Lagrangian in any natural way. In fact,  
the square of $\cC(\beta)$ is {\em not} expressible in terms of the conformally invariant Lagrangian in spite of the promising $d$-dependence of $\beta_1$ and $\beta_2$ in \rl{817}. In particular I find
\be
&&\cC_0^2(y)\propto (y^2)^2\cL(y),\nn\\
&&\cL(y)=R_{ABCD}R^{ABCD}-{4\over(d-1)} R_{AB}R^{AB}+{2\over d(d-1)} R^2,
\e{824}
which is equal to \rl{617} with the values \rl{628}, or equivalently \rl{8}, but with $d$ replaced by $d+1$. Anyway $\cL$ in \rl{824} is half-way between $\cL=C^2$ from \rl{801} and the correct expression \rl{617} with \rl{628}.

Requiring weak invariance under the special gauge transformations \rl{802} of a tensor of the same rank as $\cC(\beta)$ does not only yield $\cC(\beta)$. There are many other candidates like $\cR$ in \rl{812} with $R$ replaced by $W$ in the example after proposal 9, or $\cW$ with $R$ replaced by the traceless $C$ from \rl{801}. Maybe there exists one weakly invariant tensor that yields the correct Lagrangian through the formula \rl{824}. It would, of course, be preferable to find a tensor that is strongly invariant under the special gauge transformations  \rl{802}. However, such a tensor may not exist.

\part{Arbitrary dimensions}

 \setcounter{equation}{0}
\section{Generalizations to arbitrary even dimensions $d$}
In arbitrary even dimensions $d$ the spinning conformal particle model in \cite{Martensson:1992ax} yields after quantization the wave functions \cite{Arvidsson:2006}
\be
&&\cF_{A_1B_1C_1D_1\cdots A_2B_2C_2D_2\cdots\cdots A_sB_sC_sD_s\cdots}(y)
\e{901}
with antisymmetry in the $d/2+1$ indices $A_kB_kC_k\cdots$, $k=1,\ldots,s$, and symmetry under interchange of any two antisymmetric blocks ($A_kB_kC_kD_k\cdots$) and ($A_lB_lC_lD_l\cdots$). The conditions on $\cF$ implies that $\cF$ has the following form in terms of a gauge field $\phi$:
\be
&&\cF_{A_1B_1C_1D_1\cdots A_2B_2C_2D_2\cdots\cdots A_sB_sC_sD_s\cdots}(y)=\nn\\&&=\sum_{antisym(A_kB_kC_k\cdots)}y_{A_1}y_{A_2}\cdots y_{A_s}\dif_{B_1}\dif_{B_2}\cdots\dif_{B_s}\phi_{C_1D_1\cdots C_2D_2\cdots\cdots C_sD_s\cdots}(y),\nn\\
\e{902}
where $\phi$ is antisymmetric in the $d/2-1$ indices $C_kD_k\cdots$ and symmetric under interchange of any two numbered index block. In addition it must satisfy
\be
&&\phi_{[C_1D_1\cdots C_2]D_2\cdots\cdots C_sD_s\cdots}(y)=0.
\e{9021}
(In $d=2$ $\phi$ is just a scalar, and in $d=4$ it is a symmetric tensor.)
$\cF$ in \rl{902} is gauge invariant under the following transformations
\be
&&\phi_{C_1D_1\cdots C_2D_2\cdots\cdots C_sD_s\cdots}(y)\;\;\longrightarrow\;\;\phi_{C_1D_1\cdots C_2D_2\cdots\cdots C_sD_s\cdots}(y)+\nn\\&&+\sum_{sym(1,2,\ldots,s)}\dif_{[C_1}\varepsilon_{D_1\cdots ]C_2D_2\cdots C_3D_3\cdots\cdots\cdots C_sD_s\cdots}(y),\nn\\
&&\phi_{C_1D_1\cdots C_2D_2\cdots\cdots C_sD_s\cdots}(y)\;\;\longrightarrow\;\;\phi_{C_1D_1\cdots C_2D_2\cdots\cdots C_sD_s\cdots}(y)+\nn\\&&+\sum_{sym(1,2,\ldots,s)}y_{[C_1}\omega_{D_1\cdots] C_2D_2\cdots C_3D_3\cdots\cdots\cdots C_sD_s\cdots}(y),
\e{903}
where $\varepsilon$ and $\omega$ are arbitrary functions which are antisymmetric in each numbered index block.

$\cF$ in \rl{902} may also be written as
\be
&&\cF_{A_1B_1C_1D_1\cdots A_2B_2C_2D_2\cdots\cdots A_sB_sC_sD_s\cdots}(y)=\nn\\&&=\sum_{antisym(A_kB_kC_k\cdots)}y_{A_1}y_{A_2}\cdots y_{A_s}F_{B_1C_1D_1\cdots B_2C_2D_2\cdots\cdots B_sC_sD_s\cdots}(y),\nn\\
\e{904}
where $F$ is antisymmetric in the $d/2$ indices $B_kC_k\cdots$ and symmetric under interchange of any two numbered index block, generalizing  the  $F$ \rl{601} in $d=4$ to arbitrary even dimensions $d$. $F$ has the form
\be
&&F_{B_1C_1D_1\cdots B_2C_2D_2\cdots\cdots B_sC_sD_s\cdots}(y)=\nn\\&&=\sum_{antisym(B_kC_k\cdots)}
\dif_{B_1}\dif_{B_2}\cdots\dif_{B_s}\phi_{C_1D_1\cdots C_2D_2\cdots\cdots C_sD_s\cdots}(y),
\e{905}
where $\phi$ is the gauge field in \rl{902}. $F$ is gauge invariant only under the first transformation in \rl{903}.

Now the conformal particle models yield further conditions on the wave functions like transversality, tracelessness and second order equations for $F$ as well as the homogeneities
\be
&&(y\cdot\dif+d/2-1)\cF_{\cdots}=0\quad \Leftrightarrow\quad(y\cdot\dif+d/2-1+s)F_{\cdots}=0.
\e{906}
(see section 6 in \cite{Arvidsson:2006}). However, this implies that the Lagrangian $\cL=F^2$ satisfies
$(y\cdot\dif+d-2+2s)\cL=0$ which disagrees with the homogeneity condition \rl{505} for $s\neq1$. For $s=1$ all conditions on the wave functions are reproduced by the action principle. (This case was treated in \cite{Arvidsson:2006} for arbitrary $d$.)

Like for $s=2$ I have also for $s\geq2$ to give up some conditions that followed from the quantization of the spinning conformal particle. In fact, if I take away the traceless condition and the equation of motion there is a natural route to a consistent conformal Lagrangian theory in arbitrary even dimensions $d$.
In order for the fields in \rl{904} to enter the action principle one has to impose strong transversality on the basic gauge field $\phi$, \ie
\be
&&y^{C_1}\phi_{C_1D_1\cdots C_2D_2\cdots\cdots C_sD_s\cdots}(y)=0.
\e{907}
One of the conditions on $F$ from the wave function quantization is at least weak transversality. Requiring also strong transversality for $F$, \ie
\be
&&y^{B_1}F_{B_1C_1D_1\cdots B_2C_2D_2\cdots\cdots B_sC_sD_s\cdots}(y)=0,
\e{908}
one finds that consistency requires the following homogeneity for $\phi$ (cf the argument in section 10)
\be
&&(y\cdot\dif+d/2-s)\phi_{C_1D_1\cdots C_2D_2\cdots\cdots C_sD_s\cdots}(y)=0.
\e{909}
This in turn requires
\be
&&(y\cdot\dif+d/2)F_{B_1C_1D_1\cdots B_2C_2D_2\cdots\cdots B_sC_sD_s\cdots}(y)=0.
\e{910}
Hence, the Lagrangian $\cL=F^2$ satisfies the homogeneity condition \rl{505}. The homogeneity \rl{909} requires that $\phi$ satisfies an equation of order $\Box^s$ in arbitrary even dimensions $d$. (See section 7 in \cite{Arvidsson:2006}.)

The general ansatz for a conformally invariant Lagrangian in arbitrary even dimensions $d$ within the manifest formulation is then 
\be
&&\cL(y)=F^2+\sum_k\al_kF_k^2(y),
\e{911}
where the sum is over all possible traces, $F_k$, of $F$. Invariance under the special gauge transformations,
\be
&&\phi_{C_1D_1\cdots C_2D_2\cdots\cdots C_sD_s\cdots}(y)\;\;\longrightarrow\;\;\phi_{C_1D_1\cdots C_2D_2\cdots\cdots C_sD_s\cdots}(y)+\nn\\&&+y^2\tilde{\phi}_{C_1D_1\cdots C_2D_2\cdots\cdots C_sD_s\cdots}(y),
\e{912}
determines the constants $\al_k$ in \rl{911}.

As in section 11 I expect there exist tensors of the form
\be
&&\cW_{A_1B_1C_1D_1\cdots A_2B_2C_2D_2\cdots\cdots A_sB_sC_sD_s\cdots}(y)=\nn\\&&=\cF_{A_1B_1C_1D_1\cdots A_2B_2C_2D_2\cdots\cdots A_sB_sC_sD_s\cdots}(y)+\nn\\&&+\sum_k\beta_k\cB_k(\cF_k)_{A_1B_1C_1D_1\cdots A_2B_2C_2D_2\cdots\cdots A_sB_sC_sD_s\cdots}(y),
\e{913}
which are weakly invariant under the special gauge transformations   \rl{912}.
 $\cB_k$ are constructed analogously to those in section 11 in terms of traces, $\cF_k$, of $\cF$ in \rl{901}. ($\cW$ has the index symmetry of $\cF$.) Invariance under the special gauge transformations \rl{912} up to terms proportional to $y^2$ determines all constants $\beta_k$ in \rl{913} for which $\cB_k$ is not proportional to $y^2$. The  condition  that $\cW$ is traceless up to terms with $y^2$-factors should yield the same result.

\subsection{The spacetime formulation}
The above manifestly conformal theories should correspond to the spacetime theory that follows from the quantization of the $O(2s)$ extended supersymmetric spinning particle model   with  similar modifications as those given above. The quantization procedure in \cite{Marnelius:1988ab} is easily generalized to arbitrary even dimensions $d$. (It is sketched in appendix D in \cite{Arvidsson:2006}.) The resulting generalized Weinberg representation is also given in \cite{Bandos:2005mb}.

I expect therefore that the generalization of section 2 to arbitrary even dimensions $d$ should look  as follows: For integer spins $s$ (viewed in an abstract sense) one has field strengths or curvatures or wave functions which are tensor fields of rank $ds/2$ denoted as
\be
&&F_{\mu_1\nu_1\rho_1\cdots\mu_2\nu_2\rho_2\cdots\cdots\mu_s\nu_s\rho_s\cdots}(x),
\e{914}
with antisymmetry in the $d/2$ indices $\mu_k\nu_k\rho_k\cdots$, and symmetry under interchange of any of the numbered antisymmetric blocks. These fields also allow for the definitions of self-dual and anti-self-dual parts.  In addition $F$ satisfies
\be
&&F_{[\mu_1\nu_1\rho_1\cdots\mu_2]\nu_2\rho_2\cdots\cdots\mu_s\nu_s\rho_s\cdots}(x)=0,\nn\\
&&\dif_{[\al_1}F_{\mu_1\nu_1\rho_1\cdots]\mu_2\nu_2\rho_2\cdots\cdots\mu_s\nu_s\rho_s\cdots}(x)=0,
\e{915}
which are solved by the expression
\be
&&F_{\mu_1\nu_1\rho_1\cdots\mu_2\nu_2\rho_2\cdots\cdots\mu_s\nu_s\rho_s\cdots}(x)=\dif_{[\mu_s}\cdots\dif_{[\mu_2}\dif_{[\mu_1}\phi_{\nu_1\rho_1\cdots]\nu_2\rho_2\cdots]\cdots\nu_s\rho_s\cdots]}(x),\nn\\
\e{916}
where $\phi$ is of rank $(d/2-1)s$. It is antisymmetric in the $d/2-1$ indices $\nu_k\rho_k\cdots$, and symmetric under interchange of any two of the numbered antisymmetric blocks. It also has to satisfy the identities
\be
&&\phi_{[\nu_1\rho_1\cdots\nu_2]\rho_2\cdots\nu_3\rho_3\cdots\cdots\nu_s\rho_s\cdots}(x)=0.
\e{9161}
 The expression \rl{916} is invariant under the gauge transformations
\be
&&\phi_{\nu_1\rho_1\cdots\nu_2\rho_2\cdots\cdots\nu_s\rho_s\cdots}(x)\;\;\longrightarrow\;\;\phi_{\nu_1\rho_1\cdots\nu_2\rho_2\cdots\cdots\nu_s\rho_s\cdots}(x)+\nn\\&&+\sum_{sym\;1,\ldots,s}\sum_{antisym(\nu_1\rho_1\cdots)}\dif_{\nu_1}\varepsilon_{\rho_1\cdots\nu_2\rho_2\cdots\cdots\nu_s\rho_s\cdots}(x),
\e{917}
where the functions $\varepsilon$ is antisymmetric in each numbered block and symmetric under interchange of any two blocks. In addition to the above properties one also has the traceless condition
\be
&&\eta^{\mu_1\mu_2}F_{\mu_1\nu_1\rho_1\cdots\mu_2\nu_2\rho_2\cdots\cdots\mu_s\nu_s\rho_s\cdots}(x)=0,
\e{918}
and 
\be
&&\dif^{\mu_1}F_{\mu_1\nu_1\rho_1\cdots\mu_2\nu_2\rho_2\cdots\cdots\mu_s\nu_s\rho_s\cdots}(x)=0,
\e{9181}
The equation
\be
&&\Box F_{\mu_1\nu_1\rho_1\cdots\mu_2\nu_2\rho_2\cdots\cdots\mu_s\nu_s\rho_s\cdots}(x)=0
\e{919}
follows from the second relation in \rl{915} and \rl{9181}.
As for $d=4$ I propose to give up the equation \rl{919} and the  conditions \rl{918} and \rl{9181}. Instead one has to construct a generalized Weyl tensor $C$ by the ansatz 
\be
&&W_{\mu_1\nu_1\rho_1\cdots\mu_2\nu_2\rho_2\cdots\cdots\mu_s\nu_s\rho_s\cdots}(x)=F_{\mu_1\nu_1\rho_1\cdots\mu_2\nu_2\rho_2\cdots\cdots\mu_s\nu_s\rho_s\cdots}(x)+\nn\\&&+\sum_k\al_k B_{k\;\mu_1\nu_1\rho_1\cdots\mu_2\nu_2\rho_2\cdots\cdots\mu_s\nu_s\rho_s\cdots}(x)(F_k)
\e{920}
in complete analogy to the treatment in section 3. Tracelessness or invariance under transformations of the form 
\be
&&\phi_{\nu_1\rho_1\cdots\nu_2\rho_2\cdots\cdots\nu_s\rho_s\cdots}(x)\;\;\longrightarrow\;\;\phi_{\nu_1\rho_1\cdots\nu_2\rho_2\cdots\cdots\nu_s\rho_s\cdots}(x)+\nn\\&&+\sum_{sym\;1,\ldots,s}\sum_{\substack{antisym(\nu_1\rho_1\cdots)\\ and (\nu_2\rho_2\cdots)}}\eta_{\nu_1\nu_2}\la_{\rho_1\cdots\rho_2\cdots\nu_3\rho_3\cdots\cdots\nu_s\rho_s\cdots}(x),
\e{921}
where $\la$ is antisymmetric in each numbered index block, should uniquely  determine  the $\al$'s in the ansatz \rl{920} and for these values one should have $W=C$. The conformally invariant Lagrangian in $d$ dimensions is then given by $\cL=C^2$. It is clear that the $\phi$'s above satisfies equations of order $2s$ in arbitrary even dimensions $d$. (This is different from  \cite{Segal:2002gd} where  symmetric $\phi$'s are proposed for all dimensions $d$ with equations of the order $2s+d-4$.)

 \setcounter{equation}{0}
\section{Final discussions}
The present formulation of Lagrangian conformal field theory for higher spins is based on gauge invariant wave functions obtained from a quantization of a particular relativistic massless spinning particle model \cite{Gershun:1979fb,Howe:1988ft,Marnelius:1988ab} {\em and} the corresponding conformal spinning particle model \cite{Martensson:1992ax}. From these field strengths or curvatures I have constructed Lagrangian field theories both in a manifestly conformal form in the conformal space and the equivalent theories in spacetime. This was done for free fields with integer spins. Although some general proofs are missing the explicit construction up to spin four should be convincing. The presented construction is a direct generalization of the construction used to obtain  manifestly conformal gravity in \cite{Arvidsson:2006}. I also expect the present theory to be equivalent to the more implicit proposals by Fradkin, Linetsky and Tseytlin \cite{Fradkin:1985am,Fradkin:1989md,Fradkin:1990ps}. (They do not use field strengths.) This equivalence is natural since they also use conformal gravity as a starting point for their generalization. Since they  give Lagrangians for half-odd integer spins I expect that  the corresponding Lagrangians also exist within the present scheme. 

An interesting aspect of the present approach, which also has been an important motivating factor for this work, is the possibility to introduce interactions. This was clear to me when I realized that the manifestly conformal particle allowed for interactions with external fields provided these fields satisfy exactly the same properties as the free fields do here \cite{Arvidsson:2006,Bars:2001um}. (In this paper these conditions follow from the transversality of the manifest field strengths.) My belief in a consistent interacting theory has then increased further after the discovery of the connection to the work by Fradkin and Linetsky, since they claim the existence of  consistent cubic interactions provided an infinite set of higher spins contribute \cite{Fradkin:1989md,Fradkin:1990ps}. I expect that  interactions are possible to study also within the present scheme. Probably the manifest formulation is particular useful then.

The higher spin theory presented here is not a normal local field theory. It is a higher order field theory. The conventional wisdom is that such theories contain states with negative norms (ghosts)  in the physical state space. This would certainly be the case if there were no gauge invariances. However, since the present theories also are gauge theories the situation is perhaps not that bad. On the other hand, I have not yet analysed these aspects  and I have not studied the vast literature on the subject. Let me just point out that higher order theories may always be reduced to second order theories by means of auxiliary fields. The resulting Lagrangians are then possible to quantize although they in general are singular in Dirac's sense. (If one is unlucky second class constraints might appear which makes the situation more difficult.) In \cite{Edgren: 2006un} it was pointed out that such reductions may be done in several different ways which makes the quantization ambiguous. It was therefore emphasized that a correct quantization must be such that it is insensitive to this reduction ambiguity. Examples of consistent unitary quantum theories for higher order models are \eg the string theory in \cite{Savvidy:2003co} and the particle model in \cite{Edgren:2006un} both of which are linear theories.

Personally I believe that there exists no normal local field theory with at most second order derivatives  that describes interacting higher spin particles. Probably one is forced to consider unconventional theories like higher order field theories and/or  field theories with nonlocal interactions. The proliferation of external states which results if they are viewed as higher (infinite) order theories may be removed if one confines oneself to perturbative solutions. Such a restriction is also possible to impose for the present higher order theories if there is no other way to avoid the difficulties. A restriction to a naive perturbation theory is, however,  inconsistent \cite{Marnelius:1974rq}. A way out of this dilemma is to redefine the perturbation expansion by a redefinition of the asymptotic fields. In \cite{Eliezer:1989cr} this procedure is called localization (see section 3). However, I do not know of any general effective method to perform this. There is also the difficulty to interpret the modified theory. Anyway, in general such redefinitions seem always to imply  an increased number of asymptotic states. An example of a localized nonlocal theory is Witten's open string field theory \cite{Witten:1985cc} according to \cite{Eliezer:1989cr}.  Here closed string states appear in the loops.  Another very much less explored example, to say the least,  is given in \cite{Marnelius:1976kc,Craigie:1976ez}. There a particular nonlocal field theory is proposed as an alternative  theory to string theory for strong interactions. (The mass-spin spectrum is similar to that of the bosonic string apart from the tachyon and the degeneracy, but it is a nongeometric model  based on the old Fierz equations
with no gauge invariances.) The perturbation expansion is in principle possible to calculate exactly. What the localization in the sense of  \cite{Eliezer:1989cr} means for this model is unclear apart from requiring a modification of the asymptotic free states. (A possibly related fact is that the completeness of the naive free states  requires the introduction of a very peculiar infrared state    \cite{Craigie:1976ez}.)
 That much about unconventional quantum field theories. In view of the above aspects I believe that even theories of the type presented here, if further developed, have a chance to eventually result in a consistent (interacting) quantum field theory.

\vspace{2cm}
{\bf Acknowledgements}: It is pleasure to thank Dario Francia for discussions and Misha Vasiliev for correspondence.
\vspace{4cm}

 \begin{appendix}
%\newpage
\part{Appendices}
%\small
\section{The equivalence between the  deWit-Freedman and the Weinberg representations ($d=4$).}
The deWit-Freedman generalized curvature \cite{deWit:1979pe}, denoted $R$, may be expressed in terms of the Weinberg field strength $F$ in section 2 as follows
\be 
&&R_{\mu_1\mu_2\cdots\mu_s\nu_1\nu_2\cdots\nu_s}(x)={1\over s!}\sum_{sym(\mu_k)}F_{\mu_1\nu_1\mu_2\nu_2\cdots\mu_s\nu_s}(x),
\e{x1}
where the sum is a symmetrization over all $\mu$'s. (All symmetrizations and antisymmetrizations are without factors.) From the properties of $F$ in section 2 one may then derive the following properties of $R$ (cf \cite{deWit:1979pe}):

1) $R$ is symmetric in both $\mu$'s and $\nu$'s. This follows since $F$ is symmetric under interchange of any two antisymmetric pairs $\mu_k\nu_k$.

2) $R$ satisfies 
\be
&&R_{\mu_1\mu_2\cdots\mu_s\nu_1\nu_2\cdots\nu_s}(x)=(-1)^sR_{\mu_1\mu_2\cdots\mu_s\nu_1\nu_2\cdots\nu_s}(x).
\e{x2}
This follows since all pairs $\mu_k\nu_k$ are antisymmetric.

3) $R$ satisfies the equality
\be
&&\sum_{sym(\mu_1\mu_2\cdots\mu_s\nu_1)}R_{\mu_1\mu_2\cdots\mu_s\nu_1\nu_2\cdots\nu_s}(x)=0.
\e{x3}
This follows since one set of antisymmetric pairs in $F$ always appear in a symmetric combination with otherwise equal indices and therefore cancel.

4) The second condition in \rl{2} implies the Bianchi identities for $R$ (see (2.11) in \cite{deWit:1979pe})

5) The traceless condition \rl{3} for $F$ implies that also $R$ must be traceless:
\be
&&\eta^{\mu_1\mu_2}R_{\mu_1\mu_2\cdots\mu_s\nu_1\nu_2\cdots\nu_s}(x)=0.
\e{x4}

6) The condition \rl{31} implies obviously
\be
&&\dif^{\mu_1}R_{\mu_1\mu_2\cdots\mu_s\nu_1\nu_2\cdots\nu_s}(x)=0.
\e{x5}

7) The condition \rl{x5} and the Bianchi identities 4) imply
\be
&&\Box R_{\mu_1\mu_2\cdots\mu_s\nu_1\nu_2\cdots\nu_s}(x)=0.
\e{x6}
The last two conditions, \rl{x5} and \rl{x6}, are not in the paper by deWit and Freedman  \cite{deWit:1979pe}. They are on-shell conditions that indirectly follow from the quantization of the free $O(2s)$-extended supersymmetric particle model for which $F$ are the natural wave functions.

In a corresponding conformal treatment as given in section 2 but in terms of the deWit-Freedman curvature $R$, conditions 5)-7) have to be removed.

One may also reverse the argument: From 1)-7) one may derive the properties of the Weinberg field strength $F$ from $R$. From the relation
\be
&&F_{\mu_1\nu_1\mu_2\nu_2\cdots\mu_s\nu_s}(x)={1\over 2^s}\sum_{antisym(\mu_k\nu_k)}R_{\mu_1\mu_2\cdots\mu_s\nu_1\nu_2\cdots\nu_s}(x)\equiv\nn\\&&\equiv{1\over 2^s}R_{[\mu_1[\mu_2\cdots[\mu_s\nu_s]\cdots\nu_2]\nu_1]}(x),
\e{x7}
$F$ is antisymmetric in all pairs $\mu_k\nu_k$. From 1) it follows that $F$ is symmetric under interchange of any antisymmetric pairs $\mu_k\nu_k$ and $\mu_l\nu_l$. The first relation \rl{2} follows from the property \rl{x3} in 3).
The second relation in \rl{2} follows from the Bianchi identity 4). The traceless condition \rl{3} from \rl{x4} in 5). The equation \rl{31} from \rl{x5}, and \rl{32} from \rl{x6}.

  \setcounter{equation}{0}
\section{The explicit forms of $B_k$ in \rl{237}}
The explicit forms of the $B$-fields in \rl{237} are
\be
&B_{1\;\mu_1\nu_1\mu_2\nu_2\mu_3\nu_3\mu_4\nu_4}=&(F'_{\nu_1\nu_2\mu_3\nu_3\mu_4\nu_4}\eta_{\mu_1\mu_2})_{sym}=\nn\\&&{1\over24}\biggl(\eta_{[\mu_1[\mu_2}F'_{\nu_2]\nu_1]\mu_3\nu_3\mu_4\nu_4}+\eta_{[\mu_1[\mu_3}F'_{\nu_3]\nu_1]\mu_2\nu_2\mu_4\nu_4}+\nn\\&&+
\eta_{[\mu_1[\mu_4}F'_{\nu_4]\nu_1]\mu_2\nu_2\mu_3\nu_3}+\eta_{[\mu_2[\mu_3}F'_{\nu_3]\nu_2]\mu_1\nu_1\mu_4\nu_4}+\nn\\&&+
\eta_{[\mu_2[\mu_4}F'_{\nu_4]\nu_2]\mu_1\nu_1\mu_3\nu_3}+\eta_{[\mu_3[\mu_4}F'_{\nu_4]\nu_3]\mu_1\nu_1\mu_2\nu_2}\biggr),\nn\\
&B_{2\;\mu_1\nu_1\mu_2\nu_2\mu_3\nu_3\mu_4\nu_4}=&\half(F''_{\mu_3\nu_3\mu_4\nu_4}\eta_{\mu_1\mu_2}\eta_{\nu_1\nu_2})_{sym}={1\over48}\times \nn\\&&\biggl(\eta_{[\mu_1[\mu_2}\eta_{\nu_2]\nu_1]}F''_{\mu_3\nu_3\mu_4\nu_4}+\eta_{[\mu_1[\mu_3}\eta_{\nu_3]\nu_1]}F''_{\mu_2\nu_2\mu_4\nu_4}+\nn\\&&+\eta_{[\mu_1[\mu_4}\eta_{\nu_4]\nu_1]}F''_{\mu_2\nu_2\mu_3\nu_3}+\eta_{[\mu_2[\mu_3}\eta_{\nu_3]\nu_2]}F''_{\mu_1\nu_1\mu_4\nu_4}+\nn\\&&+\eta_{[\mu_2[\mu_4}\eta_{\nu_4]\nu_2]}F''_{\mu_1\nu_1\mu_3\nu_3}+\eta_{[\mu_3[\mu_4}\eta_{\nu_4]\nu_3]}F''_{\mu_1\nu_1\mu_2\nu_2}\biggr),\nn\\
&B_{3\;\mu_1\nu_1\mu_2\nu_2\mu_3\nu_3\mu_4\nu_4}=&(F''_{\nu_2\nu_3\mu_4\nu_4}\eta_{\mu_1\mu_2}\eta_{\nu_1\mu_3})_{sym}=-{1\over96}\times\nn\\&&\biggl(\eta_{[\mu_1[\mu_2}\eta_{\nu_2][\mu_3}F''_{\nu_3]\nu_1]\mu_4\nu_4}+\eta_{[\mu_1[\mu_3}\eta_{\nu_3][\mu_2}F''_{\nu_2]\nu_1]\mu_4\nu_4}+\nn\\&&+
\eta_{[\mu_2[\mu_1}\eta_{\nu_1][\mu_3}F''_{\nu_3]\nu_2]\mu_4\nu_4}+\eta_{[\mu_1[\mu_2}\eta_{\nu_2][\mu_4}F''_{\nu_4]\nu_1]\mu_3\nu_3}+\nn\\&&+
\eta_{[\mu_1[\mu_4}\eta_{\nu_4][\mu_2}F''_{\nu_2]\nu_1]\mu_3\nu_3}+\eta_{[\mu_2[\mu_1}\eta_{\nu_1][\mu_4}F''_{\nu_3]\nu_2]\mu_3\nu_3}+\nn\\&&+
\eta_{[\mu_1[\mu_3}\eta_{\nu_3][\mu_4}F''_{\nu_4]\nu_1]\mu_2\nu_2}+\eta_{[\mu_1[\mu_4}\eta_{\nu_4][\mu_3}F''_{\nu_3]\nu_1]\mu_2\nu_2}+\nn\\&&+
\eta_{[\mu_3[\mu_1}\eta_{\nu_1][\mu_4}F''_{\nu_4]\nu_3]\mu_2\nu_2}+\eta_{[\mu_2[\mu_3}\eta_{\nu_3][\mu_4}F''_{\nu_4]\nu_2]\mu_1\nu_1}+\nn\\&&+
\eta_{[\mu_2[\mu_4}\eta_{\nu_4][\mu_3}F''_{\nu_3]\nu_2]\mu_1\nu_1}+\eta_{[\mu_3[\mu_2}\eta_{\nu_2][\mu_4}F''_{\nu_4]\nu_3]\mu_1\nu_1}\biggr),\nn\\
&B_{4\;\mu_1\nu_1\mu_2\nu_2\mu_3\nu_3\mu_4\nu_4}=&(F''_{\nu_2\nu_3\nu_1\nu_4}\eta_{\mu_1\mu_2}\eta_{\mu_3\mu_4}-F''_{\nu_2\nu_3\nu_1\nu_4}\eta_{\mu_1\mu_3}\eta_{\mu_2\mu_4})_{sym}=
{1\over48}\times\nn\\&&\biggl(\eta_{[\mu_4[\mu_1}F''_{\nu_1][\mu_2\nu_4][\mu_3}\eta_{\nu_3]\nu_2]}+
\eta_{[\mu_4[\mu_2}F''_{\nu_2][\mu_1\nu_4][\mu_3}\eta_{\nu_3]\nu_1]}
+\nn\\&&+
\eta_{[\mu_4[\mu_1}F''_{\nu_1][\mu_3\nu_4][\mu_2}\eta_{\nu_2]\nu_3]}+
\eta_{[\mu_4[\mu_3}F''_{\nu_3][\mu_1\nu_4][\mu_2}\eta_{\nu_2]\nu_1]}
+\nn\\&&+
\eta_{[\mu_4[\mu_2}F''_{\nu_2][\mu_3\nu_4][\mu_1}\eta_{\nu_1]\nu_3]}+
\eta_{[\mu_3[\mu_4}F''_{\nu_4][\mu_1\nu_3][\mu_2}\eta_{\nu_2]\nu_1]}\biggr),\nn
\e{a1}
\be
&B_{5\;\mu_1\nu_1\mu_2\nu_2\mu_3\nu_3\mu_4\nu_4}=&(G''_{\nu_1\nu_2\nu_3\nu_4}\eta_{\mu_1\mu_2}\eta_{\mu_3\mu_4})_{sym}=\nn\\&&{1\over48}\biggl(\eta_{[\mu_1[\mu_2}\eta_{[\mu_3[\mu_4}G''_{\nu_4]\nu_3]\nu_2]\nu_1]}+\nn\\&&+\eta_{[\mu_1[\mu_3}\eta_{[\mu_2[\mu_4}G''_{\nu_4]\nu_2]\nu_3]\nu_1]}+
\eta_{[\mu_1[\mu_4}\eta_{[\mu_2[\mu_3}G''_{\nu_3]\nu_2]\nu_4]\nu_1]}\biggr),\nn\\
&B_{6\;\mu_1\nu_1\mu_2\nu_2\mu_3\nu_3\mu_4\nu_4}=&\half(F'''_{\nu_3\nu_4}
\eta_{\mu_1\mu_2}\eta_{\nu_1\nu_2}\eta_{\mu_3\mu_4})_{sym}={1\over192}\times\nn\\&&\biggl(   
\eta_{[\mu_1[\mu_2}\eta_{\nu_2]\nu_1]}\eta_{[\mu_3[\mu_4}F'''_{\nu_4]\nu_3]}+\eta_{[\mu_1[\mu_3}\eta_{\nu_3]\nu_1]}\eta_{[\mu_2[\mu_4}F'''_{\nu_4]\nu_2]}+\nn\\&&+
\eta_{[\mu_1[\mu_4}\eta_{\nu_4]\nu_1]}\eta_{[\mu_2[\mu_3}F'''_{\nu_3]\nu_2]}+\eta_{[\mu_2[\mu_3}\eta_{\nu_3]\nu_2]}\eta_{[\mu_1[\mu_4}F'''_{\nu_4]\nu_1]}+\nn\\&&+
\eta_{[\mu_2[\mu_4}\eta_{\nu_4]\nu_2]}\eta_{[\mu_1[\mu_3}F'''_{\nu_3]\nu_1]}+\eta_{[\mu_3[\mu_4}\eta_{\nu_4]\nu_3]}\eta_{[\mu_1[\mu_2}F'''_{\nu_2]\nu_1]}\biggr),\nn\\
&B_{7\;\mu_1\nu_1\mu_2\nu_2\mu_3\nu_3\mu_4\nu_4}=&(F'''_{\nu_3\nu_4}\eta_{\mu_1\mu_2}\eta_{\nu_1\mu_3}\eta_{\nu_2\mu_4})_{sym}=-{1\over192}\times\nn\\&&\biggl(\eta_{[\mu_1[\mu_2}\eta_{\nu_2][\mu_3}\eta_{\nu_3][\mu_4}F'''_{\nu_4]\nu_1]}  +
\eta_{[\mu_1[\mu_3}\eta_{\nu_3][\mu_2}\eta_{\nu_2][\mu_4}F'''_{\nu_4]\nu_1]}  +\nn\\&&+
\eta_{[\mu_1[\mu_3}\eta_{\nu_3][\mu_4}\eta_{\nu_4][\mu_2}F'''_{\nu_2]\nu_1]}  +
\eta_{[\mu_1[\mu_2}\eta_{\nu_2][\mu_4}\eta_{\nu_4][\mu_3}F'''_{\nu_3]\nu_1]}  +\nn\\&&+
\eta_{[\mu_1[\mu_4}\eta_{\nu_4][\mu_3}\eta_{\nu_3][\mu_2}F'''_{\nu_2]\nu_1]}  +
\eta_{[\mu_1[\mu_4}\eta_{\nu_4][\mu_2}\eta_{\nu_2][\mu_3}F'''_{\nu_3]\nu_1]}  +\nn\\&&+
\eta_{[\mu_2[\mu_3}\eta_{\nu_3][\mu_1}\eta_{\nu_1][\mu_4}F'''_{\nu_4]\nu_2]}  +
\eta_{[\mu_2[\mu_1}\eta_{\nu_1][\mu_3}\eta_{\nu_3][\mu_4}F'''_{\nu_4]\nu_2]}  +\nn\\&&+
\eta_{[\mu_2[\mu_1}\eta_{\nu_1][\mu_4}\eta_{\nu_4][\mu_3}F'''_{\nu_3]\nu_2]}  +
\eta_{[\mu_2[\mu_4}\eta_{\nu_4][\mu_1}\eta_{\nu_1][\mu_3}F'''_{\nu_3]\nu_2]}  +\nn\\&&+
\eta_{[\mu_3[\mu_4}\eta_{\nu_4][\mu_2}\eta_{\nu_2][\mu_4}F'''_{\nu_4]\nu_3]}  +
\eta_{[\mu_3[\mu_2}\eta_{\nu_2][\mu_1}\eta_{\nu_1][\mu_4}F'''_{\nu_4]\nu_3]}
\biggr),            \nn\\
&B_{8\;\mu_1\nu_1\mu_2\nu_2\mu_3\nu_3\mu_4\nu_4}=&\half U(\eta_{\mu_1\mu_2}\eta_{\nu_1\nu_2}\eta_{\mu_3\mu_4}\eta_{\nu_3\nu_4})_{sym}=\nn\\&&{1\over96}\biggl(  \eta_{[\mu_1[\mu_2}\eta_{\nu_2]\nu_1]} \eta_{[\mu_3[\mu_4}\eta_{\nu_4]\nu_3]} +  
 \eta_{[\mu_1[\mu_3}\eta_{\nu_3]\nu_1]} \eta_{[\mu_2[\mu_4}\eta_{\nu_4]\nu_2]} +             \nn\\&&+
  \eta_{[\mu_1[\mu_4}\eta_{\nu_4]\nu_1]} \eta_{[\mu_2[\mu_3}\eta_{\nu_3]\nu_2]}\biggr),\nn\\
&B_{9\;\mu_1\nu_1\mu_2\nu_2\mu_3\nu_3\mu_4\nu_4}=&U(\eta_{\mu_1\mu_2}\eta_{\nu_1\mu_3}\eta_{\nu_2\mu_4}\eta_{\nu_3\nu_4})_{sym}=-{1\over48}\times\nn\\&&\biggl( \eta_{[\mu_1[\mu_2}\eta_{\nu_2][\mu_3} \eta_{\nu_3][\mu_4}\eta_{\nu_4]\nu_1]}+  \eta_{[\mu_1[\mu_3}\eta_{\nu_3][\mu_2} \eta_{\nu_2][\mu_4}\eta_{\nu_4]\nu_1]}+            \nn\\&&+\eta_{[\mu_1[\mu_3}\eta_{\nu_3][\mu_4} \eta_{\nu_4][\mu_2}\eta_{\nu_2]\nu_1]}\biggr). 
\e{a2}
The notations might be a little confusing, particularly in $B_4$. What is meant is simply an antisymmetrization in $\mu_k\nu_k$ for $k=1,2,3,4$.

  \setcounter{equation}{0}
\section{The explicit expressions for $M_i$ in  \rl{322}.}
The $M_i$-expressions in \rl{322} are explicitly given by
\be
&M_{1\;\mu_1\nu_1\mu_2\nu_2\mu_3\nu_3\mu_4\nu_4}(x)=&P_{\mu_1\nu_1\mu_2\nu_2}\Lambda_{\mu_3\nu_3\mu_4\nu_4}(x)+
P_{\mu_3\nu_3\mu_4\nu_4}\Lambda_{\mu_1\nu_1\mu_2\nu_2}(x)+\nn\\
&&P_{\mu_1\nu_1\mu_3\nu_3}\Lambda_{\mu_2\nu_2\mu_4\nu_4}(x)+
P_{\mu_2\nu_2\mu_4\nu_4}\Lambda_{\mu_1\nu_1\mu_3\nu_3}(x)+\nn\\
&&P_{\mu_2\nu_2\mu_3\nu_3}\Lambda_{\mu_1\nu_1\mu_4\nu_4}(x)+
P_{\mu_1\nu_1\mu_4\nu_4}\Lambda_{\mu_2\nu_2\mu_3\nu_3}(x),\nn\\
&M_{2\;\mu_1\nu_1\mu_2\nu_2\mu_3\nu_3\mu_4\nu_4}(x)=&\Box\biggl(\eta_{[\mu_1[\mu_2}\eta_{\nu_2]\nu_1]}\Lambda_{\mu_3\nu_3\mu_4\nu_4}(x)+\nn\\&&
\eta_{[\mu_3[\mu_4}\eta_{\nu_4]\nu_3]}\Lambda_{\mu_1\nu_1\mu_2\nu_2}(x)+\nn\\
&&\eta_{[\mu_1[\mu_3}\eta_{\nu_3]\nu_1]}\Lambda_{\mu_2\nu_2\mu_4\nu_4}(x)+\nn\\&&
\eta_{[\mu_2[\mu_4}\eta_{\nu_4]\nu_2]}\Lambda_{\mu_1\nu_1\mu_3\nu_3}(x)+\nn\\
&&\eta_{[\mu_2[\mu_3}\eta_{\nu_3]\nu_2]}\Lambda_{\mu_1\nu_1\mu_4\nu_4}(x)+\nn\\&&
\eta_{[\mu_1[\mu_4}\eta_{\nu_4]\nu_1]}\Lambda_{\mu_2\nu_2\mu_3\nu_3}(x)\biggr),\nn\\
&M_{3\;\mu_1\nu_1\mu_2\nu_2\mu_3\nu_3\mu_4\nu_4}(x)=&\Box\biggl(\eta_{[\mu_2[\mu_3}\eta_{\nu_3][\mu_4}\Lambda_{\nu_4]\nu_2]\mu_1\nu_1}(x)+\nn\\&&\eta_{[\mu_2[\mu_4}\eta_{\nu_4][\mu_3}\Lambda_{\nu_3]\nu_2]\mu_1\nu_1}(x)+\nn\\&&
\eta_{[\mu_3[\mu_2}\eta_{\nu_2][\mu_4}\Lambda_{\nu_4]\nu_3]\mu_1\nu_1}(x)+\nn\\&&\eta_{[\mu_3[\mu_1}\eta_{\nu_1][\mu_4}\Lambda_{\nu_4]\nu_3]\mu_2\nu_2}(x)+\nn\\&&
\eta_{[\mu_1[\mu_3}\eta_{\nu_3][\mu_4}\Lambda_{\nu_4]\nu_1]\mu_2\nu_2}(x)+\nn\\&&\eta_{[\mu_1[\mu_4}\eta_{\nu_4][\mu_3}\Lambda_{\nu_3]\nu_1]\mu_2\nu_2}(x)+\nn\\&&
\eta_{[\mu_1[\mu_2}\eta_{\nu_2][\mu_4}\Lambda_{\nu_4]\nu_1]\mu_3\nu_3}(x)+\nn\\&&\eta_{[\mu_1[\mu_4}\eta_{\nu_4][\mu_2}\Lambda_{\nu_2]\nu_1]\mu_3\nu_3}(x)+\nn\\&&
\eta_{[\mu_2[\mu_1}\eta_{\nu_1][\mu_4}\Lambda_{\nu_4]\nu_2]\mu_3\nu_3}(x)+\nn\\&&\eta_{[\mu_2[\mu_1}\eta_{\nu_1][\mu_3}\Lambda_{\nu_3]\nu_2]\mu_4\nu_4}(x)+\nn\\&&
\eta_{[\mu_1[\mu_3}\eta_{\nu_3][\mu_2}\Lambda_{\nu_2]\nu_1]\mu_4\nu_4}(x)+\nn\\&&\eta_{[\mu_1[\mu_2}\eta_{\nu_2][\mu_3}\Lambda_{\nu_3]\nu_1]\mu_4\nu_4}(x)\biggr),\nn\\
&M_{4\;\mu_1\nu_1\mu_2\nu_2\mu_3\nu_3\mu_4\nu_4}(x)=&\Box\biggl(\eta_{[\mu_4[\mu_1}\eta_{[\mu_3[\mu_2}\Lambda_{\nu_2]\nu_1]\nu_3]\nu_4]}(x)+\nn\\&&\eta_{[\mu_4[\mu_1}\eta_{[\mu_2[\mu_3}\Lambda_{\nu_3]\nu_1]\nu_2]\nu_4]}(x)+\nn\\ &&
\eta_{[\mu_3[\mu_1}\eta_{[\mu_2[\mu_4}\Lambda_{\nu_4]\nu_1]\nu_2]\nu_3]}(x)\biggr),\nn
\e{aa0}
\be
&M_{5\;\mu_1\nu_1\mu_2\nu_2\mu_3\nu_3\mu_4\nu_4}(x)=&
P_{\mu_1\nu_1\mu_2\nu_2}\eta_{[\mu_3[\mu_4}\Lambda'_{\nu_4]\nu_3]}(x)+\nn\\&&
P_{\mu_3\nu_3\mu_4\nu_4}\eta_{[\mu_1[\mu_2}\Lambda'_{\nu_2]\nu_1]}(x)+\nn\\
&&P_{\mu_1\nu_1\mu_3\nu_3}\eta_{[\mu_2[\mu_4}\Lambda'_{\nu_4]\nu_2]}(x)+\nn\\&&
P_{\mu_2\nu_2\mu_4\nu_4}\eta_{[\mu_1[\mu_3}\Lambda'_{\nu_3]\nu_1]}(x)+\nn\\
&&P_{\mu_2\nu_2\mu_3\nu_3}\eta_{[\mu_1[\mu_4}\Lambda'_{\nu_4]\nu_1]}(x)+\nn\\&&
P_{\mu_1\nu_1\mu_4\nu_4}\eta_{[\mu_2[\mu_3}\Lambda'_{\nu_3]\nu_2]}(x),
\nn\\
&M_{6\;\mu_1\nu_1\mu_2\nu_2\mu_3\nu_3\mu_4\nu_4}(x)=&
\Box\biggl(\eta_{[\mu_1[\mu_2}\eta_{\nu_2]\nu_1]}\eta_{[\mu_3[\mu_4}\Lambda'_{\nu_4]\nu_3]}(x)+\nn\\&&
\eta_{[\mu_3[\mu_4}\eta_{\nu_4]\nu_3]}\eta_{[\mu_1[\mu_2}\Lambda'_{\nu_2]\nu_1]}(x)+\nn\\
&&\eta_{[\mu_1[\mu_3}\eta_{\nu_3]\nu_1]}\eta_{[\mu_2[\mu_4}\Lambda'_{\nu_4]\nu_2]}(x)+\nn\\&&
\eta_{[\mu_2[\mu_4}\eta_{\nu_4]\nu_2]}\eta_{[\mu_1[\mu_3}\Lambda'_{\nu_3]\nu_1]}(x)+\nn\\
&&\eta_{[\mu_2[\mu_3}\eta_{\nu_3]\nu_2]}\eta_{[\mu_1[\mu_4}\Lambda'_{\nu_4]\nu_1]}(x)+\nn\\&&
\eta_{[\mu_1[\mu_4}\eta_{\nu_4]\nu_1]}\eta_{[\mu_2[\mu_3}\Lambda'_{\nu_3]\nu_2]}(x)\biggr),
\nn\\
&M_{7\;\mu_1\nu_1\mu_2\nu_2\mu_3\nu_3\mu_4\nu_4}(x)=&
\Box\biggl(\eta_{[\mu_1[\mu_3}\eta_{\nu_3][\mu_4}\eta_{\nu_4][\mu_2}\Lambda'_{\nu_2]\nu_1]}(x)+\nn\\&&
\eta_{[\mu_1[\mu_4}\eta_{\nu_4][\mu_3}\eta_{\nu_3][\mu_2}\Lambda'_{\nu_2]\nu_1]}(x)+\nn\\&&
\eta_{[\mu_1[\mu_2}\eta_{\nu_2][\mu_4}\eta_{\nu_4][\mu_3}\Lambda'_{\nu_3]\nu_1]}(x)+\nn\\&&
\eta_{[\mu_1[\mu_4}\eta_{\nu_4][\mu_2}\eta_{\nu_2][\mu_3}\Lambda'_{\nu_3]\nu_1]}(x)+\nn\\
&&
\eta_{[\mu_1[\mu_2}\eta_{\nu_2][\mu_3}\eta_{\nu_3][\mu_4}\Lambda'_{\nu_4]\nu_1]}(x)+\nn\\&&
\eta_{[\mu_1[\mu_3}\eta_{\nu_3][\mu_2}\eta_{\nu_2][\mu_4}\Lambda'_{\nu_4]\nu_1]}(x)+\nn\\
&&
\eta_{[\mu_3[\mu_4}\eta_{\nu_4][\mu_1}\eta_{\nu_1][\mu_2}\Lambda'_{\nu_2]\nu_3]}(x)+\nn\\&&
\eta_{[\mu_3[\mu_1}\eta_{\nu_1][\mu_4}\eta_{\nu_4][\mu_2}\Lambda'_{\nu_2]\nu_3]}(x)+\nn\\
&&
\eta_{[\mu_4[\mu_1}\eta_{\nu_1][\mu_3}\eta_{\nu_3][\mu_2}\Lambda'_{\nu_2]\nu_4]}(x)+\nn\\&&
\eta_{[\mu_4[\mu_3}\eta_{\nu_3][\mu_1}\eta_{\nu_1][\mu_2}\Lambda'_{\nu_2]\nu_4]}(x)+\nn\\
&&
\eta_{[\mu_4[\mu_1}\eta_{\nu_1][\mu_2}\eta_{\nu_2][\mu_3}\Lambda'_{\nu_3]\nu_4]}(x)+\nn\\&&
\eta_{[\mu_4[\mu_2}\eta_{\nu_2][\mu_1}\eta_{\nu_1][\mu_3}\Lambda'_{\nu_3]\nu_4]}(x)\biggr),\nn
\e{aa1}
\be
&M_{8\;\mu_1\nu_1\mu_2\nu_2\mu_3\nu_3\mu_4\nu_4}(x)=&\biggl(
\eta_{[\mu_1[\mu_2}\eta_{\nu_2]\nu_1]}\eta_{[\mu_3[\mu_4}\eta_{\nu_4]\nu_3]}+\nn\\&&
\eta_{[\mu_1[\mu_3}\eta_{\nu_3]\nu_1]}\eta_{[\mu_2[\mu_4}\eta_{\nu_4]\nu_2]}+\nn\\&&
\eta_{[\mu_1[\mu_4}\eta_{\nu_4]\nu_1]}\eta_{[\mu_2[\mu_3}\eta_{\nu_3]\nu_2]}\biggr)\Box\Lambda''(x),
\nn\\
&M_{9\;\mu_1\nu_1\mu_2\nu_2\mu_3\nu_3\mu_4\nu_4}(x)=&
\biggl(
\eta_{[\mu_1[\mu_2}\eta_{\nu_2][\mu_3}\eta_{\nu_3][\mu_4}\eta_{\nu_4]\nu_1]}+\nn\\&&
\eta_{[\mu_1[\mu_3}\eta_{\nu_3][\mu_2}\eta_{\nu_2][\mu_4}\eta_{\nu_4]\nu_1]}+\nn\\&&
\eta_{[\mu_1[\mu_4}\eta_{\nu_4][\mu_2}\eta_{\nu_2][\mu_3}\eta_{\nu_3]\nu_1]}\biggr)\Box\Lambda''(x),
\nn\\
&M_{10\;\mu_1\nu_1\mu_2\nu_2\mu_3\nu_3\mu_4\nu_4}(x)=&
\biggl(\eta_{[\mu_3[\mu_4}\eta_{\nu_4]\nu_3]}P_{\mu_1\nu_1\mu_2\nu_2}+\nn\\&&
\eta_{[\mu_1[\mu_2}\eta_{\nu_2]\nu_1]}P_{\mu_3\nu_3\mu_4\nu_4}+\nn\\
&&\eta_{[\mu_2[\mu_4}\eta_{\nu_4]\nu_2]}P_{\mu_1\nu_1\mu_3\nu_3}+\nn\\&&
\eta_{[\mu_1[\mu_3}\eta_{\nu_3]\nu_1]}P_{\mu_2\nu_2\mu_4\nu_4}+\nn\\
&&\eta_{[\mu_1[\mu_4}\eta_{\nu_4]\nu_1]}P_{\mu_2\nu_2\mu_3\nu_3}+\nn\\&&\eta_{[\mu_2[\mu_3}\eta_{\nu_3]\nu_2]}P_{\mu_1\nu_1\mu_4\nu_4}\biggr)\Lambda''(x),\nn\\
&M_{11\;\mu_1\nu_1\mu_2\nu_2\mu_3\nu_3\mu_4\nu_4}(x)=&
\biggl(\eta_{[\mu_1[\mu_3}\eta_{\nu_3][\mu_4}\eta_{\nu_4][\mu_2}\dif_{\nu_2]}\dif_{\nu_1]}+\nn\\&&
\eta_{[\mu_1[\mu_4}\eta_{\nu_4][\mu_3}\eta_{\nu_3][\mu_2}\dif_{\nu_2]}\dif_{\nu_1]}+\nn\\&&
\eta_{[\mu_1[\mu_2}\eta_{\nu_2][\mu_4}\eta_{\nu_4][\mu_3}\dif_{\nu_3]}\dif_{\nu_1]}+\nn\\&&
\eta_{[\mu_1[\mu_4}\eta_{\nu_4][\mu_2}\eta_{\nu_2][\mu_3}\dif_{\nu_3]}\dif_{\nu_1]}+\nn\\&&
\eta_{[\mu_1[\mu_2}\eta_{\nu_2][\mu_3}\eta_{\nu_3][\mu_4}\dif_{\nu_4]}\dif_{\nu_1]}+\nn\\&&
\eta_{[\mu_1[\mu_3}\eta_{\nu_3][\mu_2}\eta_{\nu_2][\mu_4}\dif_{\nu_4]}\dif_{\nu_1]}+\nn\\&&
\eta_{[\mu_2[\mu_1}\eta_{\nu_1][\mu_4}\eta_{\nu_4][\mu_3}\dif_{\nu_3]}\dif_{\nu_2]}+\nn\\&&
\eta_{[\mu_2[\mu_4}\eta_{\nu_4][\mu_1}\eta_{\nu_1][\mu_3}\dif_{\nu_3]}\dif_{\nu_2]}+\nn\\&&
\eta_{[\mu_4[\mu_1}\eta_{\nu_1][\mu_3}\eta_{\nu_3][\mu_2}\dif_{\nu_2]}\dif_{\nu_4]}+\nn\\&&
\eta_{[\mu_4[\mu_3}\eta_{\nu_3][\mu_1}\eta_{\nu_1][\mu_2}\dif_{\nu_2]}\dif_{\nu_4]}+\nn\\&&
\eta_{[\mu_3[\mu_1}\eta_{\nu_1][\mu_2}\eta_{\nu_2][\mu_4}\dif_{\nu_4]}\dif_{\nu_3]}+\nn\\&&
\eta_{[\mu_3[\mu_2}\eta_{\nu_2][\mu_1}\eta_{\nu_1][\mu_4}\dif_{\nu_4]}\dif_{\nu_3]}\biggr)\Lambda''(x),
\e{aa2}
where $P$, $\Lambda$ and $\Lambda'$, $\Lambda''$ are defined in \rl{304}, \rl{319} and \rl{321}.

  \setcounter{equation}{0}
\section{Calculation of the quadratic condition\\ $\tilde{\cL}_2(y)=0$ in subsection 10.3.}
The quadratic condition for the invariance of the manifest Lagrangian in the $s=3$ case is (from \rl{6491})
\be
&&\tilde{\cL}_2=\tilde{F}^2+\al(\tilde{F}')^2+\beta(\tilde{F}'')^2=0.
\e{b1}
From \rl{639}-\rl{641} I find
\be
&&(\tilde{F}(y)|_0)^2=48(d+2)\tilde{L}^2+48(\tilde{L}')^2-192\tilde{M}^{AB}\tilde{M}_{BA},\nn\\
&&(\tilde{F}(y)|_y)^2=96\tilde{L}^2,\quad (\tilde{F}(y)|_0)\cdot(\tilde{F}(y)|_y)=-96\tilde{L}^2.
\e{b2}
where the functions on the right-hand sides are defined in \rl{642} and \rl{646}.
By means of these definitions  I find the relation 
\be
&&\tilde{M}^{AB}\tilde{M}_{BA}=\tilde{M}^2-\half (\tilde{L}')^2.
\e{b3}
Hence, 
\be
&\tilde{F}^2=&(\tilde{F}(y)|_0)^2+(\tilde{F}(y)|_y)^2+2(\tilde{F}(y)|_0)\cdot(\tilde{F}(y)|_y)=\nn\\&&
48\biggl(d\tilde{L}^2+3(\tilde{L}')^2-4\tilde{M}^2\biggr).
\e{b4}
By means of the expressions in \rl{645} I find
\be
&&(\tilde{F}'|_0)^2=4d\tilde{L}^2+12(d+2)(\tilde{L}')^2-32d\tilde{M}^{AB}\tilde{M}^{BA}+16(d+2)\tilde{M}^2,\nn\\
&&(\tilde{F}'(y)|_y)^2=8(\tilde{L}')^2+32\tilde{M}^2,\nn\\
&&(\tilde{F}'(y)|_0)\cdot(\tilde{F}'(y)|_y)=-8(\tilde{L}')^2-32\tilde{M}^2.
\e{b5}
This implies by means of \rl{b3}
\be
&(\tilde{F}'(y))^2=&(\tilde{F}'|_0)^2+(\tilde{F}'|_y)^2+2(\tilde{F}'(y)|_0)\cdot(\tilde{F}'(y)|_y)=\nn\\&&
4d^2\tilde{L}^2+4(7d+4)(\tilde{L}')^2-16d\tilde{M}^2.
\e{b6}
Finally I find from \rl{648}
\be
&&(\tilde{F}''|_0)^2=4(d+1)^2(\tilde{L}')^2,\nn\\
&&(\tilde{F}''|_y)^2=0,\quad(\tilde{F}''(y)|_0)\cdot(\tilde{F}''(y)|_y)=0.
\e{b7}
Condition \rl{b1} becomes therefore
\be
&\tilde{\cL}_2=&4d(12+\al d)\tilde{L}^2+4(36+\al(7d+4)+\beta(d+1)^2)(\tilde{L}')^2-\nn\\&&-16(12+\al d)\tilde{M}^2=0.
\e{b8}

  \setcounter{equation}{0}
\section{The transformation formulas  in subsection 10.4.}
The special gauge transformations \rl{663} imply for the basic field strength \rl{655},
$F\;\ra\;F+\tilde{F}$, where in turn
\be
&&\tilde{F}_{A_1B_1A_2B_2A_3B_3A_4B_4}(y)=\tilde{F}_{A_1B_1A_2B_2A_3B_3A_4B_4}(y)|_0+\tilde{F}_{A_1B_1A_2B_2A_3B_3A_4B_4}(y)|_y,\nn
\e{c0}
and
\be
&\tilde{F}_{A_1B_1A_2B_2A_3B_3A_4B_4}(y)|_0=&2\bigl( \eta_{[A_1[A_2}\dif_{[A_3}\dif_{[A_4}\tilde{\phi}_{B_4]B_3]B_2]B_1]}+\nn\\&&+\eta_{[A_1[A_3}\dif_{[A_2}\dif_{[A_4}\tilde{\phi}_{B_4]B_2]B_3]B_1]} +\nn\\&&+\eta_{[A_1[A_4}\dif_{[A_3}\dif_{[A_2}\tilde{\phi}_{B_2]B_3]B_4]B_1]}+\nn\\&&+\eta_{[A_2[A_3}\dif_{[A_4}\dif_{[A_1}\tilde{\phi}_{B_1]B_4]B_3]B_2]}+\nn\\&&+\eta_{[A_2[A_4}\dif_{[A_1}\dif_{[A_3}\tilde{\phi}_{B_3]B_1]B_4]B_2]}+\nn\\&&+\eta_{[A_3[A_4}\dif_{[A_1}\dif_{[A_2}\tilde{\phi}_{B_2]B_1]B_4]B_3]}\bigr), \nn\\
\e{c00}
\be
&\tilde{F}_{A_1B_1A_2B_2A_3B_3A_4B_4}(y)|_y=&2\bigl(y_{[A_1}\dif_{[A_2}\dif_{[A_3}\dif_{[A_4}\tilde{\phi}_{B_4]B_3]B_2]B_1]}+\nn\\&&+ y_{[A_2}\dif_{[A_3}\dif_{[A_4}\dif_{[A_1}\tilde{\phi}_{B_1]B_4]B_3]B_2]} +\nn\\&&+y_{[A_3}\dif_{[A_4}\dif_{[A_1}\dif_{[A_2}\tilde{\phi}_{B_2]B_1]B_4]B_3]} +\nn\\&&+y_{[A_4}\dif_{[A_1}\dif_{[A_2}\dif_{[A_3}\tilde{\phi}_{B_3]B_2]B_1]B_4]}\bigr).   \nn\\
\e{c1}
These expressions imply
\be
&\tilde{F}'_{B_1B_2A_3B_3A_4B_4}(y)&\equiv
\eta^{A_1A_2}\tilde{F}_{A_1B_1A_2B_2A_3B_3A_4B_4}(y)=\nn\\&&=\tilde{F}'_{B_1B_2A_3B_3A_4B_4}(y)|_0+\tilde{F}'_{B_1B_2A_3B_3A_4B_4}(y)|_y,\nn\\
\e{c2}
where
\be
&\tilde{F}'_{B_1B_2A_3B_3A_4B_4}(y)|_0=&2\biggl((d+2)\dif_{[A_3}\dif_{[A_4}\tilde{\phi}_{B_4]B_3]B_1B_2}+\nn\\&&+\eta_{B_1B_2}\dif_{[A_3}\dif_{[A_4}\tilde{\phi}_{B_4]B_3]}+\nn\\&&+\eta_{B_1[A_3}\dif_{[A_4}\dif_{B_2}\tilde{\phi}_{B_4]B_3]}+\eta_{B_2[A_3}\dif_{[A_4}\dif_{B_1}\tilde{\phi}_{B_4]B_3]}+\nn\\&&+\eta_{B_1[A_4}\dif_{[A_3}\dif_{B_2}\tilde{\phi}_{B_3]B_4]}+\eta_{B_2[A_4}\dif_{[A_3}\dif_{B_1}\tilde{\phi}_{B_3]B_4]}-\nn\\&&-\eta_{B_1[A_3}\dif_{[A_4}\dif^{A}\tilde{\phi}_{AB_2B_4]B_3]}-\nn\\&&-\eta_{B_2[A_3}\dif_{[A_4}\dif^{A}\tilde{\phi}_{AB_1B_4]B_3]}-\nn\\&&-\eta_{B_1[A_4}\dif_{[A_3}\dif^{A}\tilde{\phi}_{AB_2B_3]B_4]}-\nn\\&&-\eta_{B_2[A_4}\dif_{[A_3}\dif^{A}\tilde{\phi}_{AB_1B_3]B_4]}+\nn\\&&+\eta_{[A_4[A_3}\bigl(\Box\tilde{\phi}_{B_1B_2B_3]B_4]}+\dif_{B_1}\dif_{B_2}\tilde{\phi}_{B_3]B_4]}-\nn\\&&-\dif_{B_1}\dif^A\tilde{\phi}_{AB_2B_3]B_4]}-\dif_{B_2}\dif^A\tilde{\phi}_{AB_1B_3]B_4]}\bigr)
\biggr),\nn
\e{c201}
\be
&\tilde{F}'_{B_1B_2A_3B_3A_4B_4}(y)|_y=&2\biggl(y_{B_1}\dif_{B_2}\dif_{[A_3}\dif_{[A_4}\tilde{\phi}_{B_4]B_3]}+\nn\\&&+y_{B_2}\dif_{B_1}\dif_{[A_3}\dif_{[A_4}\tilde{\phi}_{B_4]B_3]}-\nn\\&&-y_{B_1}\dif_{[A_3}\dif_{[A_4}\dif^A\tilde{\phi}_{AB_2B_4]B_3]}-\nn\\&&-y_{B_2}\dif_{[A_3}\dif_{[A_4}\dif^A\tilde{\phi}_{AB_1B_4]B_3]}+\nn\\&&+y_{[A_3}\bigl(\Box\dif_{[A_4}\tilde{\phi}_{B_1B_2B_4]B_3]}+\dif_{B_1}\dif_{B_2}\dif_{[A_4}\tilde{\phi}_{B_4]B_3]}-\nn\\&&-\dif_{B_1}\dif_{[A_4}\dif^A\tilde{\phi}_{AB_2B_4]B_3]}
-\dif_{B_2}\dif_{[A_4}\dif^A\tilde{\phi}_{AB_1B_4]B_3]}\bigr)+\nn\\
&&+y_{[A_4}\bigl(\Box\dif_{[A_3}\tilde{\phi}_{B_1B_2B_3]B_4]}+\dif_{B_1}\dif_{B_2}\dif_{[A_3}\tilde{\phi}_{B_3]B_4]}-\nn\\&&-\dif_{B_1}\dif_{[A_3}\dif^A\tilde{\phi}_{AB_2B_3]B_4]}-\nn\\&&
-\dif_{B_2}\dif_{[A_3}\dif^A\tilde{\phi}_{AB_1B_3]B_4]}\bigr)\biggr),
\e{c3}
where
\be
&&\tilde{\phi}_{AB}(y)\equiv\eta^{CD}\tilde{\phi}_{ABCD}(y).
\e{c301}
Furthermore, I find
\be
&\tilde{F}''_{A_3B_3A_4B_4}(y)&\equiv\half\eta^{B_1B_2}\tilde{F}'_{B_1B_2A_3B_3A_4B_4}(y)=\nn\\&&=\tilde{F}''_{A_3B_3A_4B_4}(y)|_0+\tilde{F}''_{A_3B_3A_4B_4}(y)|_y,
\e{c4}
where
\be
&\tilde{F}''_{A_3B_3A_4B_4}(y)|_0=&2\biggl((d+3)\dif_{[A_4}\dif_{[A_3}\tilde{\phi}_{B_3]B_4]}+\nn\\&&+
\eta_{[A_4[A_3}\bigl(\Box\tilde{\phi}_{B_3]B_4]}-\dif^{A}\dif^{B}\tilde{\phi}_{ABB_3]B_4]}\bigr)\biggr),\nn\\
&\tilde{F}''_{A_3B_3A_4B_4}(y)|_y=&2\biggl(y_{[A_4}\bigl(\Box \dif_{[A_3}\tilde{\phi}_{B_3]B_4]}-
\dif_{[A_3}\dif^A\dif^B\tilde{\phi}_{ABB_3]B_4]}\bigr)+\nn\\&&+
y_{[A_3}\bigl(\Box\dif_{[A_4}\tilde{\phi}_{B_4]B_3]}-
\dif_{[A_4}\dif^A\dif^B\tilde{\phi}_{ABB_4]B_3]}\bigr)\biggr),
\e{c5}
and
\be
&\tilde{G}''_{B_1B_2B_3B_4}(y)&\equiv\eta^{A_3A_4}\tilde{F}'_{B_1B_2A_3B_3A_4B_4}(y)=\nn\\&&=\tilde{G}''_{B_1B_2B_3B_4}(y)|_0+\tilde{G}''_{B_1B_2B_3B_4}(y)|_y,
\e{c6}
where
\be
&\tilde{G}''_{B_1B_2B_3B_4}(y)|_0=&2\biggl(d\bigl(2\Box\tilde{\phi}_{B_1B_2B_3B_4}+\dif_{B_1}\dif_{B_2}\tilde{\phi}_{B_3B_4}+\dif_{B_3}\dif_{B_4}\tilde{\phi}_{B_1B_2}-\nn\\&&-\dif_{B_1}\dif^A\tilde{\phi}_{AB_2B_3B_4}-\dif_{B_2}\dif^A\tilde{\phi}_{AB_1B_3B_4}-\nn\\&&-\dif_{B_3}\dif^A\tilde{\phi}_{AB_1B_2B_4}-\dif_{B_4}\dif^A\tilde{\phi}_{AB_1B_2B_3}\bigr)+\nn\\&&+2\bigl(\dif_{B_1}\dif_{B_2}\tilde{\phi}_{B_3B_4}+\dif_{B_3}\dif_{B_4}\tilde{\phi}_{B_1B_2}\bigr)-\nn\\&&
-\bigl(\dif_{B_1}\dif_{B_3}\tilde{\phi}_{B_2B_4}+\dif_{B_1}\dif_{B_4}\tilde{\phi}_{B_2B_3}+\nn\\&&+\dif_{B_2}\dif_{B_3}\tilde{\phi}_{B_1B_4}+\dif_{B_2}\dif_{B_4}\tilde{\phi}_{B_1B_3})+\nn\\&&
+
\eta_{B_1B_2}\bigl(\Box\tilde{\phi}_{B_3B_4}+\dif_{B_3}\dif_{B_4}\tilde{\phi}-\nn\\&&-\dif_{B_3}\dif^A\tilde{\phi}_{AB_4}-\dif_{B_4}\dif^A\tilde{\phi}_{AB_3}\bigr)+\nn\\&&+
\eta_{B_3B_4}\bigl(\Box\tilde{\phi}_{B_1B_2}+\dif_{B_1}\dif_{B_2}\tilde{\phi}-\nn\\&&-\dif_{B_1}\dif^A\tilde{\phi}_{AB_2}-\dif_{B_2}\dif^A\tilde{\phi}_{AB_1}\bigr)+\nn\\&&+
\eta_{B_1B_3}\bigl(\dif^A\dif^B\tilde{\phi}_{ABB_2B_4}+\dif_{B_2}\dif_{B_4}\tilde{\phi}-\nn\\&&-\dif_{B_2}\dif^A\tilde{\phi}_{AB_4}-\dif_{B_4}\dif^A\tilde{\phi}_{AB_2}\bigr)+\nn\\&&+
\eta_{B_1B_4}\bigl(\dif^A\dif^B\tilde{\phi}_{ABB_2B_3}+\dif_{B_2}\dif_{B_3}\tilde{\phi}-\nn\\&&-\dif_{B_2}\dif^A\tilde{\phi}_{AB_3}-\dif_{B_3}\dif^A\tilde{\phi}_{AB_2}\bigr)+\nn\\&&+
\eta_{B_2B_3}\bigl(\dif^A\dif^B\tilde{\phi}_{ABB_1B_4}+\dif_{B_1}\dif_{B_4}\tilde{\phi}-\nn\\&&-\dif_{B_1}\dif^A\tilde{\phi}_{AB_4}-\dif_{B_4}\dif^A\tilde{\phi}_{AB_1}\bigr)+\nn\\&&+
\eta_{B_2B_4}\bigl(\dif^A\dif^B\tilde{\phi}_{ABB_1B_3}+\dif_{B_1}\dif_{B_3}\tilde{\phi}-\nn\\&&-\dif_{B_1}\dif^A\tilde{\phi}_{AB_3}-\dif_{B_3}\dif^A\tilde{\phi}_{AB_1}\bigr)\biggr),\nn
\e{c601}
\be
&\tilde{G}''_{B_1B_2B_3B_4}(y)|_y=&2\biggl(\bigl(y_{B_1}\dif_{B_2}+y_{B_2}\dif_{B_1}\bigr)\bigl(\Box\tilde{\phi}_{B_3B_4}+\nn\\&&+\dif_{B_3}\dif_{B_4}\tilde{\phi}-\dif_{B_3}\dif^A\tilde{\phi}_{AB_4}-\dif_{B_4}\dif^A\tilde{\phi}_{AB_3}\bigr)+\nn\\&&+\bigl(y_{B_3}\dif_{B_4}+y_{B_4}\dif_{B_3}\bigr)\bigl(\Box\tilde{\phi}_{B_1B_2}+\nn\\&&+\dif_{B_1}\dif_{B_2}\tilde{\phi}-\dif_{B_1}\dif^A\tilde{\phi}_{AB_2}-\dif_{B_2}\dif^A\tilde{\phi}_{AB_1}\bigr)-\nn\\&&-
y_{B_1}\bigl(\Box\dif^A\tilde{\phi}_{AB_2B_3B_4}+\dif_{B_3}\dif_{B_4}\dif^A\tilde{\phi}_{AB_2}-\nn\\&&-\dif_{B_3}\dif^A\dif^B\tilde{\phi}_{ABB_2B_4}-\dif_{B_4}\dif^A\dif^B\tilde{\phi}_{ABB_2B_3}\bigr)-\nn\\&&-
y_{B_2}\bigl(\Box\dif^A\tilde{\phi}_{AB_1B_3B_4}+\dif_{B_3}\dif_{B_4}\dif^A\tilde{\phi}_{AB_1}-\nn\\&&-\dif_{B_3}\dif^A\dif^B\tilde{\phi}_{ABB_1B_4}-\dif_{B_4}\dif^A\dif^B\tilde{\phi}_{ABB_1B_3}\bigr)-\nn\\&&-
y_{B_3}\bigl(\Box\dif^A\tilde{\phi}_{AB_2B_1B_4}+\dif_{B_1}\dif_{B_2}\dif^A\tilde{\phi}_{AB_4}-\nn\\&&-\dif_{B_1}\dif^A\dif^B\tilde{\phi}_{ABB_2B_4}-\dif_{B_2}\dif^A\dif^B\tilde{\phi}_{ABB_1B_4}\bigr)-\nn\\&&-
y_{B_4}\bigl(\Box\dif^A\tilde{\phi}_{AB_2B_3B_1}+\dif_{B_2}\dif_{B_1}\dif^A\tilde{\phi}_{AB_3}-\nn\\&&-\dif_{B_2}\dif^A\dif^B\tilde{\phi}_{ABB_1B_3}-\dif_{B_1}\dif^A\dif^B\tilde{\phi}_{ABB_2B_3}\bigr)\biggr).
\e{c7}
Then I have
\be
&\tilde{F}'''_{B_3B_4}(y)&\equiv\eta^{A_3A_4}\tilde{F}''_{A_3B_3A_4B_4}(y)=\nn\\&&=\tilde{F}'''_{B_3B_4}(y)|_0+\tilde{F}'''_{B_3B_4}(y)|_y,
\e{c8}
where
\be
&\tilde{F}'''_{B_3B_4}(y)|_0=&2\biggl((2d+1)\Box\tilde{\phi}_{B_3B_4}-\nn\\&&-(d-2)\dif^A\dif^B\tilde{\phi}_{ABB_3B_4}+(d+3)\bigl(\dif_{B_3}\dif_{B_4}\tilde{\phi}-\nn\\&&-\dif_{B_3}\dif^A\tilde{\phi}_{AB_4}-\dif_{B_4}\dif^A\tilde{\phi}_{AB_3}\bigr)+\nn\\&&+
\eta_{B_3B_4}\bigl(\Box\tilde{\phi}-\dif^A\dif^B\tilde{\phi}_{AB}\bigr)\biggr),
\nn\\
&\tilde{F}'''_{B_3B_4}(y)|_y=&2\biggl(\bigl(y_{B_3}\dif_{B_4}+y_{B_4}\dif_{B_3}\bigr)\bigl(\Box\tilde{\phi}-\dif^A\dif^B\tilde{\phi}_{AB}\bigr)-\nn\\
&&-y_{B_3}\bigl(\Box\dif^A\tilde{\phi}_{AB_4}-\dif^A\dif^B\dif^C\tilde{\phi}_{ABCB_4}\bigr)-\nn\\
&&-y_{B_4}\bigl(\Box\dif^A\tilde{\phi}_{AB_3}-\dif^A\dif^B\dif^C\tilde{\phi}_{ABCB_3}\bigr)\biggr).
\e{c9}
And finally
\be
&&\tilde{U}(y)\equiv\eta^{B_3B_4}\tilde{F}'''_{B_3B_4}(y)=8(d+1)\bigl(\Box\tilde{\phi}-\dif^A\dif^B\tilde{\phi}_{AB}\bigr).
\e{c10}
In the last two formulas
\be
&&\tilde{\phi}\equiv\eta^{AB}\tilde{\phi}_{AB}=\eta^{AB}\eta^{CD}\tilde{\phi}_{ABCD}.
\e{c11}

By means of these expressions I find the following equalities which are relevant for the linear condition \rl{667}
\be
&&F^{A_1B_1A_2B_2A_3B_3A_4B_4}\tilde{F}_{A_1B_1A_2B_2A_3B_3A_4B_4}(y)=\nn\\&&\quad =F^{A_1B_1A_2B_2A_3B_3A_4B_4}\tilde{F}_{A_1B_1A_2B_2A_3B_3A_4B_4}(y)|_0=\nn\\&&\quad =48F'^{B_1B_2A_3B_3A_4B_4}(y)\dif_{[A_3}\dif_{[A_4}\tilde{\phi}_{B_4]B_3]B_2B_1},
\e{c12}
\be
&&F'^{B_1B_2A_3B_3A_4B_4}(y)\tilde{F}'_{B_1B_2A_3B_3A_4B_4}(y)=\nn\\&&\quad =F'^{B_1B_2A_3B_3A_4B_4}(y)\tilde{F}'_{B_1B_2A_3B_3A_4B_4}(y)|_0=\nn\\&&\quad =2(d+2)F'^{B_1B_2A_3B_3A_4B_4}(y)\dif_{[A_3}\dif_{[A_4}\tilde{\phi}_{B_4]B_3]B_2B_1}+\nn\\
&&\quad +12F''^{A_3B_3A_4B_4}(y)\dif_{[A_3}\dif_{[A_4}\tilde{\phi}_{B_4]B_3]}+\nn\\
&&\quad +8G''^{B_1B_2B_3B_4}(y)\bigl(\Box\tilde{\phi}_{B_1B_2B_3B_4}+\dif_{B_1}\dif_{B_2}\tilde{\phi}_{B_3B_4}-\nn\\&&\quad -\dif_{B_1}\dif^A\tilde{\phi}_{AB_2B_3B_4}-\dif_{B_2}\dif^A\tilde{\phi}_{AB_1B_3B_4}\bigr),
\e{c13}
\be
&&F''^{A_3B_3A_4B_4}(y)\tilde{F}''_{A_3B_3A_4B_4}(y)=\nn\\&&=F''^{A_3B_3A_4B_4}(y)\tilde{F}''_{A_3B_3A_4B_4}(y)|_0=\nn\\&&=2(d+3)F''^{A_3B_3A_4B_4}(y)\dif_{[A_3}\dif_{[A_4}\tilde{\phi}_{B_4]B_3]}+\nn\\&&+8F'''^{B_3B_4}\bigl(\Box\tilde{\phi}_{B_3B_4}-\dif^A\dif^B\tilde{\phi}_{ABB_3B_4}\bigr),
\e{c14}
\be
&&G''^{B_1B_2B_3B_4}(y)\tilde{G}''_{B_1B_2B_3B_4}(y)=\nn\\&&=G''^{B_1B_2B_3B_4}(y)\tilde{G}''_{B_1B_2B_3B_4}(y)|_0=\nn\\&&=4dG''^{B_1B_2B_3B_4}(y)\bigl(\Box\tilde{\phi}_{B_1B_2B_3B_4}+\dif_{B_1}\dif_{B_2}\tilde{\phi}_{B_3B_4}-\nn\\ &&-\dif_{B_1}\dif^A\tilde{\phi}_{AB_2B_3B_4}-\dif_{B_2}\dif^A\tilde{\phi}_{AB_1B_3B_4}\bigr)+\nn\\ &&+2F'^{B_2B_3B_1B_4}(y)\dif_{[B_1}\dif_{[B_2}\tilde{\phi}_{B_3]B_4]}+\nn\\&&+16F'''^{B_3B_4}(y)\bigl(\Box\tilde{\phi}_{B_3B_4}+\dif_{B_3}\dif_{B_4}\tilde{\phi}-\dif_{B_3}\dif^A\tilde{\phi}_{AB_4}-\dif_{B_4}\dif^A\tilde{\phi}_{AB_3}\bigr)-\nn\\&&-8F'''^{B_2B_4}(y)\bigl(\Box\tilde{\phi}_{B_2B_4}-\dif^A\dif^B\tilde{\phi}_{ABB_2B_4}\bigr),
\e{c15}
\be
&&F'''^{B_3B_4}(y)\tilde{F}'''_{B_3B_4}(y)=F'''^{B_3B_4}(y)\tilde{F}'''_{B_3B_4}(y)|_0=\nn\\&&=2F'''^{B_3B_4}(y)\biggl((d-2)\bigl(\Box\tilde{\phi}_{B_3B_4}-\dif^A\dif^B\tilde{\phi}_{ABB_3B_4}\bigr)+\nn\\&&+(d+3)\bigl(\Box\tilde{\phi}_{B_3B_4}+\dif_{B_3}\dif_{B_4}\tilde{\phi}-\dif_{B_3}\dif^A\tilde{\phi}_{AB_4}-\dif_{B_4}\dif^A\tilde{\phi}_{AB_3}\bigr)+\nn\\&&+2U(y)\bigl(\Box\tilde{\phi}-\dif^A\dif^B\tilde{\phi}_{AB}\bigr),
\e{c16}
\be
&&U(y)\tilde{U}(y)=8(d+1)U(y)\bigl(\Box\tilde{\phi}-\dif^A\dif^B\tilde{\phi}_{AB}\bigr).
\e{c17}

\end{appendix}
\bibliographystyle{utphysmod2}
\bibliography{biblio}
\end{document}